\documentclass[11pt]{article}
\pdfoutput=1
\usepackage[centertags]{amsmath}
\usepackage[square, comma, sort&compress,numbers]{natbib}
\usepackage{array,multirow}
\numberwithin{equation}{section}
\usepackage{amssymb,amsfonts}
\usepackage{graphicx}
\usepackage{color}
\usepackage{mathtools,bm}
\usepackage{accents}

\usepackage{epsfig}
\usepackage{bbold}

\usepackage{wrapfig}
\usepackage{float}
\usepackage{soul}
\usepackage{tkz-euclide}
\usepackage{braket}
\usepackage{tikz,pgf}
\usetikzlibrary{shapes}
\usetikzlibrary{calc}
\usetikzlibrary{decorations.pathmorphing}
\usetikzlibrary{decorations.pathreplacing,shapes.misc}
\usetikzlibrary{positioning}
\usetikzlibrary{arrows}
\usetikzlibrary{decorations.markings}
\usetikzlibrary{shadings}

\usetikzlibrary{intersections}

\def\be{\begin{equation}}
\def\ee{\end{equation}}
\def\ba{\begin{array}}
\def\ea{\end{array}}

\def\dps{\displaystyle}
\newcommand{\half}{\frac{1}{2}}

\def\a{\tilde{1}}

\def\1{\tilde{1}}
\def\2{\tilde{2}}
\def\3{\tilde{3}}

\def\F{\;_2F_1}

\newdimen\tableauside\tableauside=1.0ex
\newdimen\tableaurule\tableaurule=0.4pt
\newdimen\tableaustep
\def\phantomhrule#1{\hbox{\vbox to0pt{\hrule height\tableaurule
width#1\vss}}}
\def\phantomvrule#1{\vbox{\hbox to0pt{\vrule width\tableaurule
height#1\hss}}}
\def\sqr{\vbox{%
\phantomhrule\tableaustep

\hbox{\phantomvrule\tableaustep\kern\tableaustep\phantomvrule\tableaustep}%
\hbox{\vbox{\phantomhrule\tableauside}\kern-\tableaurule}}}
\def\squares#1{\hbox{\count0=#1\noindent\loop\sqr
\advance\count0 by-1 \ifnum\count0>0\repeat}}
\def\tableau#1{\vcenter{\offinterlineskip
\tableaustep=\tableauside\advance\tableaustep by-\tableaurule
\kern\normallineskip\hbox
{\kern\normallineskip\vbox
{\gettableau#1 0 }%
\kern\normallineskip\kern\tableaurule}%
\kern\normallineskip\kern\tableaurule}}
\def\gettableau#1 {\ifnum#1=0\let\next=\null\else
\squares{#1}\let\next=\gettableau\fi\next}

\newtheorem{prop}{Proposition}

\newcommand{\bref}[1]{\textbf{\ref{#1}}}

\newcommand{\im}{\mathop{\mathrm{Im}}}
\newcommand{\re}{\mathop{\mathrm{Re}}}

\def\cD{\mathcal{D}}

\def\cF{\mathcal{F}}

\def\cJ{\mathcal{J}}

\def\cL{\mathcal{L}}
\def\cM{\mathcal{M}}

\def\cO{\mathcal{O}}

\def\cR{\mathcal{R}}
\def\cS{\mathcal{S}}

\def\cU{\mathcal{U}}
\def\cV{\mathcal{V}}

\numberwithin{equation}{section} \makeatletter
\@addtoreset{equation}{section}

\def\ads{AdS$_2\;$}
\def\be{\begin{equation}}
\def\ee{\end{equation}}
\def\ba{\begin{array}}
\def\ea{\end{array}}

\def\dps{\displaystyle}

\def\ba{\begin{array}}
\def\ea{\end{array}}

\def\dps{\displaystyle}

\def\l{\lambda}

\def\cft{CFT$_1$ }
\def\adscft{AdS$_2$/CFT$_1$ }

\def\bx{{\bf x}}

\newcommand{\z}[2]{z_{_{#1\,#2}}}

\def\C2{\text{C}_2}

\def\pref{C}
\def\sl2{sl(2,\mathbb{R})}
\def\wil{\widehat{W}}
\def\sJ{J}
\def\stJ{\tilde{J}}

\usepackage{jheppub}
\makeatletter
\def\@fpheader{\vspace{-.1cm}}
\makeatother

\title{\centering{Wilson networks in AdS and global conformal blocks}}

\author{Konstantin\ Alkalaev,}
\author{Andrey\ Kanoda,}
\author{Vladimir\ Khiteev}

\affiliation{I.E. Tamm Department of Theoretical Physics, \\
P.N. Lebedev Physical Institute, Leninsky ave. 53, 119991 Moscow, Russia}

\emailAdd{alkalaev@lpi.ru}
\emailAdd{kanoda.ao@phystech.edu}
\emailAdd{khiteev@lpi.ru}

\abstract{We develop the relation between gravitational  Wilson line networks, defined as a particular product of Wilson line operators averaged over the cap states, and conformal correlators in the context of the AdS$_2$/CFT$_1$ correspondence. The $n$-point $sl(2, \mathbb{R})$ comb channel global  conformal  block in CFT$_1$ is explicitly calculated by means of the extrapolate dictionary relation from the gravitational Wilson line network with $n$ boundary endpoints stretched in AdS$_2$. Remarkably, the Wilson line calculation directly yields  the conformal block in a particularly simple form which up to the leg factor  is given by  the comb function of cross-ratios. It is also found that the comb channel structure constants are expressed in terms of factorials and  triangle functions of conformal weights whose form determines fusion rules for a given 3-valent vertex. We obtain analytic expressions for the Wilson line matrix elements in AdS$_2$ which are building blocks of the Wilson line networks. We analyze general cap states and specify  those which lead to asymptotic values of the Wilson line networks interpreted as boundary correlators of CFT$_1$ primary operators.   The cases of (in)finite-dimensional $sl(2, \mathbb{R})$ modules carried by  Wilson lines are treated on equal footing that boils down to consideration of   singular submodules and their  contributions to the Wilson line matrix elements.}

\begin{document}

\maketitle
\flushbottom

\section{Introduction}

The relation between CFT$_2$ conformal blocks and  wave functions in $3d$  Chern-Simons theory is known since \cite{Witten:1988hf,Verlinde:1989ua,Labastida:1989wt}. In the context of the AdS$_3$/CFT$_2$ correspondence this relation can be rephrased as that of particular matrix elements of Wilson  networks stretched in vacuum AdS$_3$ with boundary endpoints which calculate conformal blocks  \cite{deBoer:2013vca,Ammon:2013hba,deBoer:2014sna,Hegde:2015dqh,Bhatta:2016hpz,Besken:2016ooo,Besken:2017fsj,Hikida:2017ehf,Hikida:2018eih,Hikida:2018dxe,Besken:2018zro,Bhatta:2018gjb,DHoker:2019clx,Castro:2018srf,Kraus:2018zrn,Hulik:2018dpl,Hung:2018mcn,Castro:2020smu,Alkalaev:2020yvq,Belavin:2022bib,Belavin:2023orw}.\footnote{In a wider context, the correspondence between  conformal blocks on the boundary and geodesic (Witten) diagrams in the bulk was extensively  studied in \cite{Hartman:2013mia,Fitzpatrick:2014vua,Hijano:2015rla,Fitzpatrick:2015zha,Alkalaev:2015wia,Hijano:2015qja,Hijano:2015zsa,Alkalaev:2015lca,Alkalaev:2015fbw,Banerjee:2016qca,Anous:2016kss,Alkalaev:2016rjl,Chen:2016cms,Chen:2017yze,Belavin:2017atm,Kraus:2017ezw,Alkalaev:2017bzx,Gobeil:2018fzy,Hijano:2018nhq,Alkalaev:2018qaz,Chen:2018qzm,Alkalaev:2018nik,Parikh:2019ygo,Anous:2019yku,Alkalaev:2019zhs,Chen:2019hdv,Jepsen:2019svc,Alkalaev:2020kxz,Anous:2020vtw,RamosCabezas:2020mew,Pavlov:2021lca}.}  Due to the (anti)holomorphic factorization of  $3d$ gravitational Wilson lines and CFT$_2$ correlation functions this prescription can also be naturally formulated in  the  \adscft case. 

These lower-dimensional cases of the AdS/CFT correspondence crucially rely on that the gravitational $o(d,2)$-connection  $A$  ($d=1,2$) used to build the Wilson line operators satisfies the zero-curvature condition $F(A) = dA + A\wedge A = 0$. The latter is the equation of motion in the lower-dimensional (topological) gravities in the Chern-Simons and BF formulations. Assuming $F(A)=0$ one can also extend the Wilson network/conformal block correspondence to higher dimensions \cite{Bhatta:2018gjb}.

In this paper, we restrict ourselves to the \adscft case and further develop the Wilson \ads network approach. The main object of our study is an AdS vertex function which is defined  as the Wilson matrix element of a particular combination of the Wilson line operators organized into a network  in the bulk with  endpoints which are not necessarily on the boundary. The network has the form of the comb channel diagram.  The combined operator is averaged over the  cap states belonging to $\sl2$ highest-weight (HW) or lowest weight (LW) (in)finite-dimensional modules. Imposing the $\sl2$ invariance condition on  AdS vertex functions generally fixes the cap states to be the Ishibashi states. However, depending of particular values of weights which correspond to either irreducible or reducible and to either finite- or infinite-dimensional modules the cap states can be different. We list all admissible caps states  and study their properties, all within a single framework. In particular, that allowed us to clarify the relation between the Wilson line construction in the finite-dimensional case elaborated previously in  \cite{Besken:2017fsj,Alkalaev:2020yvq,Belavin:2022bib,Belavin:2023orw} and in the infinite-dimensional case  \cite{Bhatta:2016hpz,Bhatta:2018gjb}.

In order to relate  AdS vertex functions in the bulk with  CFT correlation functions and, in particular, with conformal blocks on the boundary  we use the extrapolate dictionary  \cite{Balasubramanian:1998sn,Balasubramanian:1998de,Banks:1998dd,Harlow:2011ke}. As such the extrapolate relation  defines a CFT correlation function as the near-boundary asymptotics of a given AdS vertex function. The respective conformal block arises after stripping off the structure constants which are calculated to be particular functions of weights. Instead of calculating   the conformal blocks directly as the asymptotic AdS vertex functions we show that near the boundary the AdS vertex functions satisfy a recursion relation which expresses  an $n$-point function through  $(n-1)$-point functions. This recursion relation can be explicitly solved in terms of the 4-point functions. In this way, we reproduce the recursion relation for conformal blocks in the comb channel originally observed in purely CFT terms within the shadow formalism \cite{Rosenhaus:2018zqn}. In short,  our asymptotic recursion relation is a mere reflection of that the Wilson network operator in the comb channel is the matrix product and adding two more legs to the $(n-1)$-point comb diagram is realized by multiplying the respective Wilson matrix element by some typical matrix that extends it to the $n$-point comb diagram.   

We stress  that the extrapolate dictionary gives the asymptotic AdS vertex functions as the product of the conformal block  and the combination of the structure constants  which characterizes the comb channel. These structure constants are expressed in terms of factorials and the (modified) triangle function of the conformal weights (external and intermediate). It is the triangle function which defines the fusion rules arising as the triangle identities coming from the respective Clebsch-Gordan series associated to each 3-valent vertex of the Wilson line network. Of course, the resulting CFT correlation function is given only for particular intermediate operators since the AdS vertex function is defined with respect to particular exchange lines (no summation over intermediate operators).

In this way we find the  global \cft conformal blocks in the comb channel: (1) with any number $n$  of primary operators, (2) which are in (in)finite-dimensional modules of $\sl2\subset Vir$. Up to date, only lower-point Wilson network functions and respective conformal  blocks with $n=2,3,4,5$ were considered in the literature  \cite{Bhatta:2016hpz,Besken:2016ooo}.\footnote{At $n=2,3$  the conformal blocks are trivial so that only their leg factors contribute which are in fact the $2$-point  and $3$-point conformal correlation functions.} The resulting $n$-point block function turns out to be  remarkably simple and is given by the product of the leg factor, which guarantees correct conformal transformation properties, and the so-called comb function of conformally-invariant arguments  (cross-ratios)  introduced in \cite{Rosenhaus:2018zqn} to represent $n$-point  global conformal blocks in the comb channel: it is given in a closed-form as a simple hypergeometric-type series of $n-3$ arguments.

It should be noted that the above discussion emphasizes  that one of our tasks in this paper is to understand the capabilities of the Wilson line formulation viewed as a tool for calculating  conformal blocks. The point is that a direct calculation of  near-boundary Wilson line networks expected to yield the conformal blocks requires a lot of non-trivial (re)summations just because the respective $n$-point AdS vertex functions involves $3n-3$ independent (infinite) summations  coming from each $3j$ symbol and the cap state as well as contractions between them, while the near-boundary expression must contain $n-3$ summations, one for each cross-ratio. In particular, the $5$-point near-boundary analysis performed  in \cite{Bhatta:2016hpz} reproduces the  $5$-point conformal block of the form computed  in \cite{Alkalaev:2015fbw} which can be considered unsatisfactory in light of later developments:  the comb function of \cite{Rosenhaus:2018zqn} in the $5$-point case is the second Appell function with many nice properties known in the calculus that are not directly  seen within the old representation.

The paper is organized as follows. Section \bref{sec:f_irreps} reviews the gravitational Wilson line network construction in AdS$_2$ spacetime and sets our notation and conventions for $\sl2$ modules described in the ladder basis. Here, an AdS vertex function is   introduced as the Wilson network  matrix element with $n$ endpoints.  Following the extrapolate dictionary, in Section \bref{sec:ward}   we impose the $\sl2$ invariance condition on the AdS vertex functions that boils down to the cap state condition which we solve for various types of $\sl2$ modules. Here, we formulate the Wilson network/conformal block correspondence. Also, we introduce a weaker cap state condition which correctly captures only the near-boundary behaviour of AdS vertex functions. By this we mean that the strong condition is that the AdS vertex functions satisfy the $\sl2$ Ward identities in the bulk and, as a consequence, the asymptotic AdS vertex functions satisfy the conformal Ward identities on the boundary, while the weak condition requires the Ward identities only asymptotically. In Section \bref{sec:matrix}, aiming to find conformal blocks we calculate Wilson line elements in a closed form and analyze their asymptotic behaviour for all admissible cap states. We show that the Wilson matrix elements are generally given by the hypergeometric functions ${}_2F_1$ of complex arguments. Here, in particular, in order to demonstrate various peculiarities of our construction  we consider an example of the 2-point AdS vertex functions and respective 2-point CFT correlators. In Section  \bref{sec:n-p} we derive  the recursion relation  satisfied by asymptotic AdS vertex functions and explicitly solve it with the base of recursion being the 4-point AdS vertex function.  In Section \bref{sec:conclusion} we summarize our results and discuss possible further developments. Appendix \bref{app:hyper} collects a few explicit formulas for  $3j$ symbols, as well as contains  various special functions and their properties. Appendix \bref{app:matrix} collects various  detailed calculations. In Appendix \bref{app:conf_transf} the $SL(2, \mathbb{R})$ conformal transformation of the asymptotic AdS vertex functions are derived. Appendix \bref{sec:3-5} considers the lower-point AdS vertex functions and CFT correlation functions.

\section{Wilson line networks}
\label{sec:f_irreps}

\subsection{AdS$_2$ gravitational Wilson lines}
\label{sec:gravity}

We  consider the zero-curvature condition $dA+ A\wedge A =0$ for $\sl2$ gauge connections $A=A(\rho,z)$ on the two-dimensional manifold $\cM_2$ with local coordinates  $x^{\mu} = (\rho, z)$, where $\rho, z \in \mathbb{R}$. The zero-curvature condition can be realized dynamically as the equation of motion following from the BF action 
\be
S_{BF}[A, B] = \frac{1}{2}\int_{\cM_2}Tr B F\,,
\ee
where  $Tr$ stands for the Killing invariant form, $B$ is a scalar  field, and $F = dA + A\wedge A$.  The action describes a two-dimensional dilaton gravity with a non-zero cosmological constant \cite{Fukuyama:1985gg,Chamseddine:1989wn,Isler:1989hq} which is  the Jackiw-Teitelboim model  \cite{Teitelboim:1983ux,Jackiw:1984je}. The second equation of motion $dB +A B=0$ is not considered here.

Introducing the $\sl2$ commutation relations  
\be
[J_n, J_{m}] = (n-m)J_{n+m} \quad  \text{with}  \quad n,m  = -1,0,1\;,
\ee 
the solution of the zero-curvature condition can be cast into the form
\cite{Banados:1994tn}
\be
\label{connection}
A = e^{-\rho J_{0}}J_1dze^{\rho J_{0}} + J_{0}d\rho\,.
\ee
The associated metric of the AdS$_2$ spacetime is given by 
\be 
\label{ads2}
ds^2 = e^{2\rho}dz^2+d\rho^2\,,
\ee
where the conformal boundary lies at $\rho = \infty$.\footnote{AdS$_2$ spacetime has two conformal boundaries but the Poincare coordinates used here cover only one of them.}

The Wilson line is defined as 
\be
\label{Wilson} 
W_j[L] = \mathbb{P}e^{-\int_{L}A}\,,
\ee
where $L$ is a path from  $x_1$ to $x_2$; $\mathbb{P}$ is the path ordering operator; the index $j$ means that $A$ takes values in an (in)finite-dimensional module $\mathcal{R}_{j}$ of $\sl2$ of weight $j$. The main properties of $W_j[L]$: 
\be
\label{ft1}
\ba{l}
\dps
1)\; A\xrightarrow{}gA \,g^{-1} + gdg^{-1}\,,
\;\;
W_j[L]\xrightarrow{}g(x_2)W_j[L]g^{- 1}(x_1)\quad\text{--}\;\text{a gauge transformation}; 
\vspace{2mm}
\\
\dps
2)\; W_{j}[L_1+L_2]=W_{j}[L_2]W_{j}[L_1]\quad\text{--}\;\text{a path transitivity}; 
\vspace{2mm}
\\
\dps
3)\; W_j[L]\; \text{associated with a flat connection  depends only on the endpoints of $L$}.
\ea
\ee

\noindent In order to calculate the Wilson line  associated to the  connection \eqref{connection} one can gauge transform $A$ with respect to the gauge element $g(x) = e^{\rho J_0}$ and then use the gauge transformation property \eqref{ft1}. One finds the following form of the Wilson operator 
\be
\label{w_line}
W_j[x_1,x_2]=e^{-\rho_2 J_0}e^{z_{_{12}} J_1}e^{\rho_1 J_0}\,, 
\qquad \text{where} \quad z_{_{ij}}=z_i-z_j\,,
\ee
which proves to be convenient in further calculations.

Note that the whole consideration can be straightforwardly extended to the AdS$_3$/CFT$_2$ case  by introducing (anti)holomorphic coordinates $z, \bar z$. The gauge algebra is then $\sl2\oplus \sl2$ and the Wilson line operators  \eqref{Wilson} are to be  supplemented by anti-chiral Wilson operators which are associated with modules $\cR_{\bar j}$  of the anti-chiral algebra $\sl2$. The (anti)chiral factorization underlines the conformal block factorization of the boundary CFT$_2$ and  (anti)holomorphic dimensions $(h,\bar h)$ are  expressed in terms of  (anti)chiral weights $(j, \bar j)$.

\subsection{Group-theoretic conventions} 
\label{sec:rep_int}

The Wilson lines  \eqref{w_line}  can be combined to form a network with 3-valent vertices, see  Fig. \bref{fig:wilson}. The respective $\sl2$ modules meeting in  vertices  are related by 3-valent intertwiners, 
\be 
\label{intertwiner}
I_{j_1 j_2 j_3}:\quad\mathcal{R}_{j_2}\otimes\mathcal{R}_{j_3}\xrightarrow{}\mathcal{R}_{j_1}\,,
\ee
which are  invariant tensors from $Inv(\cR^{*}_{j_1}\otimes \cR_{j_2}\otimes \cR_{j_3})$. The intertwiners have the obvious invariance property
\be
\label{invariance} 
I_{j_1 j_2 j_3}U_{j_2} U_{j_3}=U_{j_1} I_{j_1 j_2 j_3}\,,
\ee
where $U_j$ are $SL(2,\mathbb{R})$ operators of the corresponding representations.

We will be interested in two types of $\sl2$ modules $\cR_j$ carried by Wilson lines, finite- and infinite-dimensional. Below we describe them in the ladder basis.
\begin{itemize}

\item {\it Finite-dimensional series}  $\cR_j = \cD_j$ with weights $j\in \mathbb{N}_0/2$, $\dim \cD_{j} = 2j+1$. The standard  ladder basis is given by 
\be
\label{D_basis}
\{\cD_j \ni \ket{j,m}:\,J_0\ket{j,m} = m \ket{j,m},\, m = -j,-j+1, ..., j-1, j\}\,,
\ee
where the highest-weight (HW) vector $\ket{j,j}$ is defined by  
\be
\label{hw1}
J_0\ket{j,j} = j \ket{j,j}\;,
\qquad
J_{-1} \ket{j,j} = 0\;.  
\ee 

\item {\it Negative discrete HW series}  $\cR_j = \cD^{-}_j$ with weights  $j\in\mathbb{R}$, $\dim \cD_{j} = \infty$. If  $j\in \mathbb{N}_0/2 $ then the respective module contains a singular vector so that $\cD^{-}_{j}/\cS_{-j-1} \approx \cD_j$, where  $\cS_{-j-1} \subset \cD^{-}_{j}$ is the singular subspace. The basis is given by 
\be
\label{-D_basis}
\{\cD^-_j \ni  \ket{j,m}:\, J_0\ket{j,m} = m \ket{j,m},\, m = j, j-1, j-2, ..., -\infty\}\,,
\ee
where the HW  vector $\ket{j,j}$ is defined by \eqref{hw1} and $m$ is generally non-integer. 

\end{itemize}
In both types of modules $\cR_{j}$ the action of $\sl2$ algebra is defined as 
\be 
\ba{l}
\label{sl2_action}
J_0\ket{j,m} = m\ket{j,m},
\vspace{2mm}
\\
\dps
J_1\ket{j,m} =\sqrt{(m+j)(j-m+1)}\ket{j,m-1} \equiv M(j,m-1)\ket{j,m-1},
\vspace{2mm}
\\
\dps
J_{-1}\ket{j,m} = -\sqrt{(m+j+1)(j-m)}\ket{j,m+1}\equiv -M(j,m)\ket{j,m+1}.
\ea
\ee 
Note that the zeros of the  coefficient $M(j,m)$  define the passage from $\cD^-_j$ to $\cD_j$ since they correspond to singular vectors. 
 
One can also consider  positive discrete lowest-weight (LW) series $\cD^+_j$ but due to the isomorphism $\cD^-_j \approx \cD^+_j$ and the (gauge) freedom in changing $J_1 \leftrightarrow J_{-1}$ in the $\sl2$ connection \eqref{connection} and, hence, in the Wilson line operator  \eqref{w_line}, we can choose, for definiteness, only one of $\cD^{\pm}_j$. In this respect note that $\cD_j$ is simultaneously a LW/HW module.

Decomposing the product $\cR_{j_2}\otimes\cR_{j_3}$ into the Clebsch-Gordan series one can explicitly single out the intertwiner. In terms of the bases \eqref{D_basis} or \eqref{-D_basis} the Clebsch-Gordan coefficient takes the form    
\be
\label{CG1}
\dps
\ket{j_2,m}\otimes\ket{j_3,n}=\sum_{k}\Big (\bra{j_1,k}I_{j_1 j_2 j_3}\ket{j_2,m}\otimes \ket{j_3,n} \Big)\ket{j_1,k}\,,
\ee
where the summation domain depends on the type of modules $\cR_{j_i}$. Using \eqref{D_basis} we obtain summation domain for finite-dimensional modules $\cD_{j}$:
\be
\label{sum_dom}
\sJ:= [\![-j,j]\!]\;,
\ee
where the notation $[\![-a,a]\!]$ means that $k = -a,-a+1, ...\,, a-1, a$. Quite analogously,  in the case of infinite-dimensional modules $\cD^-_j$ using \eqref{-D_basis}  we introduce 
\be
\label{sum_dom_discrete}
\sJ^-:= [\![-\infty,j]\!]\;.
\ee
By construction, the intertwiner in matrix form is the $\sl2 \approx su(1,1)$ Clebsch-Gordan coefficient. It can be expressed in terms of the $3j$ symbol \eqref{3-j}: 
\be 
\label{def_int}
\ba{l}\dps
\bra{j_1,k}I_{j_1 j_2 j_3}\ket{j_2,m}\otimes \ket{j_3,n}\equiv [I_{j_1 j_2 j_3}]^{k}{}_{mn}=(-1)^{j_1-k}
\begin{pmatrix}
j_1&j_2&j_3\\
-k&m&n 
\end{pmatrix}.
\ea
\ee

\subsection{AdS vertex functions in the comb channel}
\label{sec:gravity}

\begin{figure}
\centering
\includegraphics[scale=1.1]{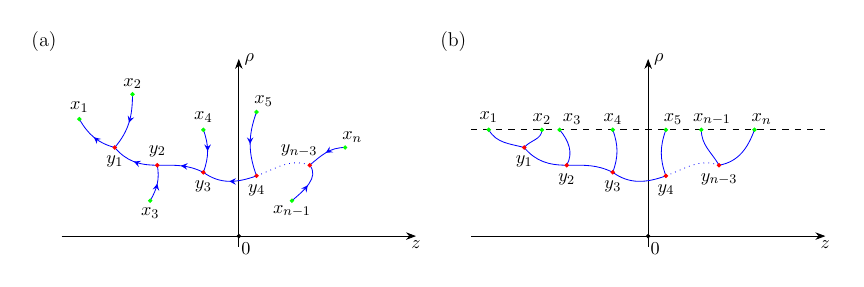}
\caption{The Wilson line network in the comb channel with arbitrary (a) and aligned (b) endpoints. Arrows indicate  orientations of  Wilson lines ($\cR_j$ and its dual $\cR^*_j$ have opposite orientations).}
\label{fig:wilson}
\end{figure}

The matrix element of the Wilson line network  can be directly  read off from the comb graph on Fig. \bref{fig:wilson} (a) by moving from left to right: 
\be
\label{mat_1}
\ba{c}
\left(W_{{j}_1}[y_1,x_1] I_{j_1 j_2 \tilde{j}_1} W_{\tilde{j}_1}[y_2, y_1] I_{\tilde{j}_1 j_3 \tilde{j}_2}\ldots W_{\tilde{j}_{{n-3}}}[y_{n-2}, y_{n-3}]I_{\tilde{j}_{{n-3}} j_{n-1} j_n}\right)
\vspace{2mm} 
\\ 
\dps
\;\times \Big(W_{{j}_2}[x_2,y_1] \ldots W_{{j}_{n-1}}[x_{n-1}, y_{n-2}]W_{{j}_n}[x_n, y_{n-2}] \Big),
\ea
\ee 
where the first line of operators describes inner edges connecting  vertices with one external edge and the second line stands for the external edges except for the first one. Note that the first Wilson operator $W_{{j}_1}[y_1,x_1]$ in the first line has a reversed order of points compared to other Wilson operators $W_{j_k}[x_k, y_{k-1}]$ in the second line. From  Fig. \bref{fig:wilson} it can be seen that the Wilson line $W_{{j}_1}[y_1,x_1]$ transfers a state from the point $y_1$ to the boundary point $x_1$ while $W_{j_k}[x_k, y_{k-1}]$ transfers a state from the boundary point $x_k$ to the vertex point $y_{k-1}$. 

The sets of endpoints and vertices   will be denoted, respectively, as 
\be
\label{coord_xy}
\ba{l}
{\bf x} = (x_1, ..., x_n)
\quad
\text{and}
\quad
{\bf y}  = (y_1, ..., y_{n-3})\;.
\ea
\ee
The endpoints ${\bf x}$ in the Wilson network  operator \eqref{mat_1} are so far  arbitrary. However, from the CFT perspective they are to be taken on the boundary. Since the conformal boundary lies  at $\rho = \infty$ it is convenient to place all  the endpoints on a hypersurface  of constant radial coordinate $\rho$, i.e. ${\bf x} = (\rho, {\bf z})$, where, eventually,  $\rho \to \infty$ (see Fig. \bref{fig:wilson} (b)), and the boundary coordinate set is 
\be
\label{bound_z}
{\bf z} = (z_1,..., z_{n})\,.
\ee  

\begin{figure}
\centering
\includegraphics[scale=1.1]{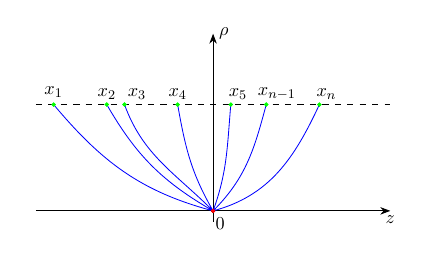}
\caption{The  Wilson line network can be equivalently represented as a graph with a single $n$-valent vertex (red dot) defined by the $n$-valent intertwiner. }
\label{fig:wilson_2}
\end{figure}

Let us now associate to  each endpoint  a particular cap state  $\ket{a_i}\in \cR_{j_i}$. Then, denoting the Wilson network operator  \eqref{mat_1} as $\wil^{\,j_1\ldots j_n}_{\tilde{j}_1\ldots\tilde{j}_{{n-3}}}({\bf x}, {\bf y})$ one can introduce  its matrix element as 
\be
\label{vert_f}
\cV_{j\tilde{j}}({\bf x}, {\bf y}) \equiv 
\bra{{a}_1} \wil^{\,j_1\ldots j_n}_{\tilde{j}_1\ldots\tilde{j}_{{n-3}}}({\bf x}, {\bf y}) \ket{{a}_2}\otimes \ket{{a}_3}\otimes \cdots\otimes \ket{{a}_n},
\ee 
which we call  an {\it AdS vertex function}. Using the intertwiner invariance property \eqref{invariance} one can directly show that the AdS vertex function is independent of positions of  the vertices $y_i \in {\bf y}$ that can be equivalently expressed by a convenient choice ${\bf y} = 0$ \cite{Bhatta:2016hpz,Besken:2016ooo,Alkalaev:2020yvq}. Since  the radial components of vertex points are zero, $\rho_i = 0$, then  it follows that any such point lies  deep inside the bulk and not on or near the boundary ($\rho \simeq \infty$), see Fig. \bref{fig:wilson_2}. Using that  ${\bf y} = 0$ we change notation for the AdS vertex function as 
\be
\cV_{j\tilde{j}}({\bf x}, {\bf y}) \to   \cV_{j\tilde{j}}(\rho,{\bf z})\,,
\ee
where ${\bf z}$ are the  boundary points \eqref{bound_z}, and $\rho$ labels the line which  will be finally pulled at (conformal) infinity, $\rho \to \infty$. 

By using the identity resolutions $\mathbb{1}=\sum_{m} \ket{j,m}\bra{j,m}$  the AdS vertex function \eqref{vert_f} can be represented as a matrix product
\be 
\label{vertex_func}
\ba{l}
\dps
\cV_{ j\tilde{j}}({\bf z},\rho)= \sum_{{\bf m},\, {\bf p}} \braket{j_1,m_1|I_{j_1 j_2 \tilde{j}_1}| j_2,m_2}\otimes\ket{\tilde{j}_1, p_1}\braket{\tilde{j}_1,p_1|I_{\tilde{j}_1 j_3 \tilde{j}_2}| j_3,m_3}\otimes\ket{\tilde{j}_2, p_2} \cdots \vspace{2mm}  \\\cdots\dps \braket{\tilde{j}_{n-3},p_{n-3}|I_{\tilde{j}_{n-3} j_{n-1} j_n}| j_{n-1},m_{n-1}}\otimes\ket{j_n,m_n}\Big(\braket{\tilde{a}_1|j_1, m_1}\braket{j_2, m_2|\tilde{a}_2}\cdots\braket{j_n, m_n|\tilde{a}_n}\Big),
\ea
\ee
where we introduced summations over  indices ${\bf m}=(m_1, ... , m_n)$ and ${\bf p}=(p_1, ... , p_{n-3})$ with  $m_i \in J_i$ and $p_k \in \tilde J_k$ in the finite-dimensional case \eqref{sum_dom}  or $m_i \in J^-_i$ and $p_k \in \tilde J^-_k$ in the  infinite-dimensional case \eqref{sum_dom_discrete},  as well as $x$-dependent cap states   
\be
\label{boundary}
\bra{\tilde{a}_1}=\bra{a_1} W_{j_1}[0,x_1]
\quad \text{and} \quad
\ket{\tilde{a}_i}=W_{j_i}[x_i,0]\ket{a_i}, 
\;\; 
i=2,..., n\,.
\ee  

Thus, the AdS vertex function is a contraction of the intertwiner matrix elements corresponding to the amputated comb diagram and particular  matrix elements of the Wilson operators  representing external legs. In fact, the amputated diagram is described by the $n$-valent intertwiner $I_{j_1 ... j_n| {\tilde j}_1 ... {\tilde j}_{n-3}}$ for external/internal modules $\cR_{j}$ and $\cR_{\tilde j}$ joined  into the comb diagram  so that the AdS vertex function takes the equivalent form \cite{Alkalaev:2020yvq}:
\be 
\label{vertex_func2}
\ba{l}
\dps
\cV_{ j\tilde{j}}({\bf z},\rho)= \bra{\tilde{a}_1} I_{j_1 ... j_n| {\tilde j}_1 ... {\tilde j}_{n-3}}\ket{\tilde{a}_2} \otimes \cdots \otimes \ket{\tilde{a}_n}\,, 
\ea
\ee
where  the $n$-valent intertwiner is given by
\be
I_{j_1 ... j_n| {\tilde j}_1 ... {\tilde j}_{n-3}} = I_{j_1j_2\tilde{j}_1} I_{\tilde j_1 j_3 \tilde j_2} \cdots  I_{\tilde j_{n-3} j_{n-1} j_{n}}\;,
\ee
see Fig. \bref{fig:wilson_2}. The generalized intertwiner invariance property directly follows from \eqref{invariance}:
\be
U_{j_1}I_{j_1 ... j_n| {\tilde j}_1 ... {\tilde j}_{n-3}} = I_{j_1 ... j_n| {\tilde j}_1 ... {\tilde j}_{n-3}}U_{j_2}\cdots U_{j_n}\;,
\ee
which can be written in the infinitesimal form as
\be 
\label{invariance_n}
(J_m)^{j_1}I_{j_1 ... j_n| {\tilde j}_1 ... {\tilde j}_{n-3}} = \sum_{i=2}^nI_{j_1 ... j_n| {\tilde j}_1 ... {\tilde j}_{n-3}}(J_m)^{j_i}\;,
\ee 
where $m=0,\pm 1$ and the superscript in $(J_m)^j$ indicates that $J_m$ is taken in $\cR_j$.

\section{Spacetime invariance and cap states}
\label{sec:ward}

The AdS/CFT correspondence is usually understood as the equality of AdS and CFT partition functions so that the correlation functions arise by  differentiating  the partition functions. On the other hand, a different way to state the correspondence is to extrapolate AdS correlation functions to the conformal boundary \cite{Balasubramanian:1998sn,Balasubramanian:1998de,Banks:1998dd,Harlow:2011ke}.      
E.g. for \ads scalar quantum fields $\hat \Phi_i(\rho_i, z_i)$ with masses $m_i$ the extrapolate  dictionary gives\footnote{Note that   the notion of spin in two dimensions is trivial and the only quantum number is given by a mass. It follows that all  \ads  (bosonic) elementary quantum fields  are exhausted by scalars with different masses. On the other hand, composite operators can have a (totally-symmetric) $o(1,1)$ tensor structure.}
\be
\label{extrapol}
\lim_{\rho \to \infty} e^{\rho \sum{\Delta_i}}\, \langle \hat \Phi_1(\rho, z_1) \cdots \hat \Phi_n(\rho, z_n) \rangle  = \langle \hat \cO_{1}(z_1) \cdots \hat \cO_{n}(z_n)\rangle \,,
\ee 
where conformal dimensions $\Delta_i$ of \cft primary operators $\hat \cO_i(z_i)$ are related to masses as $m_i^2 = \Delta_i(\Delta_i-1)$. Note that all \ads fields are placed on the hypersurface $\rho = const$ which eventually tends to the conformal boundary.

In our context the AdS vertex functions are assumed to reproduce CFT correlation functions in the way (see the relation  \eqref{vert_conf} below)  which is essentially the same as the extrapolate dictionary  relation \eqref{extrapol}. However, the AdS vertex functions are not literally AdS scalar correlation functions so  in order to draw a parallel between  the Wilson network/conformal block correspondence and the extrapolate dictionary the AdS vertex functions must be subject to particular spacetime symmetry criteria  that mimic those satisfied by AdS scalar correlation functions.

\subsection{Symmetry condition}  
\label{sec:sym_c}

We require  the AdS vertex functions to be invariant with respect to AdS$_2$ spacetime isometry  transformations:
\be
\label{st_inv}
\cV_{ j\tilde{j}}(\bx') = \cV_{ j\tilde{j}}(\bx)\;,
\qquad
\bx' = \bx'(\bx)\in SL(2, \mathbb{R})\;.
\ee    
The  infinitesimal form of the symmetry condition is given by three Ward  identities  
\be
\label{WA_wo}
\sum_{i= 1}^{n} \cJ_{m}^{(i)}\, \cV_{ j\tilde{j}}(x_1,...\,, x_i, ...\,, x_n) = 0\;, 
\qquad
m = 0, \pm 1\;,
\ee
where $\cJ_{m} = \xi_m^\mu \partial_\mu$ are the Lie derivatives along the Killing vector fields $\xi_m(x)$ of the AdS$_2$ spacetime with the metric \eqref{ads2}, 
\be
\label{L_k}
\cJ_{-1} = \partial_z\;, 
\qquad
\cJ_{0} = z\partial_z-\partial_\rho\;, 
\qquad
\cJ_{1} = z^2\partial_z-2z\partial_\rho - e^{-2\rho}\partial_z\;, 
\ee 
the superscript $i$ indicates that the derivative is taken with respect to the $i$-th coordinate. 
For general values of the radial coordinates $\rho_i$, $i=1,...,n$, the system of PDEs \eqref{WA_wo} has $2n-3$ first integrals which parameterize the $({\bf z}, \rho)$-dependence of  AdS vertex functions. However, in order to comply  with   the extrapolate dictionary   relation \eqref{extrapol}  the AdS vertex functions are to be placed  on  $\rho = const$ hypersurface [see the discussion below \eqref{coord_xy}]. In this case,   the first integrals become dependent so that there remain just $n-1$ of them. Then, one can show that on the hypersurface the AdS vertex functions are parameterized as 
\be 
\label{general_dependence}
\cV_{j\tilde{j}}({\bf x})\Big|_{\rho_1=\,...\,=\rho_n=\rho}  =\cV_{ j\tilde{j}}({\bf z},\rho) =  \cV_{ j\tilde{j}}(q_{12},...\,, q_{n-1,n})\,,
\ee
where 
\be 
\label{integrals_const}
q_{i, i+1}=(z_{i+1} - z_i)e^{\rho}\,, \qquad i = 1,...\,, n-1\,.
\ee 
This particular dependence will be explicitly  seen later when calculating the Wilson matrix elements in Section \bref{sec:matrix}.

The Ward identities \eqref{WA_wo} uniquely fix the form of the cap states $\ket{a}$ used to build the AdS vertex functions. Below we show that the \ads isometry condition boils down to the following condition imposed on the ket cap states:
\be
\label{cap_rel}
(J_1+J_{-1})\ket{a} = 0\;.
\ee 
The conjugated cap state $\bra{a}$ satisfies the same equation since $(J_{\pm})^\dagger \leftrightarrow J_{\mp}$. In fact, this condition defines  the (twisted) Ishibashi state \cite{Ishibashi:1988kg}.\footnote{The (twisted) Ishibashi states were previously used in this context in \cite{Bhatta:2016hpz,Castro:2018srf,Bhatta:2018gjb}.}  Below we list various solutions to the cap state condition depending on which particular module $\cR_j$ was chosen.

\begin{itemize}

\item In the case $j\notin\mathbb{N}_0/2$ there is a unique (up to a  normalization) vector  $\ket{a} \equiv  |j\rangle\!\rangle \in \cD^-_{j}$ \cite{Nakayama:2015mva}: 
\be 
\label{cap_states}
|j \rangle\!\rangle = \sum_{n=0}^\infty \prod_{k=1}^n \frac{(-)^n}{-4k j +4k^2 -2k} \, (J_1)^{2n}\ket{j,j}.
\ee 

\item The case  $j\in\mathbb{N}_0$ is to be considered separately because  $\cD^-_j$ contains  a  singular vector $\ket{j,-j-1}\in\cD^-_{-j-1}\subset \cD^-_j$ that  additionally generates a new solution to the cap state equation  \eqref{cap_rel}.  In other words, the kernel of $J_1+J_{-1}$ becomes two-dimensional. The cap state is $|j \rangle\!\rangle = \alpha |j \rangle\!\rangle_1+ \beta|j \rangle\!\rangle_2$, where $\alpha, \beta \in \mathbb{R}$ and the two basis cap states read   
\be 
\label{integer_sol1}
|j \rangle\!\rangle_1 = \sum_{n=0}^{j} \prod_{k=1}^n \frac{(-)^n}{-4k j +4k^2 -2k} \, (J_1)^{2n}\ket{j,j}\,, 
\ee 
\be 
\label{integer_sol2}
\hspace{7mm}|j \rangle\!\rangle_2 = \sum_{n=0}^\infty \prod_{k=1}^n \frac{(-)^n}{4k j +4k^2 +2k} \, (J_1)^{2n}\ket{j,-j-1}\,.
\ee 
The first basis cap state has finitely many terms since  $(J_1)^{2n}$ acts trivially on the HW vector $\ket{j,j}$ at $n > j$. It belongs to a finite-dimensional module,  $|j \rangle\!\rangle_1\in \cD^-_j/\cD^-_{-j-1}\approx \cD_j$. The second basis cap state  $|j \rangle\!\rangle_2 \in \cD^-_{-j-1}\subset \cD^-_j$ and,  therefore, it can be obtained from \eqref{cap_states} by $j\to -j-1$.  

\item In the case $j\in\mathbb{N}_0+\half$  the only solution is given by \eqref{integer_sol2} (see Appendix \bref{app:cap}).

\item The case of finite-dimensional modules $\cD_j$ with $j\in\mathbb{N}_0$ directly follows from the previous analysis.  The calculation here is almost the same as that for $\cD^-_j$ with $j\in\mathbb{N}_0$ and the resulting cap state is given by 
\be
\label{cs_fin}
|j \rangle\!\rangle = |j \rangle\!\rangle_1 \,.
\ee  
In the case $j \in \mathbb{N}_0 + \half$ the cap state condition has no solutions. 

\end{itemize}

An equivalent way to specify the cap states is to use the approach of Nakayama and Ooguri  \cite{Nakayama:2015mva,Nakayama:2016xvw}  which invokes the symmetry argument to localize CFT operators in the dual AdS spacetime thereby guaranteeing   the extrapolate dictionary  relation \eqref{extrapol}  (see also earlier works \cite{Miyaji:2015fia,Verlinde:2015qfa}). To provide a link between the  Nakayama-Ooguri construction  and the present Wilson line construction one introduces an \ads state           
\be 
\label{cap_and_bulk}
\hat{\Phi}(0,0)\ket{0} = |j \rangle\!\rangle\,,
\quad
\text{where $\ket{0}$ is a vacuum state}\,.
\ee 
According to the Nakayama-Ooguri construction this state satisfies the same condition \eqref{cap_rel} which now follows from adjusting AdS and CFT isometries (by the AdS/CFT correspondence one assumes that  spaces of states   of AdS and CFT theories are isomorphic).  Then, shifting this state to any point in \ads one finds one-particle state $\hat{\Phi}(\rho,z)\ket{0}$ in the space of states of the scalar theory.  This is the \ads wave function satisfying the Klein-Gordon equation. Recall that the one-particle states span an infinite-dimensional space isomorphic to $\cD^-_j$, where the weight $j$ defines the mass  $m^2 = j(j+1)$. From this perspective, the Wilson state \eqref{boundary}  
\be
\ket{\tilde{a}}=W_j[(\rho,z),0]|j \rangle\!\rangle\,,
\ee    
can  be viewed as a wave function realizing one-particle states in the \ads massive scalar theory.\footnote{Basically,  this is the $SL(2, \mathbb{R}) \approx SU(1,1)$ (non-compact, gravitational) version  of  wave functions in  $SU(2)$ (compact, non-gravitational) Chern-Simons theory   and their connection to conformal blocks \cite{Witten:1988hf,Verlinde:1989ua,Labastida:1989wt}.} The $x$-dependent vector $\ket{\tilde{a}}$ can be explicitly related to a local scalar field in the bulk  by constructing the mode expansion in terms of projections of  $\ket{\tilde{a}}$ onto the basis vectors in the respective infinite-dimensional $\sl2$ module \cite{Castro:2018srf}. In particular, by construction, $\ket{\tilde{a}}$  belongs to $\cD^-_j$ and satisfies the \ads Klein-Gordon equation with the same mass term   $m^2 = j(j+1)$ \cite{Castro:2018srf,Bhatta:2018gjb}.  Thus, going to the multi-particle states one concludes that  the AdS vertex functions of the Wilson line networks realize scalar field correlators in \ads that, in particular,  justifies the symmetry condition \eqref{WA_wo}.

The above discussion results in the following  property satisfied by the  AdS vertex function: 
\be
\label{KGV}
\left(\Box_i - m_i^2\right)  \cV_{j \tilde j}(x_1,..., x_i, ..., x_n) = 0\,,
\qquad
i = 1,..., n\,,
\ee
where $\Box_i$ is the \ads d'Alembertian evaluated in $x_i$-coordinates and $m^2_ i = j_i(j_i+1)$. We assume that the boundary points are essentially distinct, $x_i \neq x_j$, so that there are no terms  $\sim \delta(x_i - x_j)$ on the right-hand side.   In this respect, the AdS vertex functions are  similar to Wightman functions for free scalar fields but $\cV_{j \tilde j}(\bx) \neq  0$ for odd $n$.     

\subsection{Bulk cap state condition} 

Let us now derive  the cap state condition \eqref{cap_rel}   from the symmetry condition \eqref{WA_wo}. To this end, one recalls that  the Wilson line operators   are covariantly constant with respect to the endpoints, i.e.
\be
\label{cov12}
\ba{c}
\dps
\frac{\partial}{\partial x^\mu} W_j[x,y] -  W_j[x,y]A_\mu(x) = 0\;,  
\vspace{2mm}
\\
\dps
\frac{\partial}{\partial y^\mu} W_j[x,y] +  A_\mu(y) W_j[x,y]  = 0\;.
\ea
\ee
Introducing  $\varepsilon_m(x) \equiv  \xi_m^\mu A_\mu \equiv  \varepsilon^{m}_{-1}(x) J_{-1}+ \varepsilon_{0}^{m}(x) J_{0}+\varepsilon_{1}^{m}(x) J_{1}$, these relations  for $W_j[x,0]$ and $W_j[0,x]$ used to build the  AdS vertex functions can be represented in terms of the Lie derivatives \eqref{L_k}  as 
\be
\label{WAK}
\ba{l}
\dps
\cJ_{m} W_j[x,0]  = +\sum_{l=0, \pm 1} \varepsilon^m_l(x)\, J_l W_j[x,0]\;,
\vspace{2mm}
\\
\dps
\cJ_{m} W_j[0,x] = -\sum_{l=0, \pm 1} \varepsilon^m_l(x)\, W_j[0,x] J_l\;, 
\ea
\ee
where 
\be
\label{epsilon_explicit}
\varepsilon^{-1}_l(x) = \delta_{1l}\;,
\quad
\varepsilon^{0}_l(x) = \delta_{0l}\;,
\quad 
\varepsilon^1_l(x) = -(e^{-2\rho}-z^2)\delta_{1l}-2z\delta_{0l}\;.
\ee 
Then, the  left-hand side  of \eqref{WA_wo}  takes the form:
\be 
\label{Ward_proof}
\ba{c}
\dps
\sum_{i= 1}^{n} \cJ_{m}^{(i)}\cV_{ j\tilde{j}}(x_1,..., x_i, ..., x_n) = -\sum_{l=0, \pm 1}\varepsilon^{m}_l(x_1)\bra{\tilde{a}_1} (J_l)^{j_1} I_{j_1 ... j_n| {\tilde j}_1 ... {\tilde j}_{n-3}}\ket{\tilde{a}_2} \otimes \cdots \otimes \ket{\tilde{a}_n}
\vspace{2mm}
\\
\dps
\hspace{20mm} +\sum_{i= 2}^{n}\sum_{l=0, \pm 1}\varepsilon^{m}_l(x_i)\bra{\tilde{a}_1} I_{j_1 ... j_n| {\tilde j}_1 ... {\tilde j}_{n-3}}\ket{\tilde{a}_2} \otimes \cdots\otimes (J_l)^{j_i}\ket{\tilde{a}_i}\otimes\cdots \otimes \ket{\tilde{a}_n},
\ea 
\ee 
where the AdS vertex function is taken in the form   \eqref{vertex_func2}. Here,  a superscript in $(J_l)^j$ indicates that $J_l$ are taken in $\cR_j$. For $m=0$ or $m=-1$ the coefficients $\varepsilon^m_l(x)$  are $x$-independent so that one can factor them out and use the generalized intertwiner invariance property \eqref{invariance_n} to show that the right-hand side of \eqref{Ward_proof} equals zero for any cap states. In the case $m=1$ the coefficients $\varepsilon^m_l(x)$ are $x$-dependent so the previous argument does not apply that means  possible constraints to be imposed on the cap states.  In fact, since variables $x_i$ are independent, the right-hand side of \eqref{Ward_proof} equals zero iff the following relation holds
\be 
\label{cond_off}
\cJ_{1} W_j[x,0]\ket{a}  = \Big(\sum_{l=0, \pm 1} \tilde{\varepsilon}_l\, (J_l)^j\Big)\,  W_j[x,0]\ket{a},
\ee 
where $\tilde{\varepsilon}_l$ are  constants so that the right-hand side is the product of some constant $\sl2$ element and the Wilson line operator. In the sequel, the superscript in $(J_l)^j$ is omitted as we will be considering  only $\cR_j$. Then, using the relation \eqref{WAK} for $m=1$ one finds that the left-hand side of the above condition is given by 
\be 
\cJ_{1} W_j[x,0]\ket{a}  = \left[-(e^{-2\rho}-z^2)J_1-2zJ_0\right] W_j[x,0]\ket{a}.
\ee 
Combining the last two relations one finds that the coefficients $\tilde{\varepsilon}$ and the cap state $\ket{a}$ must satisfy the constraint
\be 
 \sum_{l=0, \pm 1} \tilde{\varepsilon}_l\, J_l W_j[x,0]\ket{a}  = \left[-(e^{-2\rho}-z^2)J_1-2zJ_0\right] W_j[x,0]\ket{a}.
\ee 
To solve this  equation  one commutes $J_i$ and $W_j[x,0]$ to obtain:
\be 
\ba{c}
\dps
W_j[x,0]\left[z^2e^{\rho}\tilde{\varepsilon}_{-1}J_1+e^{-\rho}\tilde{\varepsilon}_{-1}J_{-1} -2z\tilde{\varepsilon}_{-1}J_0  +e^{\rho}\tilde{\varepsilon}_1\, J_1+\tilde{\varepsilon}_0\, J_0 -ze^{\rho}\tilde{\varepsilon}_0\, J_1\right]\ket{a}  
\vspace{2mm}
\\
\dps
\hspace{12mm}= W_j[x,0]\left[(z^2e^{\rho}-e^{-\rho})J_1-2zJ_0\right] \ket{a}.
\ea
\ee 
Then, acting  with the inverse Wilson operator $W_j[0,x]$ and equating the coefficients in front of  the same powers of $z$ and $e^\rho$ one finds that the relation \eqref{cond_off} is valid provided that   
\be
\tilde{\varepsilon}_{-1} = 1\;, \qquad 
\tilde{\varepsilon}_0 = \tilde{\varepsilon}_1 = 0\;, 
\qquad (J_1+J_{-1})\ket{a}=0\;.
\ee

To summarize, we  showed that the Ward identities for the AdS vertex functions  are guaranteed by the following relation  
\be 
\label{cond_off_g}
\left(\cJ_{m} - J_{-m}\right) \ket{\tilde a} \equiv \left(\cJ_{m} - J_{-m}\right)  W_j[x,0]\ket{a}  = 0\,,
\ee 
which becomes an identity provided the cap state satisfies \eqref{cap_rel}. Note that this relation  is valid  for both (in)finite-dimensional modules.

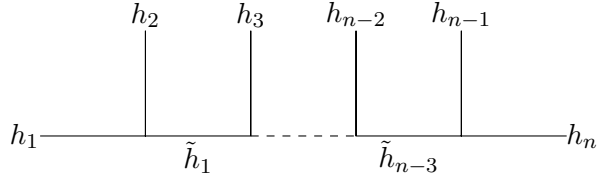
\begin{figure}
\centering
\begin{tikzpicture}[scale = 0.7]
{\draw (-2,0) -- (0,0) -- (0,2) -- (0,0) -- (2,0) -- (2,2) -- (2,0);
\draw[dashed] (2,0)-- (4,0);
\draw(4,0) -- (4,2) -- (4,0)-- (6,0)--(6,2)--(6,0)--(8,0) (-2.3, 0) node {$h_1$} (0, 2.3) node {$h_2$} (1, -0.4) node {$\tilde{h}_1$} (2, 2.3) node {$h_3$} (6, 2.3) node {$h_{n-1}$} (5, -0.4) node {$\tilde{h}_{n-3}$} (8.3, 0) node {$h_n$} (4, 2.3) node {$h_{n-2}$};}
\end{tikzpicture}
\caption{A diagram of the  $n$-point conformal block $\cF_{h \tilde{h}}({\bf z})$ in the comb channel. }
\label{fig:n}
\end{figure}

\subsection{$\sl2$ conformal blocks in the comb channel}
\label{sec:calc}

Invoking  the extrapolate dictionary relation \eqref{extrapol}  one can now explicitly relate  $n$-point global CFT$_1$ conformal blocks in the comb channel  which we denote by $\cF_{h \tilde{h}}({\bf z})$ (see Fig. \bref{fig:n}) to  the $n$-point AdS  vertex functions $\cV_{j \tilde{j}}({\bf z},\rho)$. To this end, one chooses the  cap states  $\ket{a_i}$ as in Section \bref{sec:sym_c} and relates conformal dimensions with weights as $h_ i = -j_i$ and $\tilde{h}_k = -\tilde{j}_k$.\footnote{Note that we do not impose any unitarity constraints so the conformal weights are arbitrary reals. Also, choosing  $\ket{a_i}$ as descendants of the cap states  yields correlation functions of secondary operators \cite{Alkalaev:2020yvq}.} Then, the exact relation reads 
\be
\label{vert_conf}
\lim_{\rho\to\infty}e^{-\rho\sum_{i=1}^nj_i}\,\cV_{j \tilde{j}}(\rho,{\bf z})= \pref_{j\tilde{j}}\,\cF_{h \tilde{h}}({\bf z})\,.
\ee
The normalization coefficients $\pref_{{j}\tilde{{j}}} \equiv \pref_{j_1... j_n\tilde{j}_1...\tilde{j}_{n-3}}$ are given by 
\be
\label{prefactor}
\ba{l}\dps
n=2:\quad \pref_{j_1j_2} = \frac{\delta_{j_1j_2}}{(2j_1+1)^{\half}} \,;  
\qquad 
n=3:\quad\pref_{j_1j_2j_3} = \left[\frac{(2j_1)!(2j_2)!(2j_3)!}{\Delta(j_1,j_2,j_3)}\right]^{\half};
\vspace{3mm}\\ \dps
n>3:\quad\pref_{{j}\tilde{{j}}} = \pref_{j_1j_2\tilde{j}_1}\Big[\prod_{i=1}^{n-4}\pref_{\tilde{j}_ij_{i+2}\tilde{j}_{i+1}}\Big]\pref_{\tilde{j}_{n-3}j_{n-1}j_n}\,,
\ea
\ee
where $\Delta(a,b,c) = (a+b+c+1)!(a+b-c)!(a+c-b)!(b+c-a)!$ is the modified triangle coefficient and we assume that for arbitrary real arguments the factorials are represented as $x! = \Gamma(x+1)$. Note that $C_{j_1j_2} = C_{j_1j_20}$ which means that $\pref_{j_1j_2j_3}$ defines all the normalization coefficients in \eqref{prefactor}. The pole in $\pref_{j_1j_2}$ at $j_1=-\half$  is an artefact of our choice of the normalization of the intertwiner \eqref{def_int}.\footnote{It can be eliminated  by a redefinition, but such a normalization turns out to be convenient in practice.}   

In order to have non-vanishing and real $\pref_{j_1j_2j_3}$ one restricts the weights  via triangle  inequalities:
\be
\label{triangle}
\ba{l}
\dps
1)\;\,j_1,j_2\in\mathbb{N}_0/2, j_3\in\mathbb{Z}:
\hspace{5mm}|j_1-j_2|\leq |j_3| \leq j_1+j_2\,;
\vspace{1mm} 
\\
2)\;\,\text{\hspace{0mm}in other cases}:
\hspace{15mm}j_3\leq j_1+j_2\,.
\ea
\ee 
In fact, these conditions derived here by analysing the singular points of factorials in \eqref{prefactor} completely reproduce the triangle identities coming from the Clebsch-Gordan series for $\cR_{j_1}\otimes \cR_{j_2}$ projected on $\cR_{j_3}$ depending on the choice of particular weights. Equivalently, they are encoded in the $3j$ symbol, see \eqref{3-j}.

The conformal blocks are normalized to contain the leg factors which  are particular scale prefactors depending on ${\bf z}$ that  ensure correct conformal transformation properties. With the leg factor stripped off the conformal block is a function of  cross-ratios only (the so-called bare block). On the other hand, the coefficients $\pref_{j_1j_2j_3}$ are in fact the structure constants of $3$-point correlation functions and, therefore,  $\pref_{j\tilde{j}}$ is a product of the structure constants which arise  when decomposing an $n$-point correlation function into conformal blocks in the comb channel. Thus, we conclude that the AdS Wilson line networks yield CFT correlation functions with specified structure constants expressed in terms of conformal weights \eqref{prefactor}.

\subsection{Boundary cap state conditions}
\label{sec:quasi}

Suppose now that we are not interested in interpreting the AdS vertex function as an independent and meaningful object in the bulk but only as an auxiliary function which can be used to calculate conformal blocks in the large-$\rho$ limit. It is clear then that one can choose other cap states which guarantee the Ward identities only near the boundary. Obviously,  they would satisfy some  conditions which are weaker than the cap state condition in the bulk \eqref{cap_rel}.  Indeed, requiring conformal symmetry of the AdS vertex function only at large-$\rho$  and using the arguments similar to those discussed  in the previous sections one finds a   system of asymptotic equations    
\be
\label{w_s1}
(zJ_1+J_0-j)\ket{\tilde a} =0+  \cO(e^{\rho(j-1)})\;, 
\ee
\be
\label{w_s2}
\bra{\tilde a'} (J_0 + j+zJ_1)=0+  \cO(e^{\rho(j-1)})\,, 
\ee
where the left-hand sides are given by $x$-dependent vectors in $\cR_j$ which are  sub-leading at large-$\rho$. Here,   $\ket{\tilde a} := W_j[x,0]\ket{a}$ and $\bra{\tilde a'} := \bra{a'}W_j[0,x]$, where $\ket{a}$ and $\bra{a'}$ are generally  two different vectors, $\bra{a'} \neq \ket{a}^\dagger$. Moreover, choosing $\bra{a_1}$ and $\ket{a_2}, \ket{a_3}, ...$ to be different solutions of \eqref{w_s1}-\eqref{w_s2} still guarantees the boundary Ward identities (recall that in the bulk, imposing the Ward identities  fixes the cap state (almost) uniquely).

The conditions \eqref{w_s1}-\eqref{w_s2} can be cast into the following form
\be
\label{wcsc1}
W_j[x,0](J_0 - j)\ket{a} =0+  \cO(e^{\rho(j-1)})\;, 
\ee
\be
\label{wcsc2}
\bra{a'}(J_0 + j) W_j[0,x] =0+  \cO(e^{\rho(j-1)})\,, 
\ee
which is more convenient for finding solutions. In the sequel, we call such  states as {\it quasi-Ishibashi cap states}. We are not going to describe a general solution to the system of asymptotic equations \eqref{wcsc1} and \eqref{wcsc2}, instead, below we list some simple partial solutions.      

\begin{itemize}

\item For infinite-dimensional modules $\cD^-_j$ the asymptotic equations  are solved by two (generally non-conjugated) bra and ket vectors
\be
\label{a_f}
\ba{l}
\ket{a} = \ket{j,j} + \left(J_0 - j\right)^{-1} \hat R  \ket{j,j}\,,
\vspace{2mm}
\\
\dps
\bra{a'} = \bra{j,j} \check R \left(J_0+j\right)^{-1}\,,
\ea
\ee  
where $\hat R, \check R \in \cU\left(\sl2\right)$ are some constant elements of  the universal enveloping algebra of $\sl2$. Their form is fixed by the large-$\rho$ behaviour read off from \eqref{wcsc1}-\eqref{wcsc2}.

Obviously, both the Ishibashi  states \eqref{cap_states} for $j\notin\mathbb{N}_0/2$ and \eqref{integer_sol1}-\eqref{integer_sol2} for $j\in\mathbb{N}_0$ are  partial solutions.  Yet another partial solution is given by the state  
\be 
\label{cap_states2}
|j \rangle\!\rangle = \sum_{n=0}^{\infty}(-1)^n\left[\frac{(2j)!}{(2j-n)!n!}\right]^{\half}\ket{j,j-n} \equiv e^{-J_1}\ket{j,j} \in \cD^-_j\,,
\ee 
where $j\in\mathbb{R}$ and the basis vectors $\ket{j,j-n}\in \cD^-_{j}$. In the last equality, this cap state is represented by $e^{-J_1}$ rotation of the HW vector for which reason we  name it as a {\it rotated HW state}. One may notice that it  satisfies the condition $\left(J_1-J_{-1}+2J_0\right)|j \rangle\!\rangle = 0$.

\item For finite-dimensional modules $\cD_j$ the  operator $\left(J_0+j\right)$ in  \eqref{wcsc2} has a kernel described by LW vectors (recall that $(\cD_j)^* \approx \cD_j$) so that  in this case the general solution can be represented as  
\be
\label{a_if}
\ba{l}
\ket{a} = \ket{j,j} + \left(J_0 - j\right)^{-1} \hat L \ket{j,j}\,,
\vspace{2mm}
\\
\dps
\bra{a'} = \bra{j,-j}+\bra{j,-j} \check L \left(J_0+j\right)^{-1}\,,
\ea
\ee  
with some new constant  $\hat L, \check L \in \cU\left(\sl2\right)$. One partial solution is given by \eqref{cs_fin}.  It immediately follows that the LW/HW  vectors proposed in \cite{Besken:2016ooo} as the cap states solve the above asymptotic equations (i.e.   $\hat L, \check L = 0$),
\be
\label{cap_states3}
|j \rangle\!\rangle  = \ket{j,j} \in \cD_j\,, 
\qquad j \in \mathbb{N}_0/2\,.
\ee 
\end{itemize}

We emphasize  that neither  \eqref{cap_states2} nor \eqref{cap_states3}  satisfy the cap state condition \eqref{cap_rel} in the bulk. That, in its turn, means that the respective AdS vertex function is not  $\sl2$ invariant  \eqref{st_inv}. However, if we aim at just  reproducing conformal blocks as the large-$\rho$ asymptotics then such a choice is admissible. This is not surprising since controlling  $\sl2$ invariance only near the boundary one can find more possible cap states. However, if one wants to reconstruct bulk correlators then the AdS vertex functions and the cap states are obliged to satisfy the formulated symmetry conditions. This is exactly in the spirit of the HKLL bulk reconstruction \cite{Hamilton:2005ju,Hamilton:2006az}. 

\begin{table}[t!]
\centering
\begin{tabular}{ |c|c|c|c|c|c| } 
\hline
 & $\cD^-_j$: $j\notin\mathbb{N}_0/2$ & $\cD^-_j$:  $j\in\mathbb{N}_0$ & $\cD^-_j$:  $j\in\mathbb{N}_0+\half$ & $\cD_j$:  $j\in\mathbb{N}_0$ & $\cD_j$:  $j\in\mathbb{N}_0+\half$  \\
\hline
\multirow{1}{4em}{bulk} & eq. \eqref{cap_states} & eqs. \eqref{integer_sol1}-\eqref{integer_sol2} & eq. \eqref{integer_sol2} & eq. \eqref{cs_fin} & $\varnothing$ \\ 

\hline
\multirow{1}{4em}{boundary} & eq. \eqref{cap_states2} & eq. \eqref{cap_states2} & eq. \eqref{cap_states2} & eq. \eqref{cap_states3} & \eqref{cap_states3}\\ 
\hline

\end{tabular}
\caption{The cap states. Here, the Ishibashi states are in the row ``bulk'', the quasi-Ishibashi states are in the row ``boundary''.}
\label{tab:cap}
\end{table}

For convenience, we collect the (quasi)-Ishibashi states in Table \bref{tab:cap}. We see that for particular values of weights there are no solutions to the cap state condition in the bulk \eqref{cap_rel}, though the cap state conditions near the boundary \eqref{wcsc1}-\eqref{wcsc2} always have non-trivial solutions. Presumably, the absence of solutions  means that the Ward identities imposed on the AdS vertex functions must be modified by adding spin parts. It means that the cap state condition is also modified by spin terms, for a discussion see \cite{Nakayama:2015mva,Bhatta:2018gjb}. We hope to consider this issue elsewhere.     

\section{Wilson line matrix elements and their asymptotics }
\label{sec:matrix}

In order to calculate CFT correlation functions according to the extrapolate dictionary relation  \eqref{vert_conf} one has to analyze the large-$\rho$ behaviour of the AdS vertex functions. Since a given  AdS vertex function is built  from the Wilson matrix elements then its large-$\rho$ dependence is completely defined by their asymptotic expansions. Knowing the large-$\rho$ asymptotics of the Wilson matrix elements  allows assembling  conformal correlation functions directly as their products with the respective $n$-valent intertwiner which guaranties an invariant contraction of indices in different representations.     

The key observation here is that    the  Wilson matrix elements $\bra{j,m|W_j[0,x]}j \rangle\!\rangle$ and $\langle\!\langle j|W_j[0,x] \ket{j,m}$ (we call them right and left  according to whether  the cap state is ket or bra) have different behaviour when approaching the boundary from the bulk. More precisely, the right elements are well defined near the boundary, while the left elements diverge. The Wilson matrix elements are built as infinite sums which is a natural consequence of that in the ladder basis the  cap state  is an infinite linear combination of  basis vectors, hence, this infinite summation is inherited in the Wilson matrix elements. In particular, one can see that the norm of the Ishibashi state is infinite which is yet another manifestation of the divergences occurring when considering  the Wilson line matrix elements and their contractions.\footnote{The Ishibashi state is given by a formal power series which  is a non-normalizable  coherent-type state in the respective infinite-dimensional $\sl2$ module. This is  in contrast with normalizable  states built by acting with polynomials of $J_{1}$ on the highest-weight vector.}

To cure the divergent behaviour of the left Wilson elements we find them in a closed form that  allows one to analytically continue in $\rho$ and finally expand near the boundary. Following the classification of the cap states in Table \bref{tab:cap} below we list analytical formulas for all  Wilson matrix elements $\braket{\tilde{a}|j,m} = \langle\!\langle j|W_j[0,x] \ket{j,m}$ and $\braket{j,m|\tilde{a}}  =\bra{j,m|W_j[0,x]}j \rangle\!\rangle$.   

\begin{prop} \label{prop3} Denote $q=-ze^{\rho}$. The left and right Wilson matrix elements are given by: 

\begin{itemize}

\item for the Ishibashi states in $\cD^-_j$ at $j\notin\mathbb{N}_0/2$ or in $\cD_j$ at $j\in\mathbb{N}_0$,
\be 
\label{closed_left0}
\ba{l}
\dps
\braket{\tilde{a}|j,m} = T_{jm}\, 
e^{-\rho m}\,(q+i)^{j}(q-i)^{m}\,{}_2F_1\left(-j, m-j; m+1\big|\frac{q-i}{q+i}\right),
\vspace{3mm}
\\
\dps
\braket{j,m|\tilde{a}} = (-)^{-m} \, T_{jm}\, e^{\rho m}\,(q+i)^{j-m}\,{}_2F_1\left(-j,m-j; m+1\big|\frac{q-i}{q+i}\right),
\ea
\ee 
where 
\be
T_{jm} = (-)^j \frac{j!}{m!}\left[\frac{(j+m)!}{(j-m)!(2j)!}\right]^\half;
\ee

\item for the Ishibashi states in $\cD^-_j$ at $j\in\mathbb{N}_0$,
\be 
\ba{l}
\dps
\braket{\tilde{a}|j,m} = 
\begin{cases} 
\braket{\tilde{a}|j,m} \;\text{eq. } \eqref{closed_left0},& m\geq-j\\
\braket{\tilde{a}|-j-1,m} \;\text{eq. }\eqref{closed_left0},& m<-j
\end{cases}\;,
\vspace{3mm}
\\
\dps
\braket{j,m|\tilde{a}} =
\begin{cases} 
\braket{j,m|\tilde{a}} \;\text{eq. }\eqref{closed_left0},&m\geq-j\\
\braket{-j-1,m|\tilde{a}}\;\text{eq. } \eqref{closed_left0},& m<-j
\end{cases}\;;
\ea
\ee

\item for the rotated LW/HW cap states in $\cD^-_j$,
\be 
\label{twisted_boundary_int0}
\ba{l}
\dps
\braket{\tilde{a}|j,m} = A_{jm}\,e^{-\rho m}(1+q)^{j+m}\,,
\vspace{3mm}
\\
\dps
\braket{j,m|\tilde{a}} = (-)^{j-m}A_{jm}\,e^{\rho m}(1+q)^{j-m}\,,
\ea
\ee
where  
\be
\label{coef_A}
A_{jm} = \left[\frac{(2j)!}{(j+m)!(j-m)!}\right]^\half\;;
\ee 

\item for the LW/HW states in $\cD_j$, 
\be 
\label{LHW_fin}
\ba{l}
\dps
\braket{\tilde{a}|j,m} = \,A_{jm}\,e^{-\rho m} q^{j+m}\,,
\vspace{2mm}
\\
\dps
\braket{j,m|\tilde{a}}   = (-)^{j-m}A_{jm}\,e^{\rho m} q^{j-m}\,.
\ea
\ee

\end{itemize}
\end{prop}     
The differences  of  $q$-s for $n$ points are in fact the first integrals of the AdS vertex functions \eqref{integrals_const}.  Now that we have exact expressions for the Wilson matrix elements we can analyze their leading large-$\rho$ asymptotics.  One  finds  out that regardless of what particular cap state (Ishibashi or quasi-Ishibashi) of one or another module (finite- or infinite-dimensional) is used the asymptotic Wilson matrix elements are the same (modulo signs):
\be
\label{LWMEf}
\hspace{0mm}\braket{\tilde{a}|j,m}  \approx  (-)^{\delta-j} A_{jm}\,  e^{-\rho m}q^{j+m} = (-)^{m+\delta} A_{jm}\,  z^{j+m}e^{\rho j}\,,
\ee
\be
\label{RWMEf}
\hspace{-14mm}\braket{j,m|\tilde{a}}  \approx (-)^{j-m} A_{jm}\,  e^{\rho m}q^{j-m} = A_{jm}\, z^{j-m}e^{\rho j}\,,
\ee
where the overall coefficient is given by \eqref{coef_A} 
and the parameter $\delta$ in the left Wilson matrix element distinguishes the two cases: (1) $\delta=2j$ when $\langle\!\langle j|$ is the Ishibashi states \eqref{cap_states} or \eqref{integer_sol1}-\eqref{integer_sol2};  (2) $\delta=j$ when $\langle\!\langle j|$ is the rotated LW \eqref{cap_states2} or $\delta=j$ when $\langle\!\langle j|$ is the LW vector. The symbol $\approx$ denotes keeping  leading large-$\rho$ contributions only. Also, in the case of LW/HW cap states  the subleading terms  are absent.

Let us now discuss  for which $j$ and $m$ the overall coefficient may have singular points. To this end, one notes that $A_{jm}$ can be  represented in terms of the coefficient function $M(j,m)$ of the ladder basis \eqref{sl2_action} as 
\be
\label{ajm}
A_{jm} = \frac{1}{(j-m)!}\prod_{n=0}^{j-m-1} M(j,j-n-1)\,.
\ee 
As one can see, for  $j\in\mathbb{N}_{0}/2$ and $m < -j$ one of the factors in the product becomes zero. Since the zeros correspond to singular vectors (see our comment below \eqref{sl2_action}) we conclude that the domain of $m$ is effectively restricted to $m\in[\![-j,j]\!]$ corresponding to a finite-dimensional $\cD_j$. On the other hand, $A_{jm}$ has no poles. It means that we can safely write the coefficient \eqref{coef_A} without specifying the domain of $m$ since the coefficient itself efficiently fixes this domain depending on  the weight $j$ of a given module  $\cR_j$. In other words,  the reason is that $A_{jm}$ is expressed in terms of the gamma-functions $z! = \Gamma(z+1)$ which effectively control a transition between finite- and infinite-dimensional cases: in one direction   -- through the analytic continuation of  the factorials to the  gamma-functions; in the opposite direction -- through the simple poles of the gamma-function.  

Thus, the formulae \eqref{LWMEf}, \eqref{RWMEf}, and \eqref{ajm} explicitly demonstrate the following:

\begin{prop} \label{prop2} The right and left Wilson matrix elements in $\cD_j$ can be  analytically continued to the leading asymptotics of the right and left Wilson matrix elements in $\cD^-_j$ (modulo overall signs) by extending weights  $j$ from integers to reals.
\end{prop}

\begin{prop} \label{prop1} For a given module $\cD^-_j$ or $\cD_j$ the left and right Wilson matrix elements with the corresponding  Ishibashi and quasi-Ishibashi cap states are {\it asymptotically equivalent}  (modulo overall signs). 
\end{prop}

In the next sections we calculate: (1) the Wilson matrix elements in $\cD^-_j$ with $j\notin \mathbb{N}_{0}/2$ (Sections \bref{sec:matrixR}-\bref{sec:matrixL}); (2) the Wilson matrix elements in $\cD^-_j$ with $j\in  \mathbb{N}_{0}$ (Sections \bref{sec:matrixI}); (3) the Wilson matrix elements in $\cD_j$ with $j\in  \mathbb{N}_{0}$ (the final comment in Section \bref{sec:matrixI}). The boundary consideration  of the Wilson matrix elements for the quasi-Ishibashi states of Section \bref{sec:quasi} is basically the same as that  for the true Ishibashi states  and, therefore, this analysis  is  completely relegated to Appendix \bref{app:analyt}.

\subsection{Right  Wilson matrix element}
\label{sec:matrixR}

In Appendix \bref{app:LWME} we  obtained the right Wilson matrix element in the form \eqref{RWE_fin}:
\be 
\label{RWE_fin_copy}
\ba{l}\dps
 \bra{j,m}W_j[x,0]| j \rangle\!\rangle  = \left[\frac{(2j)!(j-m)!j!^2}{(j+m)!}\right]^\half\,z^{j-m}e^{\rho j}
\vspace{2mm} 
\\
\dps
\times\sum_{n=0}^{j-m}\frac{i^n(-2j-1)!}{n!(-2j+n-1)!(j-n)!(j-m-n)!}\,{}_2F_1(-n, -j; j-n+1|-1)(-ze^{\rho})^{-n}\;.
\ea 
\ee 
This function is a polynomial in $q=-ze^{\rho}$ meaning that  its radius of convergence in $q$ is infinite.  Evaluating the sum in \eqref{RWE_fin_copy} we find that the right Wilson matrix element can be represented in a closed form \eqref{right_closed}: 
\be 
\ba{l}\dps
\label{right_closed1}
\bra{j,m}W_j[x,0]| j \rangle\!\rangle = (-)^{j-m}\frac{j!}{m!}\left[\frac{(j+m)!}{(2j)!(j-m)!}\right]^\half\,e^{\rho m}
(q+i)^{j-m}{}_2F_1\left(-j, m-j; m+1\big|\frac{q-i}{q+i}\right).
\ea 
\ee 
Expanding this function  near $|q|=\infty$ and substituting back $q=-ze^{\rho}$ one finds 
\be 
\label{bra_matrix_inf}
\bra{j,m|W_j[0,x]}j \rangle\!\rangle = \left[\frac{(2j)!}{(j+m)!(j-m)!}\right]^\half\,e^{\rho j} z^{j-m}+ \cO(e^{\rho(j-1)})\;,
\ee 
cf. \eqref{RWMEf}.

\subsection{Left  Wilson matrix element}
\label{sec:matrixL}

The left Wilson matrix element for the cap state \eqref{cap_states} in the form of the power series is  calculated in Appendix \bref{app:LWME}, where we found the expression \eqref{LWE_fin}:
\be
\label{LWE_fin2}
\ba{l}
\dps
\langle\!\langle j|W_j[0,x] \ket{j,m} = \left[\frac{(j+m)!j!^2}{(j-m)!(2j)!}\right]^\half(-z)^{m-j}e^{-\rho j}
\vspace{2mm} 
\\
\dps
\hspace{30mm}\times\sum_{n=j-m}^{\infty}(-i)^n \, \frac{{}_2F_1(-n, -j; j-n+1|-1)}{(j-n)!(n+m-j)!}\;(-ze^{\rho})^n\;.
\ea 
\ee 
Contrary to the right Wilson matrix element  this one is given by an infinite power series in variable $-ze^{\rho}$ (modulo prefactors). Thus, one is obliged to analyze the issue of convergence of the following power series    
\be
\label{psf}
q = -ze^{\rho}\;:\qquad \sum_{n} a_n q^n\,,
\qquad
\text{where}
\quad
a_n = (-i)^n\,\frac{{}_2F_1(-n, -j; j-n+1|-1)}{(j-n)!(n+m-j)!}\,.
\ee
Using e.g. the d'Alembert's ratio test one can show that the radius of convergence equals one, i.e. $|q|<1$, see Appendix \bref{app:radius}. In terms of $\rho$-coordinate one has $\rho<-\log|z|$, which means that for arbitrary $z$ the radius of convergence in $\rho$ goes to zero. Nonetheless, below we show that the function can be analytically continued past $|q|=1$ thereby making  the large-$\rho$  expansion possible. 

As shown in Appendix \bref{app:NLO} the power series \eqref{LWE_fin2} can be summed up to yield the  Wilson matrix element in a closed form \eqref{expnd_4}:
\be 
\label{closed_left}
\ba{l}
\dps
\langle\!\langle j|W_j[0,x] \ket{j,m} = (-)^j\frac{j!}{m!}\left[\frac{(j+m)!}{(j-m)!(2j)!}\right]^\half 
\vspace{2mm} 
\\
\dps
\hspace{33mm}\times e^{-\rho m}\,(q+i)^{j}(q-i)^{m}\,{}_2F_1\left(-j, m-j; m+1\big|\frac{q-i}{q+i}\right).
\ea 
\ee 

By construction, this function is still defined inside the disk $|q|=1$ (note that the factor $e^{-\rho m}$ is passive here). In order to analytically continue beyond  $|q|=1$ one notes that the second parameter of  the hypergeometric function $m-j$ is a negative integer. It means that the hypergeometric function is a polynomial in the Cayley transformed variable  $\frac{q-i}{q+i}$ with degrees running from 0 to $j-m$. Any polynomial is a holomorphic function and being originally defined in a domain it can be  analytically continued to the whole complex plane. 

Next, note that the whole expression \eqref{closed_left} is proportional to $\sum_{k=0}^{j-m}\alpha_k(q+i)^{j-k}(q-i)^{m+k}$ with some $\alpha_k$. At  $j\in\mathbb{Z}$ (then $m$ is also integer as follows  from \eqref{-D_basis}) the powers of $q+i$ and $q-i$ are integer, therefore, \eqref{closed_left} can be analytically  continued without any branch cuts. At $j\notin\mathbb{Z}$,  \eqref{closed_left}  has three branching points (coming from the prefactor $(q+i)^{j}(q-i)^{m}$): $q = i$, $q= -i$, and $q = \infty$. Choosing branch cuts along the imaginary axis $\im q$ as $(+i, +i \infty)$ and $(-i, -i \infty)$ one analytically continues the left Wilson matrix element onto the whole real axis $\re q$.

Upon analytic continuation one sets $q$ to be real again (along with $z$ and $\rho$). Finally, one can expand the analytically continued expression \eqref{closed_left} near $|q|=\infty$, substitute $q = -ze^\rho$ and obtain
\be 
\label{ket_matrix_inf2}
\langle\!\langle j|W_j[0,x] \ket{j,m} = (-)^{2j+m}\left[\frac{(2j)!}{(j-m)!(j+m)!}\right]^\half  z^{j+m} e^{\rho j} + \cO(e^{\rho (j-1)})\;,
\ee 
cf. \eqref{LWMEf}.

\subsection{Positive integer weights}
\label{sec:matrixI}

In the case of $\cD^-_j$ with $j\in\mathbb{N}_0$ there are two independent caps states $|j \rangle\!\rangle_1$  and $|j \rangle\!\rangle_2$, see \eqref{integer_sol1}, \eqref{integer_sol2}. As we discussed  below \eqref{integer_sol2} the cap state  $|j \rangle\!\rangle_2$  solves  the cap state condition for  $\cD^-_{-j-1}$
 and can be obtained from the cap state $|j \rangle\!\rangle \in \cD^-_j$ by shifting $j\to -j-1$. It follows that  the Wilson  matrix elements with  $|j \rangle\!\rangle_2$ can be directly obtained from \eqref{closed_left} and \eqref{right_closed} by using the same shift. Expanding these matrix elements near $\rho = \infty$ one obtains: 
\be 
\label{fin_repr_other}
\ba{l}
\dps
\bra{j,m}W_j[x,0]|j \rangle\!\rangle_2 \sim e^{-\rho(j+1)}+\cO(e^{\rho(-j-2)})\;,
\vspace{3mm}
\\
\dps
{}_2\langle\!\langle j|W_j[0,x]\ket{j,m} \sim e^{-\rho(j+1)}+\cO(e^{\rho(-j-2)})\;.
\ea
\ee 
Since the cap state $|j \rangle\!\rangle_2$ lies in  $\cD^-_{-j-1}$, then   it is orthogonal to $\ket{j,m}$ at $m\geq-j$. It follows that the Wilson line operators acting on the basis vectors $\ket{j,m}$ at  $m\geq-j$ are decomposed as  
\be 
\ba{l}
\dps
W_j[0,x]\ket{j,m} = \sum_{n=0}^{2j}\alpha_n\ket{j,j-n}\,,
\vspace{1mm}
\\
\dps
\bra{j,m}W_j[x,0] = \sum_{n=0}^{2j}\beta_n\bra{j,j-n}\,,
\ea
\ee 
where $\alpha_n, \beta_n$ are some coefficients. In other words, they belong to a subspace orthogonal to the cap state $|j \rangle\!\rangle_2$. Hence, the Wilson matrix elements \eqref{fin_repr_other}  equal  zero at  $m\geq-j$.

Consider now the cap state $|j \rangle\!\rangle_1$ \eqref{integer_sol1}.   The process of  calculating the right Wilson matrix element $\bra{j,m}W_j[x,0]|j \rangle\!\rangle_1$ is exactly the same as in the case  $j\notin\mathbb{N}_0/2$ \eqref{RWE_fin_copy}. To this end, we change the summation domain in \eqref{RWE_fin_copy} from $n\in[\![0,\infty]\!]$ to $n\in[\![0,2j]\!]$ according to the definition of the $|j \rangle\!\rangle_1$ \eqref{integer_sol1} and write 
\be 
\label{RWE_fin}
\ba{l}\dps
 \bra{j,m}W_j[x,0]| j \rangle\!\rangle_1 = \sum_{n=0}^{2j}\frac{(-i)^n(-2j-1)!}{n!(-2j+n-1)!}\frac{j!}{(j-n)!}\,{}_2F_1(-n, -j; j-n+1|-1)
\vspace{2mm} 
\\
\dps
\hspace{10mm} \times \sum_{p=0}^{\infty}\frac{z^p}{p!}e^{\rho (j-n)}\bra{j,m} (J_1)^{p+n}\ket{j,j} = \left[\frac{(2j)!(j-m)!}{(j+m)!}\right]^\half\,\sum_{n=0}^{j-m}z^{j-m-n}e^{\rho (j-n)}
\vspace{2mm} 
\\
\dps
\hspace{10mm} \times\frac{(-i)^n(-2j-1)!}{n!(-2j+n-1)!}\frac{j!}{(j-n)!(j-m-n)!}\,{}_2F_1(-n, -j; j-n+1|-1)\;,
\ea 
\ee 
where we used that $\bra{j,m} (J_1)^{p+n}\ket{j,j}$ is non-zero only for $p+n = j-m$. From the restriction $p\geq 0$ it follows that $n\leq j-m$ and, since $2j\geq j-m$, the summation domain becomes $n\in[\![0,j-m]\!]$. Also note that  $\bra{j,m} (J_1)^{j-m}\ket{j,j}$ equals zero if $m<-j$ because of $(J_1)^{2j+1}\ket{j,j} = 0$. One can see that the resulting expression coincides with   \eqref{RWE_fin_copy} so the right Wilson matrix element at  $j\in\mathbb{N}$ is given by \eqref{right_closed1}:
\be 
\ba{l}\dps
\label{wilson_fin_1ket}
\bra{j,m}W_j[x,0]| j \rangle\!\rangle_1 = (-)^{j-m}\frac{j!}{m!}\left[\frac{(j+m)!}{(2j)!(j-m)!}\right]^\half\,e^{\rho m}(q+i)^{j-m}{}_2F_1\left(-j, m-j; m+1\big|\frac{q-i}{q+i}\right)
\vspace{2mm} 
\\
\dps
\hspace{31mm}=\left[\frac{(2j)!}{(j+m)!(j-m)!}\right]^\half\,e^{\rho j} z^{j-m}+ \cO(e^{\rho(j-1)})\;.
\ea 
\ee 

Similarly one computes the corresponding left Wilson matrix element. Taking \eqref{LWE_fin2} and introducing a new  summation domain one obtains:
\be
\label{LWE_finite}
\ba{l}
\dps
{}_1\langle\!\langle j|W_j[0,x] \ket{j,m} = \sum_{n=0}^{2j}\frac{(-i)^n(-2j-1)!}{n!(-2j+n-1)!}\frac{j!}{(j-n)!}\,{}_2F_1(-n, -j; j-n+1|-1)
\vspace{2mm} 
\\
\dps
\hspace{10mm} \times \sum_{p=0}^{\infty}\frac{(-z)^p}{p!}e^{\rho (n-j)}\bra{j,j} (J_{-1})^n(J_1)^{p}\ket{j,m}=
\left[\frac{(j+m)!j!^2}{(j-m)!(2j)!}\right]^\half(-z)^{m-j}e^{-\rho j}
\dps
\vspace{2mm} 
\\
\dps
\hspace{30mm}\times\sum_{n=j-m}^{2j}(-i)^n \, \frac{{}_2F_1(-n, -j; j-n+1|-1)}{(j-n)!(n+m-j)!}\;(-ze^{\rho})^n\;.
\ea 
\ee 
The matrix element $\bra{j,j} (J_{-1})^n(J_1)^{p}\ket{j,m}$ is non-zero only when $p-n=m-j$. Since $p>0$ and $n\in[\![0,2j]\!]$, then $m\in[\![-j,j]\!]$,  otherwise, the matrix element is zero. After that, one repeats the same steps as in Appendix \bref{app:NLO} and expands the resulting expression  near $\rho=\infty$. Since the summation domain is finite, then the radius of convergence for $q$ is infinite and \eqref{LWE_finite} is defined for any real $\rho$ and $z$. The final answer is the same as in \eqref{closed_left},  \eqref{ket_matrix_inf2}:
\be 
\label{wilson_fin_1answ}
\ba{l}
\dps
{}_1\langle\!\langle j|W_j[0,x] \ket{j,m} = (-)^j \frac{j!}{m!}\left[\frac{(j+m)!}{(j-m)!(2j)!}\right]^\half e^{-\rho m}\,(q+i)^{j}(q-i)^{m}
\vspace{2mm} 
\\
\dps
\times \,{}_2F_1\left(-j, m-j; m+1\big|\frac{q-i}{q+i}\right)= (-)^{2j+m}\left[\frac{(2j)!}{(j-m)!(j+m)!}\right]^\half  z^{j+m} e^{\rho j} + \cO(e^{\rho (j-1)})\;,
\ea 
\ee 
cf. \eqref{RWMEf}. 

We conclude that for $j\in\mathbb{N}_0$ the matrix elements \eqref{fin_repr_other}  with $|j \rangle\!\rangle_2$ are suppressed by the matrix elements \eqref{wilson_fin_1ket} and \eqref{wilson_fin_1answ} with  $|j \rangle\!\rangle_1$. In this way, we see that the degeneracy of the cap states $|j \rangle\!\rangle = \alpha |j \rangle\!\rangle_1+ \beta|j \rangle\!\rangle_2$ \eqref{integer_sol1} is lifted and it is the cap state $|j \rangle\!\rangle_1$ which defines the boundary behaviour of the Wilson matrix elements which form conforms with the general formula \eqref{LWMEf}, \eqref{RWMEf}. Simultaneously, in the case of $j\in\mathbb{N}_0+\half$ the Wilson matrix elements are defined with respect to $|j \rangle\!\rangle_2$ only and, therefore, they decay near the boundary much faster than assumed by the extrapolate dictionary relation \eqref{vert_conf}. Therefore, in this case the boundary CFT correlation function vanishes.

The form of the Wilson matrix elements in $\cD_j$ directly follows from the above analysis. In this case, the cap state is given by  $|j \rangle\!\rangle_1\in \cD_j$ \eqref{cs_fin} and the boundary behaviour of the Wilson matrix elements in $\cD_j$ coincides with that of the Wilson matrix elements in $\cD^-_j$ with $j\in\mathbb{N}_0$, see \eqref{LWMEf}, \eqref{RWMEf}.

\subsection{Conformal invariance} 

We now show that near the boundary the Ward identities for  AdS vertex functions  go to the Ward identities for CFT correlation functions.  Using the right  matrix element $\braket{j,m|\tilde{a}}$ asymptotics \eqref{RWMEf} one directly finds how the \ads Killing generators are restricted on the boundary:    
\be 
\label{boundary_transform}
\cJ_n\braket{j,m|\tilde{a}} = \cL_n\braket{j,m|\tilde{a}}+\cO(e^{\rho(j-1)})\;,
\qquad \forall\, \bra{j,m}\;,
\ee 
where 
\be 
\cL_{n} = z^{n+1}\partial_{z} - j(n+1) z^n \,,
\qquad
n = 0, \pm1\,. 
\ee 
This is  the standard differential realization  of $\sl2$ algebra  on \cft primary fields of conformal dimension $h = -j$. The same relation holds for the left matrix element $\braket{\tilde{a}|j,m}$ asymptotics \eqref{LWMEf}, 
\be 
\label{boundary_transform_1}
\cJ_n\braket{\tilde{a}|j,m} = \cL_n\braket{\tilde{a}|j,m}+\cO(e^{\rho(j-1)})\;.
\ee 
Substituting   \eqref{boundary_transform}-\eqref{boundary_transform_1} into  the Ward identities \eqref{WA_wo} one obtains 
\be 
\label{conf_Ward}
\sum_{i= 1}^{n} \cL_{m}^{(i)}\, \cV_{ j\tilde{j}}(x_1,..., x_i, ..., x_n)\Big|_{\rho_1=\,...\,=\rho_n=\rho} =0+ \cO\big(e^{\rho\left(\sum_{i=1}^nj_i-1\right)}\big)\;, 
\ee 
where the superscript $i$ indicates that the differential operator  is taken with respect to the $i$-th coordinate. Taking the limit $\rho\to\infty$ and using the identification with CFT$_1$ correlation functions \eqref{vert_conf} one finds out that the above relation goes into the $sl(2, \mathbb{R})$ conformal Ward identities. 

In Appendix \bref{app:conf_transf} we also  show  that  $n$-point AdS  vertex functions \eqref{vertex_func} are  conformally invariant against finite transformations, i.e. they change under $SL(2, \mathbb{R}): z \to w(z)$ as
\be
\label{coftransconf}
\ba{l}
\dps
\cV_{ j\tilde{j}}(\rho,{\bf z})=\Big(\frac{\partial z}{\partial w}\Big)^{j_1}\Big|_{w=w_1}\cdots\,\,\Big(\frac{\partial z}{\partial w}\Big)^{j_n}\Big|_{w=w_n}\cV_{ j\tilde{j}}(\rho,{\bf w}) +\cO\big(e^{\rho\left(\sum_{i=1}^nj_i-1\right)}\big)\,.
\ea
\ee
Using the extrapolate dictionary relation  \eqref{vert_conf} one obtains  the conformal transformation law of $n$-point CFT correlation functions of primary operators.\footnote{Originally, this property was   established for 4-point functions \cite{Besken:2016ooo} in the case of  finite-dimensional modules (for quasi-Ishibashi states, in the present terminology). In Appendix \bref{app:conf_transf} we show this property for  $n$-point functions in  the case of infinite-dimensional modules.   See \cite{Alkalaev:2020yvq,Besken:2018zro,DHoker:2019clx} for more discussion in the present context and  \cite{Bhatta:2018gjb} for the symmetry analysis in $d$ dimensions.}

\subsection{2-point functions}
\label{sec:2pt}

Before proceeding with $n$-point functions in the next section, let us consider in some detail   the $2$-point AdS vertex function and the respective CFT correlation function  in the infinite-dimensional case. Other lower-point (near-boundary) AdS vertex functions and respective CFT correlators  are considered in Appendix \bref{sec:3-5}. 

\paragraph{AdS vertex function.} It can be found by taking $n=2$ in the definition  \eqref{vertex_func}:
\be 
\label{2pt-bulk}
\cV_{j_1 j_2}(\rho,{\bf{z}})=\sum_{m_1\in \sJ^-_1}\,\sum_{m_2 \in \sJ^-_2}[I_{j_1j_2}]^{m_1}{}_{m_2}\braket{\tilde{a}_1|j_1, m_1}\braket{j_2, m_2|\tilde{a}_2}\,,
\ee 
where  the Wilson matrix elements are given by \eqref{closed_left0} and the $2$-valent intertwiner $\in Inv\big((\cD_{j_1}^-)^*\otimes \cD_{j_2}^-\big)$ can be directly read off from \eqref{def_int} by substituting $j_3=0$:
\be 
\label{2valent_discrete}
[I_{j_1j_2}]^{m_1}{}_{m_2}=\frac{\delta_{j_1 j_2}\delta_{m_1 m_2}}{[2j_1+1]^{\half}}\,,
\qquad
m_{1,2} \in J^-_{1,2}\;.
\ee 
From the Ward identities \eqref{general_dependence} it follows that the $2$-point AdS vertex function depends only on the variable $q_{12} = q_1-q_2$ \eqref{integrals_const} . Shifting the  coordinates as $z_i \to z_i-z$ one stays with the same AdS vertex function. Choosing $z = z_2-ie^{-\rho}$  one obtains
\be
\ba{l}
\dps
\cV_{j_1 j_2}(\rho,{\bf{z}})=C_{j_1 j_2}\sum_{m_1\in
J^{-}_2}\left(\frac{j_1!}{m_1!}\right)^{2}\frac{(j_1+m_1)!}{(j_1-m_1)!(2j_1)!}\,(q_{12}+2i)^{j_1}(q_{12})^{m_1}
\vspace{2mm}
\\
\dps
\hspace{20mm}\times(-)^{-m_1}(2i)^{j_1-m_1}\,{}_2F_1\left(-j_1, m_1-j_1; m_1+1\big|\frac{q_{12}}{q_{12}+2i}\right) \,.
\ea
\ee
Making the Pfaff transformation \eqref{4kummer2f1} and decomposing the hypergeometric function into the series \eqref{def_hyper} yields 
\be
\ba{l}
\dps
\cV_{j_1 j_2}(\rho,{\bf{z}})=C_{j_1 j_2}\sum_{m_1\in
J^{-}_2}\sum_{n=0}^{j_1-m_1}(-)^{2j_1}(2i)^{2j_1-m_1}\,\left(\frac{q_{12}+2i}{2i}\right)^{m_1}(q_{12})^{m_1}
\vspace{2mm} 
\\
\dps
\hspace{20mm}\times \frac{(j_1+m_1+n)!(m_1-j_1+n-1)!j_1!}{(-j_1-1)!m_1!(2j_1)!(m_1+n)!}\frac{\left(-\frac{q_{12}}{2i}\right)^n}{n!}
\,.
\ea
\ee
Reindexing  $m_1' = j_1-m_1$, $n' = m'_1-n$ and summing over $m_1'$ by means of the generalized Newton binomial \eqref{newton} for $\Big|\frac{2i}{q_{12}+2i}\Big| < 1$ and $\Big|\frac{4}{q_{12}^2}\Big|<1$ as well as performing the analytic continuation as described in the previous section, one finally obtains 
\be
\label{2pt-discrete2}
\cV_{j_1 j_2}(\rho,{\bf{z}})=(-)^{j_1}C_{j_1 j_2} \, q^{2j_1}_{12}\F\left(-j_1,-j_1;-2j_1|-\frac{4}{q^2_{12}}\right)\,.
\ee
The large-$\rho$ asymptotics can be  obtained by sending $q_{12} = (z_1-z_2)e^{\rho} \to \infty$ and singling out the divergent $\rho$-dependent prefactor as
\be
\cV_{j_1 j_2}(\rho,{\bf{z}})\simeq (-)^{j_1}C_{j_1 j_2} \, q^{2j_1}_{12} \sim  (-)^{j_1}e^{2\rho j_1} z_{21}^{2j_1}\,,
\ee
thereby reproducing the 2-point CFT correlation function, see \eqref{2pt_discrete_final} below for more details.

Up to the constant, the $2$-point AdS vertex function just calculated  is the bulk-to-bulk  propagator in \ads \cite{Fronsdal:1974ew} on the $\rho = const$ hyperplane
\be 
\label{bb_propagator}
G_{h}(x_1,x_2) = e^{-h\sigma(x_1,x_2)}\F\left(h,\half;h+\half\,\big|e^{-2\sigma(x_1,x_2)}\right)\,,
\ee 
where $j_1=j_2 = -h$ and $\sigma(x_1,x_2)$ is the geodesic length between points $x_1$ and $x_2$. Using the metric \eqref{ads2} one can show that the geodesic length between two points on the same $\rho$-plane is given by
\be 
\label{geodesic_length}
\sigma(x_1,x_2) = \log\left(\frac{|q_{12}|\sqrt{4+q_{12}^2}+2+q_{12}^2}{2}\right)\,.
\ee 
Substituting this expression into the bulk-to-bulk propagator \eqref{bb_propagator} and making the quadratic transformation \eqref{quadratic} one obtains the $2$-point AdS vertex function \eqref{2pt-discrete2} up to the prefactor $\pref_{j_1 j_2}$. In fact, here we reproduced  the 2-point result obtained in \cite{Castro:2018srf} in a different setup.

\paragraph{Near-boundary calculation.} If one is interested in studying  asymptotic AdS vertex functions only, then one directly considers  \eqref{2pt-bulk} at  large-$\rho$, i.e. when   the Wilson matrix elements are given by the asymptotics \eqref{LWMEf}. In fact, it is the  $n$-point Wilson network near the boundary that we analyze in Section \bref{sec:n-p} and the exact computation in the bulk is a future task. Summing over  $m_2$ along with reindexing  $k = j_1-m_1$ yields
\be 
\label{newton_inter}
\cV_{j_1 j_2}(\rho,{\bf{z}})=(-)^{j_1}  \frac{\delta_{j_1 j_2}}{{[2j_1+1]^{\half}}}e^{2\rho j_1}\sum_{k=0}^{\infty}{\frac{(2j_1)!}{k!(2j_1-k)!}}\,(-z_1)^{2j_1-k}z_2^{k}+\cO(e^{\rho(2j_1-1)})\,.
\ee 
Using the generalized Newton theorem \eqref{newton} one  finally obtains 
\be 
\label{2pt_discrete_final}
\lim_{\rho\to\infty}e^{-2\rho j_1} \cV_{j_1 j_2}(\rho,{\bf{z}})
=(-)^{j_1}\frac{\delta_{j_1j_2}\,z_{_{21}}^{2j_1}}{[2j_1+1]^{\half}}\equiv \frac{C}{(z_1-z_2)^{2h}}\,,
\ee 
where in the last equality we used $j_1=j_2 = -h$  and introduced  the normalization constant $C = (-)^{h}[1-2h]^{-\half}$ that through the identification \eqref{vert_conf} gives us  the  2-point CFT correlation function. Note that in the case of the cap states chosen as  (rotated) LW/HW vectors the sign $(-)^{j_1}$ does not appear in the $2$-point AdS vertex function.

So far we have been controlling only the large-$\rho$ behaviour of the Wilson  matrix elements. However, these also depend on $z$-variables and combining them together  brings to light the issue of convergence in $z$-space.  Indeed, for the series \eqref{newton_inter} to be convergent the boundary points must be ordered as $z_1>z_2$. The same ordering prescription 
\be
z_1 >z_2 > \ldots >z_n
\ee
persists for higher-point AdS vertex functions that  yields a particular  OPE ordering for CFT correlation functions giving rise to the conformal block decomposition in the comb channel. Remarkably, the convergence and ordering issues are absent for the AdS vertex functions in the finite-dimensional modules since all  functions in this case are polynomials both in  $z$ and $\rho$ coordinates.  

Finally, note that since $(z-1)! \equiv  \Gamma(z)$ has simple poles in $z=0,-1,-2,\ldots$ one finds out that for  $2j_1\in\mathbb{N}_0$  all terms in \eqref{newton_inter} with $k>2j_1$ vanish. Of course, this is just a direct consequence of  Proposition \bref{prop2} since the AdS vertex function is built as the product of left and right Wilson elements. Thus, the  infinite series  for (half-)integer weights keeps only finitely many terms. This yields  the 2-point function for  finite-dimensional modules which can be equivalently calculated using the HW/LW vectors as the cap states. The only  difference is given by the overall sign that follows from Proposition \bref{prop1}. This observation extends to the $n$-point case that allows us to simplify all calculations by using  finite-dimensional irreps and then analytically continue in weights. We use this trick in the next section.

\section{Higher-point  functions and recursion relations}
\label{sec:n-p}

As discussed in the previous section, by virtue of Propositions \bref{prop2} and \bref{prop1} we can restrict our consideration to finite-dimensional modules. The $n$-point conformal blocks will be calculated by solving the recursion relation satisfied by the asymptotic AdS vertex functions (though, for finite-dimensional modules these functions are exact in $\rho$). The base of recursion is given by  the 4-point conformal block which along with 3-point and 5-point blocks is calculated in Appendix \bref{sec:3-5}.  

\subsection{Recursive construction of  AdS vertex functions}
\label{sec:rec_prel}

Here and later, we use the following notation for  coordinate sets:
\be
{\bf z} = \{z_1, ..., z_{n} \}\;,
\qquad
{\bf z'} = \{z_1, ..., z_{n-1} \}\;,
\qquad
{\bf z''} = \{z_1, ..., z_{n-3} \}\;.
\ee
According to the matrix expression  of the AdS vertex functions \eqref{vertex_func}, the  $(n-1)$-point  function for a set of finite-dimensional modules $\cD_j$ and $\cD_{\tilde j}$ can be represented as 
\be 
\label{n-1p_def}
\ba{l}
\dps
\cV_{ j\tilde{j}}(\rho,{\bf z'})= \sum_{\substack{\{m_i\in \sJ_{i}\}_{i=1,..., n-1}\\ 
\{p_i\in\stJ_i\}_{i = 1,..., n-4}}}[I_{j_1j_2\tilde{j}_1}]^{m_1}{}_{m_2p_1}\dots[I_{\tilde{j}_{n-5}j_{n-3} \tilde{j}_{n-4}}]^{p_{n-5}}{}_{m_{n-3}p_{n-4}}
\vspace{2mm}
\\
\times[I_{\tilde{j}_{n-4}j_{n-2} j_{n-1}}]^{p_{n-4}}{}_{m_{n-2}m_{n-1}}\braket{\tilde{a}_1|j_1, m_1}\dots \braket{j_{n-2}, m_{n-2}|\tilde{a}_{n-2}}\braket{j_{n-1}, m_{n-1}|\tilde{a}_{n-1}}\,,
\ea
\ee
and the $n$-point  function as
\be 
\label{n-p_def}
\ba{l}
\dps
\cV_{ j\tilde{j}}(\rho,{\bf z})= \sum_{\substack{\{m_i\in \sJ_{i}\}_{i=1,..., n}
\\ 
\{p_i\in\tilde{J}_{i}\}_{i=1,..., n-3}}}[I_{j_1j_2\tilde{j}_1}]^{m_1}{}_{m_2p_1}\dots[I_{\tilde{j}_{n-5}j_{n-3} \tilde{j}_{n-4}}]^{p_{n-5}}{}_{m_{n-3}p_{n-4}}
\vspace{2mm}
\\
\times[I_{\tilde{j}_{n-4}j_{n-2} \tilde{j}_{n-3}}]^{p_{n-4}}{}_{m_{n-2}p_{n-3}}\braket{\tilde{a}_1|j_1, m_1}\dots \braket{j_{n-2}, m_{n-2}|\tilde{a}_{n-2}}\braket{j_{n-1}, m_{n-1}|\tilde{a}_{n-1}}
\vspace{2mm}
\\\dps
\times[I_{\tilde{j}_{n-3}j_{n-1} j_n}]^{p_{n-3}}{}_{m_{n-1}m_n}\braket{j_n, m_n|\tilde{a}_n}\,.
\ea
\ee
The form of these two expressions suggests that the $n$-point AdS vertex function  can be built from the $(n-1)$-point AdS vertex function by adding new elements. First of all, we observe that the terms present in the first lines of both expressions coincide. Then, comparing the second lines we see that the intertwiner $[I_{\tilde{j}_{n-4}j_{n-2} j_{n-1}}]^{p_{n-4}}{}_{m_{n-2}m_{n-1}}$ in  \eqref{n-1p_def} is changed to $[I_{\tilde{j}_{n-4}j_{n-2} \tilde{j}_{n-3}}]^{p_{n-4}}{}_{m_{n-2}p_{n-3}}$ in \eqref{n-p_def}. Finally, the third line is present only in \eqref{n-p_def} and contains  $[I_{\tilde{j}_{n-3}j_{n-1} j_n}]^{p_{n-3}}{}_{m_{n-1}m_n}\braket{j_n, m_n|\tilde{a}_n}$. 

That brings us to the idea of splitting expressions \eqref{n-1p_def} and \eqref{n-p_def} into two parts by singling out the terms contained in  the first and second  lines. Indeed,  let us    define the auxiliary matrix element 
\be
\label{C_aux}
\ba{c}
\dps 
\mathbb{C}_{p_{n-4}}(\rho,{\bf z''})\;\; = \sum_{\substack{\{m_i\in \sJ_i\}_{i=1,..., n-3}\\ \{p_i\in \stJ_i\}_{i=1,...,n-5}}}[I_{j_1j_2\tilde{j}_1}]^{m_1}{}_{m_2p_1}\dots[I_{\tilde{j}_{n-5}j_{n-3} \tilde{j}_{n-4}}]^{p_{n-5}}{}_{m_{n-3}p_{n-4}}
\vspace{2mm}
\\
\dps
\hspace{30mm}\times\braket{\tilde{a}_1|j_1, m_1}\dots \braket{j_{n-3}, m_{n-3}|\tilde{a}_{n-3}},
\ea
\ee
which is the same for both expressions. It has one external index in $\cD_{\tilde{j}_{n-4}}$ module. Of course, this matrix element  has less dummy summation indices  since it describes just a sub-diagram (the same for both diagrams) connected to other edges by additional summations.  

\begin{figure}
\centering
\includegraphics[width=1.0\linewidth]{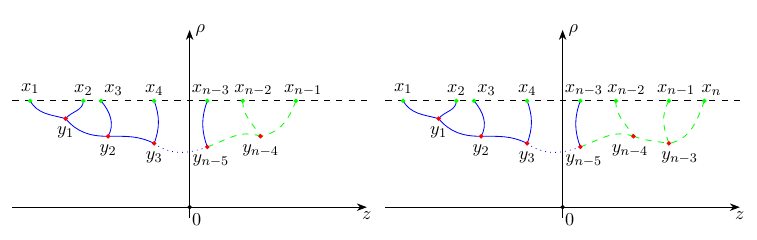} 
\caption{From $n-1$ points (left) to $n$ points (right). The blue graph stands for the auxiliary matrix element $\mathbb{C}_{p_{n-4}} \in \cD_{\tilde{j}_{n-4}}$  \eqref{C_aux}. Dotted green lines denote those parts of the diagrams which are affected by adding one more endpoint.} 
\label{fig_cut}
\end{figure}

The technical reason for such a factorization is that the last two legs of the $(n-1)$-point Wilson line network carrying $\cD_{j_{n-2}}$ and $\cD_{j_{n-1}}$ are connected to the rest of the diagram only by  intertwiner $I_{\tilde{j}_{n-4} j_{n-2} j_{n-1}}$ which arises in the second line of \eqref{n-1p_def}. Extending $(n-1)$-point diagram to $n$-point diagram by adding two more legs (one intermediate $\cD_{\tilde{j}_{n-3}}$  and one external $\cD_{j_n}$) results in replacing  $I_{\tilde{j}_{n-4} j_{n-2} j_{n-1}}$ by $I_{\tilde{j}_{n-4}j_{n-2} \tilde{j}_{n-3}}$ which correspond to replacing external $\cD_{j_{n-1}}$ by internal $\cD_{\tilde{j}_{n-3}}$ and adding one more intertwiner $I_{\tilde{j}_{n-3}j_{n-1} j_n}$ arsing in the third line of \eqref{n-p_def}. This procedure is depicted on Fig. \bref{fig_cut}.  In what follows we explicitly single out $\mathbb{C}_{p_{n-4}}(\rho,{\bf z''})$ in both  AdS vertex functions that allows us to find out that they are recursively related.  

\subsection{$n-1$ points}

Using  \eqref{C_aux} the $(n-1)$-point function \eqref{n-1p_def} can be cast into the form  
\be
\label{n-1p+aux}
\ba{l}
\dps \cV_{j_1 \cdots j_{n-1}\tilde{j}_1 \cdots \tilde{j}_{n-4}}(\rho,{\bf z'}) = \hspace{-5mm}
\sum_{\substack{{\{m_i\in J_i\}}_{i = n-2,n-1}\\ p_{n-4}\in \stJ_4 }}\;\mathbb{C}_{p_{n-4}}(\rho,{\bf z''})[I_{\tilde{j}_{n-4} j_{n-2} j_{n-1}}]^{p_{n-4}}{}_{m_{n-2} m_{n-1}}
\vspace{0mm}  
\\ 
\dps
\hspace{61mm}\times \braket{j_{n-2}, m_{n-2}|\tilde{a}_{n-2}}\braket{j_{n-1}, m_{n-1}|\tilde{a}_{n-1}}.
\ea
\ee
To simplify further calculations we make use of  the following coordinate $SL(2, \mathbb{R})$ transformation
\be 
w(z) = \frac{\z{n-3}{n-1}(z-z_{n-1})}{\z{n-3}{n-2}(z-z_{n-2})}\,,
\label{n-transfrom}
\ee 
which maps points $z_i\in {\bf z'}$ into 
\be 
\label{fixed_p}
\ba{l}\dps
w_i := w(z_{_i}) = \frac{\z{n-3}{n-1}\z{i}{n-1}}{\z{n-3}{n-2}\z{i}{n-2}}, 
\quad 
i = 1,2,..., n-4\,;\quad w(z_{n-3}) = 1\,;
\vspace{2mm}
\\
\dps
 w\big(z_{n-2}+\frac{\z{n-3}{n-1}\z{n-2}{n-1}}{\z{n-3}{n-2}}{\omega^{-1}}\big)=\omega+\frac{\z{n-3}{n-1}}{\z{n-3}{n-2}}=\omega+\cO(1)\,;\quad w(z_{n-1}) = 0,
\ea
\ee
where we introduced the large coordinate parameter $\omega\to\infty$ to regularize the pole $z = z_{n-2}$ in \eqref{n-transfrom}. We want to simplify \eqref{n-1p+aux} without calculating explicitly the auxiliary matrix element $\mathbb{C}_{p_{n-4}}({\bf z''})$. The most
efficient way to do that is to write the AdS vertex function in $w$-coordinates, take the last two points $w_{n-2}$ and $w_{n-1}$ as $\infty$ and $0$ because the Kronecker symbols arising in the corresponding Wilson matrix elements will drastically simplify the intertwiner and then make an inverse transformation to \eqref{n-transfrom}  to obtain the simplified AdS vertex function in $z$-coordinates. Namely, the last two Wilson matrix elements \eqref{LHW_fin} in $w$-coordinates are given by 
\be 
\label{n-Wilson}
\ba{l}
\braket{j_{n-2},m_{n-2}|\tilde{a}_{n-2}}=e^{\rho j_{n-2}}\delta_{-j_{n-2}, m_{n-2}}\omega^{2j_{n-2}}+\cO(\omega^{2j_3-1})\,,
\vspace{2mm}\\\dps
\quad\braket{j_{n-1},m_{n-1}|\tilde{a}_{n-1}}=e^{\rho j_{n-1}}\delta_{j_{n-1} ,m_{n-1}}\,,
\ea
\ee 
where we used that the prefactor $A_{jm}$ trivializes: $A_{jj} = A_{j-j} = 1$. In the first matrix element we kept only the leading asymptotics in $\omega$. Indeed, one can see that the AdS vertex function  \eqref{vertex_func} is a polynomial in $\omega$ of powers running from $0$ to $2j_{n-2}$ and the leading contribution at $\omega\xrightarrow{}\infty$ is provided by the maximal power ($j_{n-2}\in\mathbb{N}$). The Kronecker delta in the second matrix element appeared because the right Wilson matrix element \eqref{LHW_fin} is non-zero for $z=0$ only when $m=j$. Then, using the Wilson matrix elements \eqref{n-Wilson}, resolving the Kronecker deltas by summing over $m_{n-2}$ and $m_{n-1}$ and using the $3j$ symbol \eqref{s3-j}\footnote{Actually, there is no need for the intertwiner ($\sim 3j$ symbol) with arbitrary arguments \eqref{general-intertwiner}. Indeed, resolving the Kronecker deltas in the corresponding Wilson matrix elements one is left with a much more simple $3j$ symbol \eqref{s3-j}.} one eventually finds the $(n-1)$-point AdS vertex functions in $w$-coordinates:
\be
\label{VC_eq}
\ba{l}\dps
\dps \cV_{j_1 \cdots j_{n-1}\tilde{j}_1 \cdots \tilde{j}_{n-4}}(\rho,{\bf w'}) = e^{\rho (j_{n-2}+j_{n-1})}\,\omega^{2j_{n-2}}\,\mathbb{C}_{j_{n-1}-j_{n-2}}(\rho,{\bf w''})
\vspace{2mm}
\\ 
\dps
\hspace{34mm}\times\, \pref_{\tilde{j}_{n-4}j_{n-2}j_{n-1}}\left[\frac{(j_{n-1}+\tilde{j}_{n-4}-j_{n-2})!(j_{n-2}+\tilde{j}_{n-4}-j_{n-1})!}{(2\tilde{j}_{n-4})!}\right]^{\half},
\ea
\ee
where  the $w$-coordinate sets are  ${\bf w'} = (w_1,..., w_{n-4}, 1, \omega, 0)$ and ${\bf w''} = (w_1, ..., w_{n-4}, 1)$ and the structure constant $\pref_{\tilde{j}_{n-4}j_{n-2}j_{n-1}}$ is given by \eqref{prefactor}. For the sake of simplicity, here and later we suppress the subleading  terms $\cO(\omega^{2j-1})$. Now, making  the inverse coordinate transformation  
\be 
\label{inverse_transf}
z(w) = \frac{-z_{n-2}\z{n-3}{n-1}w + z_{n-1}\z{n-3}{n-2}}{-\z{n-3}{n-1}w + \z{n-3}{n-2}}
\ee 
and using the conformal transformation law \eqref{coftransconf} as well as taking the limit $\omega\to\infty$ we get   
\be
\label{n-1_pt_vert}
\ba{l}\dps
\cV_{j_1 \cdots j_{n-1}\tilde{j}_1 \cdots \tilde{j}_{n-4}}(\rho,{\bf{z'}})=\lim_{\omega\to\infty}\bigg[\big(\z{n-1}{n-2}\z{n-3}{n-1}\z{n-3}{n-2}\big)^{\sum_{k=1}^{n-1}j_{k}}\,\z{n-3}{n-2}^{-2j_{n-1}}\,\z{n-1}{n-2}^{-2j_{n-3}}\,\big(\omega\z{n-1}{n-3}\big)^{-2j_{n-2}}
\vspace{2mm}
\\
\dps
\times\prod_{i = 1}^{n-4}\big(\z{n-1}{n-3}w_i({\bf{z'}}) + \z{n-3}{n-2}\big)^{-2j_i}\cV_{j_1 \cdots j_{n-1}\tilde{j}_1 \cdots \tilde{j}_{n-4}}(\rho,{\bf{w'}}({\bf{z'}}))\bigg] = \mathbb{C}_{j_{n-1}-j_{n-2}}(\rho,{\bf{w''}}({\bf{z'}}))
\vspace{2mm}
\\
\dps
\times e^{\rho (j_{n-2}+j_{n-1})}\Big(\z{n-1}{n-2}\z{n-3}{n-1}\z{n-3}{n-2}\Big)^{\sum_{k=1}^{n-1}j_{k}}\,\z{n-3}{n-2}^{-2j_{n-1}}\,\z{n-1}{n-2}^{-2j_{n-3}}\,\z{n-1}{n-3}^{-2j_{n-2}}
\vspace{2mm}
\\
\dps
\times\prod_{i = 1}^{n-4}(\z{n-1}{n-3}w_i({\bf{z'}}) + \z{n-3}{n-2})^{-2j_i}\pref_{\tilde{j}_{n-4}j_{n-2}j_{n-1}}\left[\frac{(j_{n-1}+\tilde{j}_{n-4}-j_{n-2})!(j_{n-2}+\tilde{j}_{n-4}-j_{n-1})!}{(2\tilde{j}_{n-4})!}\right]^{\half}\vspace{2mm},
\ea
\ee
where ${\bf{w'}}({\bf{z'}}) = (w_1({\bf{z'}}),..., w_{n-4}({\bf{z'}}), 1, \omega, 0)$. The power-law  prefactors in the first equality are the Jacobians from \eqref{coftransconf}. 

\subsection{$n$ points} 

The $n$-point case  is considered along the same lines. Substituting the auxiliary  element \eqref{C_aux} into the $n$-point AdS vertex function  \eqref{n-p_def} one obtains 
\be
\ba{l}
\label{n_point_with_aux}
\dps \cV_{j_1 \cdots j_n\tilde{j}_1 \cdots \tilde{j}_{n-3}}(\rho,{\bf{z}}) 
= \hspace{-5mm}\sum_{\substack{\{m_i\in \sJ_i\}_{i = n-2,n-1,n}\\ \{p_i\in\stJ_i\}_{i = n-4,n-3}}}\hspace{-3mm}\mathbb{C}_{p_{n-4}}(\rho,{\bf z''})\, [I_{\tilde{j}_{n-4}\, j_{n-2}\,\tilde{j}_{n-3}}]^{p_{n-4}}{}_{m_{n-2} p_{n-3}}
[I_{\tilde{j}_{n-3}\,j_{n-1}\, j_n}]^{p_{n-3}}{}_{m_{n-1} m_n}\vspace{2mm} 
\\ 
\dps
\hspace{50mm}\times\, \braket{j_{n-2}, m_{n-2}|\tilde{a}_{n-2}}\braket{j_{n-1}, m_{n-1}|\tilde{a}_{n-1}}\braket{j_n, m_n|\tilde{a}_n}\,.
\ea
\ee
Omitting the details of the calculation (which is essentially the same as in the $(n-1)$-point case, see Appendix \bref{app:details}), we represent  the final result 
$$
\ba{l}
\dps \cV_{j_1 \cdots j_n\tilde{j}_1 \cdots \tilde{j}_{n-3}}(\rho,{\bf{z}}) = e^{\rho (j_{n-2}+j_{n-1}+j_n)}
\big(\z{n-1}{n-2}\z{n-3}{n-1}\z{n-3}{n-2}\big)^{\sum_{k=1}^n j_k}\,\z{n-3}{n-2}^{-2j_{n-1}}\,\z{n-1}{n-2}^{-2j_{n-3}}\,\big(\z{n-1}{n-3}\big)^{-2j_{n-2}}
\vspace{2mm}
\\ 
\dps
(\z{n-1}{n-3}w_n({\bf{z}})+ \z{n-3}{n-2})^{-2j_n}\prod_{i = 1}^{n-4}(\z{n-1}{n-3}w_i({\bf{z}}) + \z{n-3}{n-2})^{-2j_i}\hspace{-0mm}\sum_{k\in K_n}\hspace{-0mm}\mathbb{C}_{\tilde{j}_{n-3}-j_{n-2}-k}(\rho,{\bf w''}({\bf{z}}))\,
\ea
$$
\vspace{-3mm}
\be 
\label{n_pt_vert}
\ba{l}
\dps
\times w_n^{k+j_n+j_{n-1}-\tilde{j}_{n-3}}({\bf{z}})
\frac{(j_{n-1}-j_n-\tilde{j}_{n-3})_k(j_{n-2}-\tilde{j}_{n-4}-\tilde{j}_{n-3})_k}{(-2\tilde{j}_{n-3})_k k!}\pref_{\tilde{j}_{n-3}j_{n-1}j_n}\pref_{\tilde{j}_{n-4}j_{n-2}\tilde{j}_{n-3}}
\vspace{2mm}  
\\ 
\dps
\times \left[\frac{(\tilde{j}_{n-4}-\tilde{j}_{n-3}+j_{n-2}+ k)!(\tilde{j}_{n-4}+\tilde{j}_{n-3}-j_{n-2}-k)!}{(2\tilde{j}_{n-4})!}\right]^{\half},
\ea
\ee
where the summation domain is $K_n:=[\![0,j_n-j_{n-1}+\tilde{j}_{n-3}]\!]$.

\subsection{Recursion relation} 
\label{sec:rec}

The previously obtained expressions for the AdS vertex functions  demonstrate  that  a given   $n$-point AdS vertex function \eqref{n_pt_vert} can be represented in terms of  $(n-1)$-point AdS vertex functions \eqref{n-1_pt_vert} with running last external weight. Indeed, expressing the auxiliary matrix element  $\mathbb{C}$ form the relation \eqref{VC_eq} (which contains no summations) through the $(n-1)$-point AdS vertex function and substituting it into \eqref{n_pt_vert} (which contains just one summation over $k$) along with the change of the  summation index $k$ we find the following relation 
\be
\label{rec_rel1}
\ba{l}
\dps
\cV_{j_1 \cdots j_n\tilde{j}_1 \cdots \tilde{j}_{n-3}}(\rho,{\bf{z}}) 
=   \cV_{\tilde{j}_{n-3}j_{n-1}j_n}(\rho,z_{n-2},z_{n-1},z_n)\,z_{_{n-2\,n-1}}^{-2\tilde{j}_{n-3}}\hspace{-0mm}\sum_{k_n\in K_n}e^{\rho (-2 \tilde{j}_{n-3}+k_n)}\frac{\pref_{\tilde{j}_{n-4}j_{n-2}\tilde{j}_{n-3}}}{\pref_{\tilde{j}_{n-4}j_{n-2}\tilde{j}_{n-3}-k_n}}
\vspace{2.5mm}  
\\\dps
\times\Big(\frac{z_{_{n n-1}}z_{_{n-2\,n-1}}}{z_{_{n n-2}}}\Big)^{k_n}\frac{(j_{n-1}-j_n-\tilde{j}_{n-3})_{k_n}(j_{n-2}-\tilde{j}_{n-4}-\tilde{j}_{n-3})_{k_n}}{k_n!(-2\tilde{j}_{n-3})_{k_n}}\,\cV_{j_1 \cdots j_{n-2}(\tilde{j}_{n-3}-k_n)\tilde{j}_1 \cdots \tilde{j}_{n-4}}(\rho,{\bf z}),
\ea 
\ee 
where  the summation domain is given by 
\be
\label{Kn}
K_n:=[\![0,j_n-j_{n-1}+\tilde{j}_{n-3}]\!]\;.
\ee
The 3-point AdS vertex function $\cV_{\tilde{j}_{n-3}j_{n-1}j_n}$ on the right-hand side  just encodes power-law prefactors in  $z$. Here, we assume that the base of recursion is $n=5$ which means that the first non-trivial recursion relations expresses the 5-point function on the left-hand side in terms of 4-point functions on the right-hand side.\footnote{This is the recursion relation \eqref{4pt_double1}.} Note that at  $n=3$ this recursion relation breaks up. However, it is  still valid  at  $n=4$, but  needs replacing   $\tilde{j}_{n-4}$ with $j_1$. This is why the more convenient base of recursion is $n=5$. Here, we emphasize again that the recursion relation holds for asymptotic AdS vertex functions  which coincide with their bulk expressions only for finite-dimensional modules.  
 
In order to solve  the recursion relation \eqref{rec_rel1}  we reorganize  it as  
\be
\label{rec_scheme}
\ba{l}
\dps
\cV^{(n)}_{j_n}
=\sum_{k_n\in K_n}\beta_{n,k_n}\,\gamma_{n,k_n,j_n}\,\cV^{(n-1)}_{\tilde{j}_{n-3}-k_n}\,,
\ea 
\ee 
where 
\be 
\label{abc_factors}
\ba{l}\dps
\beta_{n,k_n} = e^{\rho (j_{n-1}-\tilde{j}_{n-3}+k_n)}(2\tilde{j}_{n-3})!
\left[{\frac{(\Delta(\tilde{j}_{n-4}, j_{n-2}, \tilde{j}_{n-3}-k_n)!}{(2\tilde{j}_{n-3}-2k_n)!}}\right]^{\half}\left[{\frac{(2j_{n-1})!}{\Delta(\tilde{j}_{n-4},j_{n-2},\tilde{j}_{n-3})}}\right]^{\half}
\vspace{2.5mm}  
\\
\dps
\hspace{28mm}\times \frac{(j_{n-2}-\tilde{j}_{n-4}-\tilde{j}_{n-3})_{k_n}}{k_n!(-2\tilde{j}_{n-3})_{k_n}}\,
\Big(\frac{z_{_{n-2\,n-1}}z_{_{n\, n-1}}}{z_{_{n\, n-2}}}\Big)^{k_n+j_{n-1}-\tilde{j}_{n-3}},
\vspace{2.5mm}  
\\\dps
\gamma_{n,k_n,j_n} = e^{\rho j_n}\left[{\frac{{(2j_n)!}}{\Delta(\tilde{j}_{n-3},j_{n-1}, j_n)}}\right]^{\half}(-j_n+j_{n-1}-\tilde{j}_{n-3})_{k_n}\,z_{_{n\,n-1}}^{j_n}z_{_{n\,n-2}}^{j_n}z_{_{n-2\,n-1}}^{-j_n}\,.
\ea 
\ee 
This form comes from using explicit expressions  for the $3$-point AdS vertex function and structure constants in \eqref{rec_rel1}. The coefficient in the recursion relation \eqref{rec_scheme} is split in two parts: (1) the factor $\beta_{n,k_n}$ which is independent of $j_n$; (2) the  factor $\gamma_{n,k_n,j_n}$ which is dependent on $j_n$. All the vertex functions and  coefficients in \eqref{rec_scheme} still  depend on other weights and coordinates but here we indicate only those which are relevant for the recursion procedure. 

At the next step of recursion, $n \to n+1$, the relation \eqref{rec_scheme}  reads 
\be
\label{rec_scheme+1} 
\cV^{(n+1)}_{j_{n+1}} = \,\sum_{k_{n+1}\in K_{n+1}}\beta_{n+1,k_{n+1}}\,\gamma_{n+1,k_{n+1},j_{n+1}}\,\cV^{(n)}_{\tilde{j}_{n-2}-k_{n+1}}\,.
\ee 
Substituting the $n$-point AdS vertex function \eqref{rec_scheme} into the right-hand side of \eqref{rec_scheme+1} one obtains 
\be 
\label{n+1_from_n-1}
\cV^{(n+1)}_{j_{n+1}} = \,\sum_{k_n\in K_n}\sum_{k_{n+1}\in K_{n+1}}\hspace{-5mm}\beta_{n,k_n} \, \beta_{n+1,k_{n+1}} \, \gamma_{n,k_n,\tilde{j}_{n-2}-k_{n+1}}\,\gamma_{n+1,k_{n+1}, j_{n+1}}\,\cV^{(n-1)}_{\tilde{j}_{n-3}-k_n}\,.
\ee 
Considering $n=4$ as the base of recursion one can directly see that the final recursive solution is given  in terms of summing over 4-point AdS vertex functions with running last external weight,
\be 
\label{n_vert_schem}
\cV_{j_1 \cdots j_n\tilde{j}_1 \cdots \tilde{j}_{n-3}}
=
\sum_{\{k_j\in K_j\},j=5,...,n}\beta_{n,k_n}\,\gamma_{n,k_n,j_n}\,\cV_{j_1j_2j_3(\tilde{j}_2-k_5)\tilde{j}_1}
\prod_{i=5}^{n-1}\beta_{i,{k_i}}\,\gamma_{i,{k_i},\tilde{j}_{i-2}-k_{i+1}}\,,
\ee 
where the summation domains are 
\be
\label{KI}
\ba{l}
\dps
K_5:=[\![0, \min(\tilde{j}_{1}-j_{3}+\tilde{j}_{2}, \tilde{j}_{3}-j_{4}+\tilde{j}_{2})]\!]\,,
\vspace{1mm} 
\\
K_i:=[\![0, \min(\tilde{j}_{i-4}-j_{i-2}+\tilde{j}_{i-3}-k_{i-1}, \tilde{j}_{i-2}-j_{i-1}+\tilde{j}_{i-3})]\!]\,, \quad i = 6,..., n-1\,, 
\vspace{1mm} 
\\
K_n:=[\![0, \min(\tilde{j}_{n-4}-j_{n-2}+\tilde{j}_{n-3}-k_{n-1}, j_n-j_{n-1}+\tilde{j}_{n-3})]\!]\,.
\ea
\ee 

\paragraph{Recursive solution.} In order to explicitly evaluate the products and sums in \eqref{n_vert_schem} it is convenient to use the obvious identity $\beta_{n,k_n}\prod_{i=5}^{n-1}\beta_{i,{k_{i}}} = \beta_{5,k_5}\prod_{i=5}^{n-1}\beta_{i+1,{k_{i+1}}} $ and rewrite \eqref{n_vert_schem} as 
\be 
\label{n_mod_vert_schem}
\cV_{j_1 \cdots j_n\tilde{j}_1 \cdots \tilde{j}_{n-3}}
=
\sum_{\{k_j\in K_j\},j=5,...,n}\,\gamma_{n,k_n,j_n}\,\beta_{5,k_5}\cV_{j_1j_2j_3(\tilde{j}_2-k_5)\tilde{j}_1}
\prod_{i=5}^{n-1}\beta_{i+1,{k_{i+1}}}\,\gamma_{i,{k_i},\tilde{j}_{i-2}-k_{i+1}}\,.
\ee 
The right-hand side of this  expression can be schematically  represented as $\sum \gamma (\beta \cV) (\prod \beta \gamma)$. In what follows we calculate each factor in this formula separately.  Since the $\gamma$-factor here is explicitly  given by \eqref{abc_factors} then we focus on the $\beta \cV$-factor and the $\prod \beta \gamma$-factor.   

\paragraph{$\beta \cV$-factor.} From Appendix \bref{app:4pt} we know that the $4$-point AdS vertex function $\cV_{j_1 j_2 j_3 (\tilde{j_2}-k_5)\tilde{j}_1}$ which is present in  \eqref{n_mod_vert_schem} is given in terms of the hypergeometric function, see \eqref{final4pt}. Using the hypergeometric series \eqref{def_hyper} one represents it as 
\be 
\ba{l}\dps
\label{4pt_end}
\cV_{j_1 j_2 j_3 (\tilde{j_2}-k_5)\tilde{j}_1}(\rho,{\bf z})= \left[\frac{(2j_1)!(2j_2)!(2j_3)!(2\tilde{j_2}-2k_5)!}{\Delta(j_1,j_2,\tilde{j}_1)\Delta(j_3,\tilde{j_2}-k_5,\tilde{j}_1)}\right]^{\half}z_{_{12}}^{j_1+j_2}z_{_{34}}^{j_3+\tilde{j_2}-k_5}
\left(\frac{z_{_{23}}}{z_{_{13}}}\right)^{j_2-j_1}
\hspace{-1mm} 
\left(\frac{z_{_{24}}}{z_{_{23}}}\right)^{\tilde{j_2}-k_5-j_3}
\vspace{2.5mm}  
\\
\dps
\hspace{20mm}\times\, e^{\rho(j_1+j_2+j_3+\tilde{j_2}-k_5)}(2\tilde{j}_1)!\sum_{k_4\in K_4} \frac{(j_2-j_1-\tilde{j}_1)_{k_4}(j_3-\tilde{j}_1-\tilde{j_2}+k_5)_{k_4}}{(-2\tilde{j}_1)_{k_4}k_4!}\,\chi_4^{k_4-\tilde{j}_1}\,,
\ea 
\ee 
where we introduced the cross-ratio $\chi_4$ belonging to the set 
\be
\label{ratio}
\chi_n=\frac{\z{n-3}{ n-2}\z{n-1}{ n}}{\z{n-3}{ n-1}\z{n-2}{ n}},\quad n=4,5,6,...\;.
\ee 
The summation domain in \eqref{4pt_end} is  
\be
\label{K4}
K_4 := [\![0, \min(\tilde{j}_1+j_1-j_2, \tilde{j}_1-j_3+\tilde{j_2}-k_5)]\!].
\ee 
Substituting  $\beta_{5,k_5}$ given by \eqref{abc_factors} we find the $(\beta \cV)$-factor:
\be
\label{second_factor}
\ba{l}\dps
\beta_{5,k_5}\,\cV_{j_1j_2j_3(\tilde{j}_2-k_5)\tilde{j}_1} = 
\sum_{k_4\in K_4}(2\tilde{j}_2)!
\left[{\frac{(\Delta(\tilde{j}_1, j_3, \tilde{j}_2-k_5)!}{(2\tilde{j}_2-2k_5)!}}\right]^{\half}\left[{\frac{(2j_4)!}{\Delta(\tilde{j}_1,j_3,\tilde{j}_2)}}\right]^{\half}
\frac{(j_{3}-\tilde{j}_1-\tilde{j}_2)_{k_5}}{k_5!(-2\tilde{j}_2)_{k_5}}\,
\ea
\ee
$$
\ba{l}
\times\Big(\frac{z_{_{34}}z_{_{54}}}{z_{_{53}}}\Big)^{k_5+j_4-\tilde{j}_2}\left[\frac{(2j_1)!(2j_2)!(2j_3)!(2\tilde{j_2}-2k_5)!}{\Delta(j_1,j_2,\tilde{j}_1)\Delta(j_3,\tilde{j_2}-k_5,\tilde{j}_1)}\right]^{\half}z_{_{12}}^{j_1+j_2}z_{_{34}}^{j_3+\tilde{j_2}-k_5}\left(\frac{z_{_{23}}}{z_{_{13}}}\right)^{j_2-j_1}\left(\frac{z_{_{24}}}{z_{_{23}}}\right)^{\tilde{j_2}-k_5-j_3}
\vspace{2mm}  
\\
\dps
\hspace{30mm}\times e^{\rho(j_1+j_2+j_3+j_4)}(2\tilde{j}_1)!\chi_4^{k_4-\tilde{j}_1} \frac{(j_2-j_1-\tilde{j}_1)_{k_4}(j_3-\tilde{j}_1-\tilde{j_2}+k_5)_{k_4}}{(-2\tilde{j}_1)_{k_4}k_4!}
\vspace{2mm}
\\ 
\dps
= \sum_{k_4\in K_4}(2\tilde{j}_1)!(2\tilde{j}_2)!\left[\frac{(2j_1)!(2j_2)!(2j_3)!(2j_4)!}{\Delta(j_1,j_2,\tilde{j}_1)\Delta(\tilde{j}_1,j_3,\tilde{j}_2)}\right]^{\half}
\left(\frac{z_{_{13}}z_{_{12}}}{z_{_{23}}}\right)^{j_1}\left(\frac{z_{_{12}}z_{_{32}}}{z_{_{3 1}}}\right)^{j_2}\left(\frac{z_{_{23}}z_{_{43}}}{z_{_{4 2}}}\right)^{j_3}
\vspace{2.5mm}  
\\
\dps
\hspace{30mm}\times e^{\rho(j_1+j_2+j_3+j_4)}\chi_4^{k_4-\tilde{j}_1}\chi_5^{k_5-\tilde{j}_2}
\frac{(j_2-j_1-\tilde{j}_1)_{k_4}(j_3-\tilde{j}_1-\tilde{j_2})_{k_4+k_5}}{(-2\tilde{j}_1)_{k_4}k_4!(-2\tilde{j}_{2})_{k_5}k_5!}\;.
\ea
$$
\paragraph{$\prod \beta \gamma$-factor.} By means of  \eqref{abc_factors}  the $\beta\gamma$-product is explicitly calculated to be   
\be
\label{first_factor}
\ba{l}
\dps
\beta_{i+1,{k_{i+1}}}\,\gamma_{i,{k_i},\tilde{j}_{i-2}-k_{i+1}} = (2\tilde{j}_{i-2})!
\left[{\frac{(\Delta(\tilde{j}_{i-3}, j_{i-1}, \tilde{j}_{i-2}-k_{i+1})!}{(2\tilde{j}_{i-2}-2k_{i+1})!}}\right]^{\half}\left[{\frac{(2j_i)!}{\Delta(\tilde{j}_{i-3},j_{i-1},\tilde{j}_{i-2})}}\right]^{\half}
\vspace{2.5mm}  
\\
\dps
\times e^{\rho j_i}\frac{(j_{i-1}-\tilde{j}_{i-2}-\tilde{j}_{i-3})_{k_{i+1}}}{k_{i+1}!(-2\tilde{j}_{i-2})_{k_{i+1}}}\,
\Big(\frac{z_{_{i-1\,i}}z_{_{i+1\,i}}}{z_{_{i+1\, i-1}}}\Big)^{k_{i+1}+j_i-\tilde{j}_{i-2}}\left[{\frac{{(2\tilde{j}_{i-2}-2k_{i+1})!}}{\Delta(\tilde{j}_{i-3},j_{i-1}, \tilde{j}_{i-2}-k_{i+1})}}\right]^{\half}
\vspace{2mm}
\\
\dps
\times(j_{i-1}-\tilde{j}_{i-2}-\tilde{j}_{i-3}+k_{i+1})_{k_i}
\Big(\frac{z_{_{i\,i-1}}z_{_{i\,i-2}}}{z_{_{i-2\, i-1}}}\Big)^{\tilde{j}_{i-2}-k_{i+1}}\Big(\frac{z_{_{i-1\,i}}z_{_{i+1\,i}}}{z_{_{i+1\, i-1}}}\Big)^{j_i} = (2\tilde{j}_{i-2})!
\vspace{2.5mm}  
\\
\dps\times\left[{\frac{(2j_i)!}{\Delta(\tilde{j}_{i-3},j_{i-1},\tilde{j}_{i-2})}}\right]^{\half}\Big(\frac{z_{_{i+1\,i}}z_{_{i-1\, i-2}}}{z_{_{i+1\, i-1}}z_{_{i\,i-2}}}\Big)^{k_{i+1}-\tilde{j}_{i-2}}\Big(\frac{z_{_{i-1\,i}}z_{_{i+1\,i}}}{z_{_{i+1\, i-1}}}\Big)^{j_i}\frac{(j_{i-1}-\tilde{j}_{i-2}-\tilde{j}_{i-3})_{k_i+k_{i+1}}}{k_{i+1}!(-2\tilde{j}_{i-2})_{k_{i+1}}}\,,
\ea
\ee 
where we used the obvious identity  
\be
(j_{i-1}-\tilde{j}_{i-2}-\tilde{j}_{i-3}+k_{i+1})_{k_i}(j_{i-1}-\tilde{j}_{i-2}-\tilde{j}_{i-3})_{k_{i+1}} = (j_{i-1}-\tilde{j}_{i-2}-\tilde{j}_{i-3})_{k_i+k_{i+1}}\;.
\ee 
\paragraph{Closed-form formula.} Now that we have all ingredients we can write down the final answer. Substituting the factors \eqref{abc_factors}, \eqref{second_factor}, and \eqref{first_factor}  into  the recursive solution \eqref{n_mod_vert_schem} one finds the $n$-point AdS vertex function in the form: 
\be
\label{n_pt_ver}
\ba{l}
\dps
\cV_{j_1 \cdots j_n\tilde{j}_1 \cdots \tilde{j}_{n-3}}(\rho,{\bf z})=e^{\rho\sum_{i=1}^n j_i}\prod_{i=1}^{n-3}(2\tilde{j}_i)!\Bigg[{\frac{ \prod_{i=1}^{n}(2j_i)!}{{\Delta(j_1,j_2,\tilde{j}_1)\displaystyle\prod_{i=3}^{n-2}\Delta(\tilde{j}_{i-2},j_i,\tilde{j}_{i-1})\Delta(\tilde{j}_{n-3},j_{n-1},j_n)}}}\Bigg]^{\half}
\vspace{2mm} 
\\
\dps
\hspace{47.5mm}\times\prod_{l=2}^{n-1}\Big(\frac{z_{_{l+1\,l}}z_{_{l-1\,l}}}{z_{_{l+1\,l-1}}}\Big)^{j_l}\Big(\frac{z_{_{12}}z_{_{13}}}{z_{_{23}}}\Big)^{j_1}\Big(\frac{z_{_{n\,n-1}}z_{_{n\,n-2}}}{z_{_{n-2\,n-1}}}\Big)^{j_n}
\vspace{2mm}
\\
\dps
\times\sum_{\{k_i\in K_i\}, i=4,...,n}\Bigg[\chi^{-\tilde{j}_1+k_4}_4\frac{(j_2-j_1-\tilde{j}_1)_{k_4}(j_3-\tilde{j}_1-\tilde{j}_2)_{k_4+k_5}}{k_4!(-2\tilde{j}_1)_{k_4}}\chi^{-\tilde{j}_{n-3}+k_n}_n\frac{(j_{n-1}-j_n-\tilde{j}_{n-3})_{k_n}}{k_n!(-2\tilde{j}_{n-3})_{k_n}}
\vspace{2mm}
\\
\dps
\hspace{23mm}\times \prod_{i=5}^{n-1}\bigg(\chi^{-\tilde{j}_{i-3}+k_i}_i\frac{(j_{i-1}-\tilde{j}_{i-2}-\tilde{j}_{i-3})_{k_i+k_{i+1}}}{k_i!(-2\tilde{j}_{i-3})_{k_i}}\bigg)
\Bigg],
\ea 
\ee 
where the summation domains are given by 
\be
\label{comb_dom}
\ba{l}
\dps
K_4 := [\![0, \min(-j_2+j_1+\tilde{j}_1, -j_3+\tilde{j}_1+\tilde{j_2})]\!]\,,
\vspace{1mm} 
\\
K_i:=[\![0, \min(-j_{i-2}+\tilde{j}_{i-3}+\tilde{j}_{i-4}-k_{i-1}, -j_{i-1}+\tilde{j}_{i-2}+\tilde{j}_{i-3})]\!]\,, \quad i = 5,..., n-1\,, 
\vspace{1mm} 
\\
K_n:=[\![0, \min(-j_{n-2}+\tilde{j}_{n-3}+\tilde{j}_{n-4}-k_{n-1}, -j_{n-1}+j_n+\tilde{j}_{n-3})]\!]\,.
\ea
\ee 
By introducing the structure constants \eqref{prefactor} and  the $n$-point leg factor as
\be \dps
\mathcal{L}_{j_1 \dots j_n}({\bf z}) = \prod_{l=2}^{n-1}\Big(\frac{z_{_{l+1\,l}}z_{_{l-1\,l}}}{z_{_{l+1\,l-1}}}\Big)^{j_l}\Big(\frac{z_{_{12}}z_{_{13}}}{z_{_{23}}}\Big)^{j_1}\Big(\frac{z_{_{n\,n-1}}z_{_{n\,n-2}}}{z_{_{n-2\,n-1}}}\Big)^{j_n},
\ee 
the expression  \eqref{n_pt_ver} can be represented in terms of  the  comb function (see \eqref{comb}, the last line; the summation domain in the comb function \eqref{comb-domains}  matches the domain \eqref{comb_dom}): 
\be
\label{comb_fin}
\ba{l}
\dps
\cV_{j_1 \cdots j_n\tilde{j}_1 \cdots \tilde{j}_{n-3}}(\rho,{\bf z})=e^{\rho\sum_{i=1}^n j_i}\,C_{j\tilde{j}}\;\mathcal{L}_{j_1 \dots j_n}({\bf z}) \left(\prod_{k=1}^{n-3}\chi_{k+3}^{-\tilde{j}_k}\right)
\vspace{2.5mm}  
\\
\dps
\hspace{10mm}\times F_{n-3} \left[\begin{array}{cccc}
     j_2-j_1-\tilde{j}_1,&j_3-\tilde{j}_1-\tilde{j}_2,&...\,,&j_{n-1}-j_n-\tilde{j}_{n-3} \\
     -2\tilde{j}_1, &-2\tilde{j}_2,&...\,, &-2\tilde{j}_{n-3} \end{array};\chi_4,...\,,\chi_n\right]\,.
\ea 
\ee 
Up to the structure constants this expression is  the $n$-point global conformal block in the comb channel \cite{Rosenhaus:2018zqn}.

\section{Conclusion}
\label{sec:conclusion}

In this work we have  been studying  the \adscft correspondence as the Wilson network/conformal block correspondence. Having  defined the AdS$_2$ vertex functions as  the Wilson network operators averaged over the cap states we have calculated the $n$-point global conformal blocks in the comb channel as well as the respective structure constants through the extrapolate dictionary relation. In particular, our results include: 

\begin{itemize}

\item Extrapolate dictionary relation for the AdS$_2$ vertex functions and CFT$_1$ correlation functions. In this paper, contrary to the previous studies in \cite{Bhatta:2016hpz,Besken:2017fsj,Bhatta:2018gjb,Alkalaev:2020yvq}, we have shifted the focus from just calculating  conformal blocks on the boundary to the study of  AdS$_2$ vertex functions -- bulk-to-bulk gravitational Wilson line networks -- as independent tools to probe the geometry related to local scalar fields in the bulk. In this context, our study is the first step towards extending the 2-point analysis in Ref. \cite{Castro:2018srf} to any number of points $n$.

\item Cap states in (in)finite-dimensional modules in Table \bref{tab:cap} and analytic formulas for the Wilson matrix elements in Proposition \bref{prop3}. Also, we have formulated  Propositions \bref{prop2} and \bref{prop1} about their large-$\rho$ asymptotics which demonstrate  a universal behaviour of the Wilson matrix elements in different $\sl2$ modules near the \ads boundary. In particular, it is shown that among this  family of  AdS$_2$ vertex functions having the same boundary behaviour there is just one member which is $\sl2$ invariant in the bulk.

\item Explicit calculation of $n$-point global  conformal blocks via Wilson line networks. We have shown that the AdS$_2$ vertex functions near the boundary satisfy a simple recursion relation that can be explicitly solved to give conformal blocks in the form previously known on the CFT$_1$ side. 

\end{itemize}

\noindent These constructs can be straightforwardly extended to the case of AdS$_3$/CFT$_2$ in the spirit of Refs. \cite{Bhatta:2016hpz,Besken:2017fsj,Bhatta:2018gjb,Alkalaev:2020yvq,Belavin:2022bib,Belavin:2023orw} just because of the (anti)chiral factorization underlying both the $\sl2\times \sl2$ Chern-Simons theory and the boundary CFT$_2$. The three-dimensional case is important on its own because the gravitational connections may have non-trivial holonomies and  the respective Wilson lines may probe various topological defects such as the BTZ black hole. The only essential change in studying the AdS$_3$ vertex functions  is to choose different cap states defined by $\sl2 \oplus \sl2$ algebra of global (conformal) isometries which were  analyzed for the respective infinite modules  in \cite{Nakayama:2015mva, Bhatta:2016hpz,Castro:2018srf}.

It would be interesting to extend  the study  of 2-point AdS$_3$ vertex functions  \cite{Castro:2018srf,Castro:2020smu} to the multipoint case by finding exact analytic expressions and  formulating  HHKL construction as well as considering the topological defects in the bulk. Exact formulas for the Wilson matrix elements obtained in this paper can be useful in achieving this objective. A sample calculation has been given here in Section \bref{sec:2pt} in the case of the 2-point \ads vertex function  which expectedly reproduced the well-known form of the bulk-to-bulk propagator \cite{Castro:2018srf}.

Yet another interesting  area of research is to elaborate  the Wilson network approach in AdS$_3$ spacetimes with conformal boundaries being genus-$g$ Riemann surfaces (the present case is $g=0$) that would allow one to formulate analogous near-boundary  recursion relations and find closed-form formulas for genus-$g$ global conformal blocks. The interest towards genus-$g$ Wilson networks, both boundary-to-boundary and bulk-to-bulk, is related to the study of non-vacuum AdS geometries  with (dis)connected genus-$g$ conformal boundaries, see e.g.  \cite{Guica:2015zpf,Guica:2016pid,Henneaux:2019sjx}. Up to date, definition and analysis of the Wilson matrix network in gravitational (non-compact) CS theory with a genus-$g$ boundary conditions  as well as conformal blocks on genus-$g$ Riemann surfaces are poorly understood apart from a notable exception of $g=1$ case given by the thermal AdS$_3$ and boundary torus CFT$_2$. In particular, considering the boundary Wilson matrix elements as the tool to compute  conformal blocks, one may hope that they both will have the similar  recursive structure that will allow one to find concise analytic expression by analogy with the comb function of \cite{Rosenhaus:2018zqn} at $g=0$ and the necklace function recently found in \cite{Alkalaev:2023evp} at $g=1$.

The present Wilson line networks can be considered in other channels beyond the comb channel.\footnote{In the context of the geodesic networks in AdS$_3$ non-comb channels were considered e.g. in \cite{Banerjee:2016qca}.} Apart from the cap states which remain the same, the respective Wilson line operators can be  directly defined by means of  $n$-valent intertwiners of different topologies (i.e. those that are obtained by tensoring $sl(2)$ modules in different orders). They are analogous to the $n$-valent intertwiner introduced in Section \bref{sec:gravity} for the AdS vertex function in the comb channel. Note that the check of  the Ward identities in the bulk crucially depends on that the $n$-valent intertwiner is $sl(2)$ invariant that results in the same cap state condition. Thus, the only (essential) difference between different channels will be an order of summations with one or another intertwiner. On the CFT side this exactly matches different  OPE channels. It would be interesting to compute CFT blocks as near-boundary AdS vertex functions in non-comb channels and compare the resulting functions with those described in \cite{Fortin:2020zxw}, where  global $n$-point conformal blocks  were considered  by formulating the rule set which allows one to constructively compute global blocks in arbitrary channels as  hypergeometric-type series.\footnote{In the comb channel case this procedure reproduces the comb function of \cite{Rosenhaus:2018zqn}.}

Finally, thinking about higher-dimensional extensions of the Wilson line construction one notes that gravity in AdS$_{d+1}$ for $d \geq 3$ is not topological so that applications of the gravitational connections subject to the zero-curvature condition are therefore limited (see however, \cite{Bhatta:2018gjb}). Nevertheless, it would be interesting to study bulk-to-bulk Wilson networks within the frame formulations of higher-dimensional gravity  and, in the future,   higher-spin theory. Recall that the unfolded formulation of higher-spin theory naturally operates in terms of zero-curvature conditions imposed 0-form and 1-form fields taking values in the higher-spin algebra \cite{Bekaert:2004qos}.  Coming back to lower dimensions, such higher-spin theories identified with  $sl(n)\oplus sl(n)$ Chern-Simons theories were considered in the present  context in \cite{Ammon:2013hba,deBoer:2013vca,deBoer:2014sna,Hegde:2015dqh}. Developing the full-fledged Wilson line approach for any $n$ would result, in particular, in finding general  $sl(n)$ global conformal blocks arising as the large-$c$ limit of $W_n$ CFT (for $n=3$ see \cite{Fateev:2011qa,Besken:2016ooo}).

\vspace{2mm}

\noindent \textbf{Acknowledgements.}
We are  grateful to  Semyon Mandrygin and Mikhail Pavlov for useful discussions.

\appendix

\section{$3j$ symbol and special functions}
\label{app:hyper}

\paragraph{$3j$ symbol and intertwiner.} The $3j$ symbol is defined as the matrix element of the 3-valent intertwiner (see e.g. \cite{varshalovich})
\be 
\label{3-j}
\ba{l}
\dps
\begin{pmatrix}j_a&j_b&j_c\\a&b&c\end{pmatrix} = (-1)^{j_{a}-j_{b}-c}\delta_{a+b+c, 0}\sqrt{\frac{(j_{a}+j_{b}-j_{c})!(j_{a}-j_{b}+j_{c})!(j_{b}-j_{a}+j_{c})!}{(j_{a}+j_{b}+j_{c}+1)!}}
\ea
\ee
$$
\times\sum_{m\in M}\frac{(-1)^{m}\sqrt{(j_{a}+a)!(j_{a}-a)!(j_{b}+b)!(j_{b}-b)!(j_{c}+c)!(j_{c}-c)!}}{m!(j_{c}-j_{b}+m+a)!(j_{c}-j_{a}+m-b)!(j_{a}+j_{b}-j_{c}-m)!(j_{a}-m-a)!(j_{b}-m+b)!}\,,
$$
where $M:= [\![\max(0, b+j_{a}-j_{c}, -a-j_{c}+j_{b}),\min(b+j_{b}, j_{a}-a,j_{a}+j_{b}-j_{c})]\!]$. Note that parameters $j_a,j_b,j_c$ satisfy the triangle inequalities $|j_a-j_b|\leq j_c\leq j_a+j_b$ and $j_a+j_b+j_c\in\mathbb{Z}$. At $a=j_a$ the $3j$ symbol is drastically simplified, 
\be 
\label{s3-j}
\ba{l}
\dps
\begin{pmatrix}j_a&j_b&j_c\\j_a&b&c\end{pmatrix} = (-1)^{j_{a}-j_{b}-c}\delta_{j_a+b+c, 0}
\vspace{2mm}\\\dps
\hspace{15mm} \times \sqrt{\frac{(2j_a)!(-j_{a}+j_{b}+j_{c})!(j_a+j_b+c)!(j_c-c)!}{(j_{a}+j_{b}+j_{c}+1)!(j_{a}-j_{b}+j_{c})!(j_{a}+j_{b}-j_{c})!(-j_a+j_b-c)!(j_c+c)!}}\,.
\ea\ee
The $3$-valent intertwiner for finite-dimensional modules is given by \cite{HOLMAN19661}
$$ 
\ba{l}
\label{general-intertwiner}\dps
[I_{j_1j_2j_3}]^{m_1}{}_{m_2m_3} = \sum_{l = [\![-j_1,j_1]\!]}(-1)^{j_1-m_1}\delta_{l,-m_1}\begin{pmatrix}j_1&j_2&j_3\\l&m_2&m_3\end{pmatrix} = \delta_{m_1,m_2+m_3}(-1)^{j_2+m_1-m_3}
\vspace{2mm}
\\
\dps
\times\Bigg[\frac{\Gamma(1-j_1+j_2+j_3)\Gamma(j_1+j_2-j_3+1)\Gamma(1+j_1-j_2+j_3)\Gamma(1+m_1+j_1)\Gamma(-m_1+j_1+1)}{\Gamma(j_1+j_2+j_3+2)}\Bigg]^{\half}
\ea
$$
\be
\ba{l}
\dps
\times\sum_{k\in K}\frac{\sqrt{\Gamma(-m_3+j_3+1)\Gamma(-m_2+j_2+1)\Gamma(1+m_2+j_2)\Gamma(1+m_3+j_3)}}{\Gamma(-m_2-j_1+j_3+k+1)\Gamma(-m_1+j_3-j_2+k+1)}\vspace{2mm}
\\
\dps
\times (-1)^{k}\frac{1}{k!\Gamma(m_1+j_1-k+1)\Gamma(1+m_2+j_2-k)\Gamma(j_1+j_2-j_3-k+1)}\; ,
\ea
\ee 
where $K:= [\![\max(0, m_2+j_1-j_3, m_1-j_3+j_2),\min(m_2+j_2, m_1+j_1,j_1+j_2-j_3)]\!]$ for $j_1, j_2, j_3 \in \mathbb{N}/2$. In \cite{HOLMAN19661} it was shown  that the 3-valent intertwiner for  discrete series $\cD^{\pm}_j$ is the analytical continuation of \eqref{general-intertwiner} to any real weights $j_1, j_2, j_3$. Throughout the paper we denote $\Gamma(z+1) = z!$ for any $z$. 

\paragraph{Binomial expansion.} The  generalized Newton  binomial  is given by 
\be
\label{newton}
\dps(x+y)^{a}=\sum_{k=0}^{\infty}\frac{\Gamma(a+1)}{k!\Gamma(a-k+1)}\,x^{k}y^{a-k}\,, 
\ee
where $a\in \mathbb{R}$ and  $|x|<|y|$.

\paragraph{Hypergeometric function.} The hypergeometric series  ${}_2F_1(a, b; c| z)$ is given by 
\be
\label{def_hyper}
{}_2F_1(a_1, a_2; c| z)=\sum_{m=0}^{\infty}\frac{(a_1)_m (a_2)_m}{(c)_m}\frac{z^m}{m!}\,,
\ee
where $(p)_k={\Gamma(p+k)}/{\Gamma(p)}$ is the Pochhammer symbol and  $|z|<1$. The series  can be  analytically continued to the whole complex plane with the branch cut $(1,\infty)$ by means of  the Euler integral representation.   The hypergeometric function at different  (small/large) values of $z$ can be related by the    Pfaff transformation 
\be
\label{4kummer2f1}
{}_2F_1(a_1, a_2; c|z) =(1-z)^{-a_2}\,{}_2F_1(c-a_1, a_2; c| \frac{z}{z-1})\,.
\ee
At  $c = a_1-a_2+1$ the hypergeometric function satisfies the quadratic relation
\be 
\ba{l}
\label{quadratic}
\dps
{}_2F_1(a_1, a_2; a_1-a_2+1|z)
\vspace{2mm}
\\\dps
=(1-\sqrt{z})^{-2a_1}\F\left(a_1,a_1-a_2+\half;2a_1-2a_2+1\, \big|-\frac{4\sqrt{z}}{(1-\sqrt{z})^2}\right)\,.
\ea
\ee 
At $z=-1$ and $c = a_1-a_2+1$ the hypergeometric function can be represented as
\be 
\label{hyper_-1}
{}_2F_1(a_1, a_2; a_1-a_2+1|-1)=\frac{2^{-a_1}\sqrt{\pi}(a_1-a_2)!}{(\frac{a_1}{2}-a_2!)(\frac{a_1-1}{2})!}\,.
\ee 

\paragraph{Lemma.} Using \eqref{def_hyper} and \eqref{4kummer2f1} one can prove the following relation:
\be
\label{lemma}
\ba{l}
\dps\sum_{t\in T_1}\frac{(j_a-n)!w^{t}}{(j_b-j_c+n+t)!(j_a-n-t)!(j_a+j_c-j_b-t)!t!} = \frac{1}{(j_a+j_c-j_b)!(j_b-j_c+n)!}
\vspace{2mm}
\\ 
\dps
\times\,{}_2F_1(n-j_a, j_b-j_c-j_a; j_b-j_c+n + 1| w) =\frac{(1-w)^{j_a+j_c-j_b}}{(j_a+j_c-j_b)!(j_b-j_c+n)!}
\vspace{2mm}
\\ 
\dps
\times\,{}_2F_1(j_b-j_c+j_a+1, j_b-j_c-j_a; j_b-j_c+n + 1| \frac{w}{w-1})
\vspace{2mm}
\\ 
\dps
=\frac{(1-w)^{j_c-j_b+j_a}}{(j_a+j_b-j_c)!}\sum_{t\in T_2} \frac{ (j_a+j_b-j_c+t)!(-\frac{w}{w-1})^t}{(j_a-j_b+j_c-t)!(j_b-j_c+n+t)!t!}\,,
\ea
\ee
where $j_a,\,j_b,\,j_c\in \mathbb{N}/2$ and $n\in \mathbb{Z}$,  the summation domains are
\be
\ba{l}
T_1:=[\![\max(0,-j_{b}+j_c),\min(j_a-n, j_a+j_c-j_b)]\!]\;,
\vspace{2mm}
\\
\dps
T_2:=  [\![\max(0,j_c-j_a-j_b, j_c-j_b-n), j_a+j_c-j_b]\!]\;. 
\ea
\ee

\paragraph{Appell function.} The second Appell function is defined as 
\be 
F_2 \left[\begin{array}{ccc}
     a_1,&b, &a_2  \\
     &c_1, c_2 \end{array};z_1,z_2\right] = \sum_{m_1,m_2 =0}^{\infty}\frac{(a_1)_{m_1}(b)_{m_1+m_2}(a_2)_{m_2}}{(c_1)_{m_1}(c_2)_{m_2}}\frac{z_1^{m_1}}{m_1!}\frac{z_2^{m_2}}{m_2!}\,.
     \label{def_appell}
\ee 
Similarly to \eqref{4kummer2f1} the second Appell function satisfies the following identity  
\be
F_2 \left[\begin{array}{ccc}
     a_1,&b_1, &a_2  \\
     &c_1, c_2 \end{array};z_1,z_2\right] = (1-z_1)^{-b_1}F_2 \left[\begin{array}{ccc}
     c_1-a_1,&b_1, &a_2  \\
     &c_1, c_2 \end{array};\frac{z_1}{z_1-1},-\frac{z_2}{z_1-1}\right]\label{1kummerf2}\;,\ee
\be\hspace{13mm}=(1-z_1-z_2)^{-b_1}F_2 \left[\begin{array}{ccc}
     c_1-a_1,&b_1, &c_2-a_2  \\
     &c_1, c_2 \end{array};\frac{z_1}{z_1+z_2-1},\frac{z_2}{z_1+z_2-1}\right]\,.\label{3kummerf2}
\ee
Note that the second identity can be obtained from the first one by using the symmetry property, $F_2 \left[\begin{array}{ccc}
     a_1,&b_1, &a_2  \\
     &c_1, c_2 \end{array};z_1,z_2\right] = F_2 \left[\begin{array}{ccc}
     a_2,&b_1, &a_1  \\
     &c_2, c_1 \end{array};z_2,z_1\right]$, and making the transformation \eqref{1kummerf2} twice. 
\paragraph{Comb function.} Both the hypergeometric function  and the second  Appell function can be viewed as  particular cases of the comb function \cite{Rosenhaus:2018zqn}:
\be 
\label{comb}
\ba{l}
F_N \left[\begin{array}{ccccc}
     a_1,&b_1,&...\,,&b_{N-1}, &a_2 \\
     &c_1,&...\,, &c_N \end{array};z_1,...\,,z_N\right] 
\vspace{2mm}
\\ 
\dps
= \sum_{m_1,...,m_N=0}^{\infty}\frac{(a_1)_{m_1}(b_1)_{m_1+m_2}\dots(b_{N-1})_{m_{N-1}+m_N}(a_2)_{m_N}}{(c_1)_{m_1}\dots(c_N)_{m_N}}\frac{z_1^{m_1}}{m_1!}\cdots \frac{z_n^{m_N}}{m_N!}
\vspace{2mm}
\\ 
\dps
\equiv \sum_{m_1,...,m_N=0}^{\infty} z_1^{m_1}\frac{(a_1)_{m_1}(b_1)_{m_1+m_2}}{(c_1)_{m_1}m_1!}\,
z_N^{m_N}\frac{(a_2)_{m_N}}{(c_N)_{m_N}m_N!}
\,\prod_{i=2}^{N-1}\bigg(z_i^{m_i}\frac{(b_i)_{m_{i}+m_{i+1}}}{(c_i)_{m_i}m_i!}\bigg).
    
\ea 
\ee 
The summation domain can change depending on particular values of parameters. We are interested in the  case of negative integer parameters $a,b,c\in\mathbb{Z}_-$ subjected to the following restrictions: $a_1>c_1, b_i>c_i, b_i>c_{i+1}, a_2>c_N$, where $i=1,...,N-1$, which are in fact the triangle inequalities \eqref{triangle}.    The zeros of the Pochhammer symbols in \eqref{comb} truncate  the summation domain from  infinite to finite one:
\be 
\label{comb-domains}
\ba{l}
\dps
\;m_1\in[\![0, \min(-a_1, -b_1)]\!]\;,
\\
\dps
\;m_i\in[\![0, \min(-b_{i-1}-m_{i-1}, -b_i)]\!]\;,\quad i=2,..., N-1\;,
\\
\dps
m_N\in[\![0, \min(-b_{N-1}-m_{N-1},-a_2)]\!]\;.
\ea
\ee 

Note that the comb function satisfies the following Pfaff-type identity 
\be 
\ba{l}
\label{comb_pfaff}
F_{N} \left[\begin{array}{ccccc}
     a_{1},&b_{1},&...\,,&b_{N-1}, &a_{2} \\
     &c_{1},&...\,, &c_{N} \end{array};z_{1},...\,,z_{N}\right]
\vspace{2mm}
\\ 
\dps
=(1-z_1)^{-b_1}F_{N} \left[\begin{array}{ccccc}
     c_1-a_{1},&b_{1},&...\,,&b_{N-1}, &a_{2} \\
     &c_{1},&...\,, &c_{N} \end{array};\frac{z_{1}}{z_1-1},-\frac{z_{2}}{z_1-1},z_3 ...\,,z_{N}\right]\,,
\ea
\ee 
which generalizes those for the hypergeometric and Appell functions given  above. Another identity for the comb function is given by
\be 
\ba{l}
\label{double_pfaff_comb}
F_{N} \left[\begin{array}{ccccc}
     a_{1},&b_{1},&...\,,&b_{N-1}, &a_{2} \\
     &c_{1},&...\,, &c_{N} \end{array};z_{1},...\,,z_{N}\right]=(1-z_1)^{-b_1}(1-z_N)^{-b_{N-1}}
\vspace{2mm}
\\ 
\dps
 \times F_{N}\left[\begin{array}{ccccc}
     c_1-a_1,&b_1,&...\,,&b_{N-1}, &c_N-a_2 \\
     &c_1,&...\,, &c_{N} \end{array};\frac{z_{1}}{z_1-1},-\frac{z_2}{z_1-1},z_3 ...,z_{N-2}\,,-\frac{z_{N-1}}{z_N-1},\frac{z_N}{z_N-1}\right]\,.
\ea
\ee 
This can be shown by using  the following symmetry property of the comb function, 
\\
$F_{N} \left[\begin{array}{ccccc}
     a_{1},&b_{1},&...\,,&b_{N-1}, &a_{2} \\
     &c_{1},&...\,, &c_{N} \end{array};z_{1},...\,,z_{N}\right] = F_{N} \left[\begin{array}{ccccc}
     a_{2},&b_{N-1},&...\,,&b_{1}, &a_{1} \\
     &c_{N},&...\,, &c_{1} \end{array};z_{N},...\,,z_{1}\right]$,  and making the transformation  \eqref{comb_pfaff} twice. In this paper we do not use the transformations \eqref{comb_pfaff} and \eqref{double_pfaff_comb}, but they may be useful in calculations similar to those done  in Appendix \bref{sec:3-5}. It would be interesting to find other identities (including linear, quadratic, cubic, etc.)  

\section{Detailed calculations}
\label{app:matrix}

\subsection{Solving the cap state condition}
\label{app:cap}

It the case $j \in \mathbb{N}_0+\half$ the only solution to the cap state condition \eqref{cap_rel} is given by  $|j \rangle\!\rangle_2$ from \eqref{integer_sol2}. To show this  one considers the  ansatz  
\be 
\label{half_int_anz}
\ket{a} = \sum_{n=0}^{2j}f_n(J_1)^{n}\ket{j,j}\;.
\ee 
The condition \eqref{cap_rel} imposed on  \eqref{half_int_anz} gives the relation
\be
\ba{l}\dps
(J_1+J_{-1})|\alpha \rangle\!\rangle = \sum_{n=0}^{2j}f_n\left[(J_1)^{n+1}+J_{-1}(J_1)^{n}\right]\ket{j,j}
\vspace{2mm}
\\
\dps
 = \sum_{n=0}^{2j-2}\left[f_n+(2j-n-1)(-n-2)f_{n+2}\right](J_1)^{n+1}\ket{j,j}
-2jf_1\ket{j,j}+f_{2j-1}(J_1)^{2j}\ket{j,j}=0\;,
\ea
\ee
which is equivalent to the following system
\be 
\label{chains}
f_1=0\;,
\qquad
f_{2j-1} = 0\;,
\qquad
f_{n+2} = \frac{1}{(n+2)(2j-n-1)}f_{n}\;.
\ee
One can see that $f_1=f_3=\ldots=f_{2n+1}=0$ and $f_{2j-1} = f_{2j-3} = \ldots = f_{2j-2n-1}=0$, with $n = 0,1,2,... \,$. Indices in the first chain of equations are odd while indices in the second one are even for $j\in \mathbb{N}_0+\half$, therefore, the coefficient $f_i=0$ for $i=0,1,...,2j$. Note that for $j\in \mathbb{N}_0$ the ansatz \eqref{half_int_anz} is the same  and the two chains of equations in \eqref{chains} coincide giving rise to  a non-trivial solution $|j \rangle\!\rangle_1$ \eqref{integer_sol1}. 

\subsection{Another form of the Ishibashi state} 
\label{app:LWME}

In order to calculate the Wilson  matrix elements we  represent the Ishibashi state  \eqref{cap_states} in a more convenient form (the conjugated state $\langle\!\langle j|$ is obtained by replacing $J_1 \to J_{-1}$). Recall that for notational convenience we replace all the  gamma-functions  by  factorials according to  $\Gamma(x+1) = x!$. First, consider the following chain of identities 
\be
\ba{l}\dps
\label{even_part}
\prod_{k=1}^n \frac{1}{-4k j +4k^2 -2k} = 2^{-2n}\prod_{k=1}^n\frac{1}{(-j +k -\half)k} = \frac{(-j-\half)!}{2^nn!2^n(-j+n-\half)!}
\vspace{2mm} 
\\
\dps
= \frac{2^n(n-\half)!2^n(-j+n-1)!(-2j-1)!}{(-\half)!(2n)!(-j-1)!(-2j+2n-1)!} = 2^{2n}\frac{(-2j-1)!}{(2n)!(-2j+2n-1)!}\frac{(-\half)!j!}{(-n-\half)!(j-n)!}
\vspace{2mm} 
\\
\dps
=\frac{(-2j-1)!}{(2n)!(-2j+2n-1)!}\frac{(-\half)!j!}{\sqrt{\pi}(j-2n)!}{}_2F_1(-2n, -j; j-2n+1|-1)\;,
\ea
\ee
where we used the relation \eqref{hyper_-1}  along with  the obvious relation $(2n)! = \prod_{k=1}^n2k (2k-1) = 2^{2n}\prod_{k=1}^nk (k-\half) = 2^{2n}n!\frac{(n-\half)!}{(-\half)!} $. Using the hypergeometric series  \eqref{def_hyper} one can show that
\be 
\ba{l}\dps
\label{odd_part}
\frac{(-2j-1)!}{(2n+1)!(-2j+2n)!}\frac{j!}{(j-2n-1)!}{}_2F_1(-2n-1, -j; j-2n|-1)
\vspace{2mm} 
\\
\dps
= \frac{(-2j-1)!}{(2n+1)!(-2j+2n)!} \sum_{k=0}^{2n+1}(-1)^{k}\frac{j!(-2n-2+k)!(-j-1+k)!}{(-2n-2)!(-j-1)!(j-2n-1+k)!k!}\\\dps
=-\frac{(-2j-1)!}{(-2j+2n)!((-j-1)!)^2} \sum_{k=0}^{2n+1}(-1)^{k}\frac{(-j-1+k)!(-j+2n-k)!}{(2n+1-k)!k!}\;.
\ea
\ee 
The $k$-th term here  reads
\be 
(-1)^{k}\frac{(-j-1+k)!(-j+2n-k)!}{(2n+1-k)!k!}\;,
\ee 
while ($2n+1-k$)-th term equals
\be 
(-1)^{2n+1-k}\frac{(-j+2n-k)!(-j-1+k)!}{k!(2n+1-k)!} = -(-1)^{k}\frac{(-j+2n-k)!(-j-1+k)!}{i!(2n+1-k)!}\;.
\ee 
These two types of  terms with $k = 0,...,n$  cancel each other in \eqref{odd_part} so that 
\be 
\sum_{k=0}^{2n+1}(-1)^{k}\frac{(-j-1+k)!(-j+2n-k)!}{(2n+1-k)!k!}=0\;.
\ee
The prefactor in \eqref{odd_part} is finite for $j\in \mathbb{R}$, whence, the whole expression \eqref{odd_part} equals zero, i.e.
\be 
\label{odd_part_final}
\frac{(-2j-1)!}{(2n+1)!(-2j+2n)!}\frac{j!}{(j-2n-1)!}{}_2F_1(-2n-1, -j; j-2n|-1)\equiv 0\;.
\ee
One can notice that \eqref{odd_part_final} is obtained from \eqref{even_part} by replacing $2n$ with $2n+1$. It means that  we  rewrite the coefficient in the Ishibashi state \eqref{cap_states} in the form  \eqref{even_part} and  add the zero expressed through the hypergeometric series. Then, we  use the identity \eqref{odd_part_final} to obtain
\be 
\ba{l}\dps
| j \rangle\!\rangle = \sum_{n=0}^{\infty}\frac{(-)^n(-2j-1)!}{(2n)!(-2j+2n-1)!}\frac{j!}{(j-2n)!}\,{}_2F_1(-2n, -j; j-2n+1|-1)(J_1)^{2n}\ket{j,j}
\vspace{2mm} 
\\
\dps
-\sum_{n=0}^{\infty}\frac{(-)^{n+\half}(-2j-1)!}{(2n+1)!(-2j+2n)!}\frac{ j!}{(j-2n-1)!}\,{}_2F_1(-2n-1, -j; j-2n|-1)(J_1)^{2n+1}\ket{j,j}.
\ea
\ee 
Unifying the two sums above into a single sum one represents the Ishibashi state as\footnote{The cap states discussed in \cite{Bhatta:2018gjb} satisfy the equation $(J_1-J_{-1})\ket{a} = 0$ which differs from the cap state condition \eqref{cap_rel} by a sign. The difference for solutions results in an additional $i^{2n}$ factor in \eqref{Ishibashi_new}. In fact, in this Appendix we demonstrate how to relate the two forms of the cap states given in \cite{Nakayama:2015mva} and \cite{Bhatta:2018gjb}  as the Bessel and the hypergeometric functions of generators, respectively. It turns out that the hypergeometric form is much more convenient for our purposes.}
\be
\label{Ishibashi_new}
| j \rangle\!\rangle = \sum_{n=0}^{\infty}\frac{(-i)^n(-2j-1)!}{n!(-2j+n-1)!}\frac{j!}{(j-n)!}\,{}_2F_1(-n, -j; j-n+1|-1)(J_1)^n\ket{j,j}.
\ee 
Note that despite the presence of $i^n$ the coefficients here are still real since all imaginary terms sum up to zero.  Using this new form of the Ishibashi state the right Wilson matrix element can be written as
\be 
\label{RWE_fin}
\ba{l}\dps
 \bra{j,m}W_j[x,0]| j \rangle\!\rangle = \sum_{n=0}^{\infty}\frac{(-i)^n(-2j-1)!}{n!(-2j+n-1)!}\frac{j!}{(j-n)!}\,{}_2F_1(-n, -j; j-n+1|-1)
\vspace{2mm} 
\\
\dps
\times\bra{j,m}e^{z J_1}e^{\rho J_0}(J_1)^n\ket{j,j} = \left[\frac{(2j)!(j-m)!j!^2}{(j+m)!}\right]^\half\,z^{j-m}e^{\rho j}
\vspace{2mm} 
\\
\dps
\times\sum_{n=0}^{j-m}\frac{i^n(-2j-1)!}{n!(-2j+n-1)!(j-n)!(j-m-n)!}\,{}_2F_1(-n, -j; j-n+1|-1)(-ze^{\rho})^{-n}\;.
\ea 
\ee
The left Wilson matrix element for the conjugated Ishibashi state can now be rewritten as
\be 
\label{LWE_fin}
\ba{l}
\dps
\langle\!\langle j|W_j[0,x] \ket{j,m} = \sum_{n=0}^{\infty}\frac{(-i)^n(-2j-1)!}{n!(-2j+n-1)!}\frac{j!}{(j-n)!}\,{}_2F_1(-n, -j; j-n+1|-1)
\vspace{2mm} 
\\
\dps
\hspace{10mm}\times\bra{j,j}(J_{-1})^ne^{-\rho J_0}e^{-z J_1}\ket{j,m} = \left[\frac{(j+m)!j!^2}{(j-m)!(2j)!}\right]^\half(-z)^{m-j}e^{-\rho j}
\vspace{2mm} 
\\
\dps
\hspace{20mm}\times\sum_{n=j-m}^{\infty}(-i)^n \, \frac{{}_2F_1(-n, -j; j-n+1|-1)}{(j-n)!(n+m-j)!}\;(-ze^{\rho})^n\;.
\ea 
\ee 
These expressions  are further analyzed in Sections \bref{sec:matrixR}, \bref{sec:matrixL}, and Appendices \bref{app:rightW}-\bref{app:NLO}. Using the cap state which  coefficients are expressed in terms of the  hypergeometric function turns out to be useful when finding the Wilson matrix elements in a closed form.

\subsection{Proving a closed-form of the right Wilson matrix element}
\label{app:rightW}

The calculation of the right Wilson matrix element \eqref{RWE_fin} proceeds by representing  the hypergeometric coefficients as \eqref{def_hyper} and changing  $n=-k+j-m$:
\be 
\ba{l}\dps
 \bra{j,m}W_j[x,0]| j \rangle\!\rangle = e^{\rho m}\left[\frac{(j-m)!}{(j+m)!(2j)!}\right]^\half\,\sum_{t=0}^{j-m}\sum_{k=0}^{t-j+m}(-iq)^k
\vspace{2mm} 
\\
\dps
\hspace{30mm}\times(i)^{j-m}(-)^{m}\frac{j!(k-j+m-1+t)!(m+j+k)!}{t!(j-t)!(-j-1)!(m+k+t)!k!}\;,
\ea 
\ee 
where, for convenience, we introduced a new variable $q=-ze^{\rho}$. Representing the sum over $k$ as the hypergeometric series \eqref{def_hyper} and making the Pfaff transformation \eqref{4kummer2f1} yields
\be 
\ba{l}\dps
 \bra{j,m}W_j[x,0]| j \rangle\!\rangle = \left[\frac{(j-m)!}{(j+m)!(2j)!}\right]^\half\,e^{\rho m}\,\sum_{t=0}^{j-m}i^{j-m}(-)^{m}\frac{j!(-j+m-1+t)!(m+j)!}{t!(j-t)!(-j-1)!(m+t)!}\
\vspace{2mm} 
\\
\dps
\hspace{40mm}\times(1+iq)^{j-m-t}{}_2F_1\left(-j+m+t, t-j; m+t+1\big|\frac{iq}{1+iq}\right).
\ea 
\ee 
In its turn the hypergeometric function in the second line can be represented as \eqref{def_hyper} with a summation parameter $k$. Then, by changing  $k = -n+j-m-t$ one obtains
\be 
\ba{l}\dps
 \bra{j,m}W_j[x,0]| j \rangle\!\rangle = \left[\frac{(j-m)!}{(j+m)!(2j)!}\right]^\half\,e^{\rho m}\sum_{n=0}^{j-m}\sum_{t=0}^{j-m-n}i^{j-m}(-)^{m+t}\
\vspace{2mm} 
\\
\dps
\hspace{30mm}\times(1+iq)^{n}(iq)^{j-m-n-t}\frac{(-1-n)!(m+j)!(-m-1-n)!}{(-j-1)!^2(j-n)!t!(j-m-n-t)!}\;.
\ea 
\ee 
After using the generalized Newton binomial \eqref{newton} and representing the sum over $n$ as the hypergeometric function by means of  \eqref{def_hyper} one finally finds:
\be 
\ba{l}\dps
\label{right_closed}
 \bra{j,m}W_j[x,0]| j \rangle\!\rangle = (-)^{j-m}\frac{j!}{m!}\left[\frac{(j+m)!}{(2j)!(j-m)!}\right]^\half\,e^{\rho m}
\vspace{2mm} 
\\
\dps
\hspace{30mm}\times(q+i)^{j-m}{}_2F_1\left(-j, m-j; m+1\big|\frac{q-i}{q+i}\right).
\ea 
\ee 
Note that the summation domains on each step of the calculation are finite. Thus, the variable $q$ is not restricted by the condition of convergence so the right Wilson matrix elements are well-defined for any real $q$, in particular, near $|q|=\infty$.

\subsection{Radius of convergence}
\label{app:radius}

Let us use  the ratio test to find the radius of convergence  of the power series \eqref{psf}:
\be 
\label{radius1}
R = \lim_{n\to\infty}\left|\frac{a_n}{a_{n+1}}\right| = \lim_{n\to\infty}\left|\frac{(n+m-j+1)}{(j-n)}\frac{{}_2F_1(-n, -j; j-n+1|-1)}{{}_2F_1(-n-1, -j; j-n|-1)}\right|,
\ee 
which by means of the identity \eqref{hyper_-1} can be cast into the form 
\be
R = \lim_{n\to\infty}\left|\frac{(n+m-j+1)}{2}\frac{(\frac{n}{2}-\half)!(\frac{n}{2}-j-1)!}{(\frac{n}{2})!(\frac{n}{2}-j-\half)!}\right|\;.
\ee
Then, using   the Stirling's approximation $n!\sim \sqrt{2\pi n}\,(n/e)^{n}$
one finds 
\be
\ba{c}
\dps
R = \lim_{n\to\infty}\left|\frac{(n+m-j+1)}{2}\left(\frac{n}{2e}\right)^{\frac{n}{2}-\half}e^{-\half}\left(\frac{n}{2e}\right)^{\frac{n}{2}-1-j}e^{-j-1}\left(\frac{n}{2e}\right)^{-\frac{n}{2}+\half+j}e^{j+\half}\left(\frac{n}{2e}\right)^{-\frac{n}{2}}\right|
\vspace{2mm}
\\
\dps
=\lim_{n\to\infty}\frac{(n+m-j+1)}{n}=1\;.
\ea
\ee
Thus, the radius of convergence does not depend on $m$ and $j$ so that any left Wilson matrix element converges for $|q|<1$, where $q = -z e^\rho$.

\subsection{Proving a closed-form of the left Wilson matrix element} 
\label{app:NLO}

Below we make a few  resummations  that allows us to find a closed-form formula for the left Wilson matrix element \eqref{LWE_fin}. Rewriting the hypergeometric coefficient  as \eqref{def_hyper} with a new summation  parameter $t$ and changing $n=k+t$ one finds:
\be 
\ba{l}\dps
\label{expnd_0}
\langle\!\langle j|W_j[0,x] \ket{j,m} = \left[\frac{(j+m)!}{(j-m)!(2j)!}\right]^\half q^{m-j}e^{-\rho m}
\vspace{2mm} 
\\
\dps
\hspace{30mm}\times \sum_{t=0}^\infty\sum_{k\in K}(-)^t\,\frac{(k+t)!(-j-1+t)!(-j+k-1)!}{t!k!(-j-1)!^2(k+m-j+t)!}\,(iq)^{k+t}\;,
\ea 
\ee 
where $K = [\![\max(0,j-m-t), \infty]\!]$.
Representing the sum over $k$ as the hypergeometric series with $|q|<1$ and making the Pfaff transformation \eqref{4kummer2f1} yields 
\be 
\ba{l}\dps
\label{expnd_1}
\langle\!\langle j|W_j[0,x] \ket{j,m} = \left[\frac{(j+m)!}{(j-m)!(2j)!}\right]^\half q^{m-j}e^{-\rho m}(1-iq)^{j}
\vspace{2mm} 
\\
\dps
\hspace{13mm}\times\sum_{t=0}^\infty\frac{(-j-1+t)!}{(-j-1)!(m-j+t)!}(-iq)^{t}\;{}_2F_1\left(-j, m-j; m-j+t+1\big|\frac{iq}{iq-1}\right).
\ea 
\ee 
The argument of the hypergeometric coefficient  here satisfies $|\frac{iq}{iq-1}|<1$  for any real $q$. It follows that one can again represent it as \eqref{def_hyper} with a new parameter $k$:
\be 
\ba{l}\dps
\label{expnd_2}
\langle\!\langle j|W_j[0,x] \ket{j,m} = \left[\frac{(j+m)!}{(j-m)!(2j)!}\right]^\half q^{m-j}e^{-\rho m}(1-iq)^{j}
\vspace{2mm} 
\\
\dps
\hspace{5mm}\times\sum_{k=0}^{j-m}\sum_{t=j-m-k}^\infty\frac{(k+m-j-1)!(-j+k-1)!j!}{k!(m-j-1)!(-j-1)!(k+m-j+t)!(j-t)!}(iq)^{k+t}(iq-1)^{-k}\,.
\ea 
\ee 
After changing  $t = s-k-m+j$ and using the generalized Newton binomial  \eqref{newton} (with $|q|<1$) one  sums over $s$ to  obtain:
\be 
\ba{l}\dps
\label{expnd_3}
\langle\!\langle j|W_j[0,x] \ket{j,m} = \left[\frac{(j+m)!}{(j-m)!(2j)!}\right]^\half i^{j-m}e^{-\rho m}(1-iq)^{j}(1+iq)^{m}
\vspace{2mm} 
\\
\dps
\hspace{30mm}\times\sum_{k=0}^{j-m}\frac{j!(k+m-j-1)!(-j+k-1)!}{k!(m-j-1)!(-j-1)!(k+m)!}(1-iq)^{-k}(iq+1)^k\;.
\ea 
\ee 
Finally, one sums over  $k$ in the second line  to obtain the hypergeometric series. The sum is finite and, hence, converges for any $q\in \mathbb{R}$ (possible poles $q=\pm i$ are not on the real axis). Thus, we have a closed-form formula for the left Wilson matrix element:
\be 
\ba{l}\dps
\label{expnd_4}
\langle\!\langle j|W_j[0,x] \ket{j,m} = (-)^j\frac{j!}{m!}\left[\frac{(j+m)!}{(j-m)!(2j)!}\right]^\half e^{-\rho m}
\vspace{2mm} 
\\
\dps
\hspace{30mm}\times(q+i)^{j}(q-i)^{m}\,{}_2F_1\left(-j, m-j; m+1\big|\frac{q-i}{q+i}\right).
\ea 
\ee 
The resulting expression is real despite the presence of complex-valued  arguments that can be explicitly seen by complex conjugation.  In Section \bref{sec:matrixL} we analytically continue this function past $|q|=1$.

\subsection{Wilson matrix elements for quasi-Ishibashi states}
\label{app:analyt}

\paragraph{Rotated HW state.} In this case, the calculation is much easier than that for the Ishibashi states.  To this end,  one represents the rotated HW cap state \eqref{cap_states2} as
\be
\label{NO_cs}
|j \rangle\!\rangle  = \sum_{n=0}^\infty \frac{(-)^n}{n!} \, (J_1)^{n} \ket{j,j}\,.
\ee  
The conjugated state $\langle\!\langle j|$ is obtained by replacing $J_1 \to J_{-1}$. The right Wilson matrix element is given by
\be 
\ba{l}\dps
\label{rotated_right}
\bra{j,m|W_j[0,x]}j \rangle\!\rangle = \sum_{n=0}^\infty \frac{(-)^n}{n!} \,  \bra{j,m}e^{zJ_1}e^{\rho J_0}(J_1)^{n}\ket{j,j} 
\vspace{2mm}
\\
\dps
= \left[\frac{(2j)!(j-m)!}{(j+m)!}\right]^\half \;\sum_{n=0}^{j-m} \frac{(-)^n}{n!} \,  e^{\rho (j-n)}\frac{z^{j-n-m}}{(j-n-m)!}
\vspace{2mm}
\\
\dps
= \left[\frac{(2j)!}{(j+m)!(j-m)!}\right]^\half e^{\rho m}(ze^{\rho}-1)^{j-m}
\equiv(-)^{j-m}\left[\frac{(2j)!}{(j+m)!(j-m)!}\right]^\half e^{\rho m}(1+q)^{j-m} \;,
\ea
\ee 
where in the last line we used the (standard) Newton binomial \eqref{newton}.
Its large-$\rho$ asymptotics reads
\be 
\bra{j,m|W_j[0,x]}j \rangle\!\rangle = \left[\frac{(2j)!}{(j+m)!(j-m)!}\right]^\half z^{j-m} e^{\rho j} + \cO(e^{\rho (j-1)})\;,
\ee 
cf. \eqref{RWMEf}

The left Wilson matrix elements is calculated to be 
\be 
\ba{l}\dps
\label{rotated_left0}
\langle\!\langle j|W_j[0,x]\ket{j,m} = \sum_{n=0}^\infty \frac{(-)^n}{n!} \,  \bra{j,j}(J_{-1})^{n}e^{-\rho J_0}e^{-zJ_1}\ket{j,m} 
\vspace{2mm}
\\
\dps
\hspace{30mm}=\sum_{n=j-m}^\infty(-)^n \,\frac{q^{n+m-j}}{n!(n+m-j)!} e^{-\rho m}\bra{j,j}(J_{-1})^{n}(J_1)^{n+m-j}\ket{j,m} 
\vspace{2mm}
\\
\dps
\hspace{30mm}=\left[\frac{(2j)!}{(j+m)!(j-m)!}\right]^{\half}e^{-\rho m}\sum_{t=0}^{\infty}\frac{(j+m)!}{t!(j+m-t)!}\,q^{t} \;,
\ea
\ee
where we changed the summation parameter $n = t-m+j$ when going from the second line to the third line  in which  we used the notation $q$ \eqref{psf}. The resulting sum has a finite radius of convergence, $|q|<1$, whence,  the matrix element is well-defined for small $|q|$ only.  

In order to find the large-$\rho$ asymptotics we analytically continue \eqref{rotated_left0} as follows. For small $|q|$, using the generalized Newton binomial \eqref{newton} one can  explicitly sum over $t$ to obtain the following   closed-form  expression: 
\be 
\label{twisted_boundary_int}
\langle\!\langle j|W_j[0,x]\ket{j,m} = \left[\frac{(2j)!}{(j+m)!(j-m)!}\right]^\half e^{-\rho m}(1+q)^{j+m}\,.
\ee 
This function  is analytically continued  to the whole complex $q$-plane avoiding branch points $q = -1$ and $q = \infty$, after which  $q$ is set to be real again (along with $z$ and $\rho$).  Choosing the branch cut as $(-1,i\infty)$, it follows that the branch cut restricts variables as $\rho \in \mathbb{R}$ and $z \in (-\infty,e^{-\rho})\cup(e^{-\rho},\infty)$ (indeed, avoiding  the branch cut at some $\rho\in\mathbb{R}$ implies the constraint $z\neq e^{-\rho}$, which becomes $z\neq0$ at $\rho = \infty$).\footnote{Note that the restriction $z\neq 0$ is an artefact of our  choice of the bulk point $y = 0$ (see the discussion below \eqref{vert_f}). In fact, different choices of $y$ yield other  deleted neighbourhoods.} Then, substituting $q = -ze^\rho$ into the analytically continued function \eqref{twisted_boundary_int} and expanding it near $\rho = \infty$ yields 
\be 
\label{twisted_boundary_fin}
\langle\!\langle j|W_j[0,x]\ket{j,m} = (-)^{j+m}\left[\frac{(2j)!}{(j-m)!(j+m)!}\right]^{\half}z^{j+m} e^{\rho j} + \cO(e^{\rho(j-1)})\;,
\ee 
cf. \eqref{LWMEf}. 

\paragraph{HW cap state.} This is the simplest case since the cap state is just the HW vector \eqref{cap_states3}. The conjugated cap state is just the LW weight vector. The left/right Wilson matrix elements are easily found to be 
\be 
\label{HLW1}
\ba{l}
\dps
\; \; \;\;\langle\!\langle j|W_j[0,x]\ket{j,m}
= (-)^{j+m}\left[\frac{(2j)!}{(j-m)!(j+m)!}\right]^\half z^{j+m}e^{\rho j}\;,
\ea
\ee 
\be 
\label{HLW2}
\ba{l}
\dps
\bra{j,m}W_j[x,0]|j \rangle\!\rangle  
= \left[\frac{(2j)!}{(j-m)!(j+m)!}\right]^\half z^{j-m}e^{\rho j}\;.
\ea
\ee 
These relations are exact in $\rho$.

Finally, note that the finite-dimensional matrix elements \eqref{HLW1}, \eqref{HLW2} can also be obtained from the rotated HW cap state analysis. Indeed, our consideration below \eqref{twisted_boundary_int} applies for $j+m \notin \mathbb{N}_0$. When $j+m \in \mathbb{N}_0$ the summation domain of $t$ in \eqref{rotated_left0} becomes finite as the binomial coefficient $\frac{(j+m)!}{t!(j+m-t)!}$  equals zero for $t>j+m$. This means that there is no need to analytically continue function \eqref{twisted_boundary_int} and one can directly expand near $\rho= \infty$. The condition $j+m \in \mathbb{N}_0$ along with the obvious relation $j-m\in \mathbb{N}_0$ \eqref{-D_basis} leads to $j\in \mathbb{N}_0/2$, i.e. in this case we effectively  deal with finite-dimensional modules. Note, however, that the singular submodule is not seen in this picture since the respective matrix elements are sub-leading. As we learned in Section \bref{sec:matrixI} the singular cap state contributions  are suppressed  near the boundary. On the other hand, the quasi-Ishibashi states correctly capture only near-boundary effects.

\subsection{Rearranging the $n$-point AdS vertex function}
\label{app:details}

For the sake of simplicity, in the $n$-point case we make the same transformation \eqref{n-transfrom} as in the $(n-1)$-point case. Recalling that the inverse transformation \eqref{inverse_transf} produces the Jacobians in the AdS vertex function \eqref{coftransconf} one observes that using the same coordinate map for the $n$-point and $(n-1)$-point AdS vertex functions yields almost the same  Jacobian factors. The only difference is that the $n$-point AdS vertex function gets an additional  $\Big(\frac{\partial z}{\partial w}\Big)^{j_n}\Big|_{w = w_n}$ because of the $n$-th point $w(z_n)$. Since  we aim to represent the $n$-point AdS vertex  function in terms of $(n-1)$-point ones by means of a recursion relation, then there is no need to consider coinciding factors in both expressions. Substituting \eqref{n-Wilson} and \eqref{general-intertwiner} into \eqref{n_point_with_aux} and resolving the Kronecker deltas by summing over $m_{n-1}$, $m_{n-2}$, $p_{n-3}$ and $p_{n-4}$ results in
$$
\ba{l}
\dps \cV_{j_1 \cdots j_n\tilde{j}_1 \cdots \tilde{j}_{n-3}}({\bf{w}}) = \hspace{-7mm}\sum_{m_n\in[\![-j_n,\tilde{j}_{n-3}-j_{n-1}]\!]}\hspace{-10mm}\omega^{2j_{n-2}}\mathbb{C}_{j_{n-1}+m_n-j_{n-2}}({\bf{w''}})\vspace{2mm} w_n^{j_n-m_n}  \left[\frac{(2j_n)!(2j_{n-1})!(2j_{n-2})!}{(\tilde{j}_{n-4}+j_{n-1}+m_n-j_{n-2})!}\right]^{\half} 
\\
\dps
\times\left[\frac{(\tilde{j}_{n-4}-j_{n-1}-m_n+j_{n-2})!(\tilde{j}_{n-3}+\tilde{j}_{n-4}-j_{n-2})!(\tilde{j}_{n-3}+j_n-j_{n-1})!}{(\tilde{j}_{n-3}-\tilde{j}_{n-4}+j_{n-2})!(\tilde{j}_{n-4}+j_{n-2}-\tilde{j}_{n-3})!(\tilde{j}_{n-4}+j_{n-2}+\tilde{j}_{n-3}+1)!(-\tilde{j}_{n-3}+j_n+j_{n-1})!}\right]^{\half}
\ea
$$
\be
\ba{l}\dps
\times\left[\frac{1}{(\tilde{j}_{n-3}+j_{n-1}+j_n+1)!(\tilde{j}_{n-3}+j_{n-1}-j_n)!}\right]^{\half} \frac{(-j_n-m_n-1)!}{(-\tilde{j}_{n-3}-j_{n-1}-m_n-1)!(\tilde{j}_{n-3}-j_{n-1}-m_n)!}\;. 
\ea
\ee
Changing $m_n = -k - j_{n-1}+\tilde{j}_{n-3}$ and using \eqref{prefactor} one finds 
\be
\ba{l}
\dps \hspace{-8mm}\cV_{j_1 \cdots j_n\tilde{j}_1 \cdots \tilde{j}_{n-3}}({\bf{w}}) = \hspace{-7mm}\sum_{k\in[\![0,j_n-j_{n-1}+\tilde{j}_{n-3}]\!]}\hspace{-3mm}\mathbb{C}_{\tilde{j}_{n-3}-j_{n-2}-k}({\bf{w''}})w_n^{j_n+k+j_{n-1}-\tilde{j}_{n-3}}
\vspace{2mm}
\\ 
\dps
\times \pref_{\tilde{j}_{n-4}j_{n-2}\tilde{j}_{n-3}}\pref_{\tilde{j}_{n-3}j_{n-1}j_n}
\frac{(j_{n-1}-j_n-\tilde{j}_{n-3})_k(j_{n-2}-\tilde{j}_{n-4}-\tilde{j}_{n-3})_k}{(-2\tilde{j}_{n-3})_k k!}\omega^{2j_{n-2}}
\vspace{2mm}
\\ 
\dps
\times\left[\frac{(\tilde{j}_{n-4}-\tilde{j}_{n-3}+j_{n-2}+ k)!(\tilde{j}_{n-4}+\tilde{j}_{n-3}-j_{n-2}-k)!}{(2\tilde{j}_{n-4})!}\right]^{\half}\,.
\ea
\ee
Applying the inverse transformation \eqref{inverse_transf} and taking the limit $\omega \to \infty$,
\be 
\ba{l}\dps
\cV_{j_1 \cdots j_n\tilde{j}_1 \cdots \tilde{j}_{n-3}}({\bf{z}})=\lim_{\omega\to\infty}\bigg[\big(\z{n-1}{n-2}\z{n-3}{n-1}\z{n-3}{n-2}\big)^{\sum_{k=1}^nj_{k}}\,\z{n-3}{n-2}^{-2j_{n-1}}\,\z{n-1}{n-2}^{-2j_{n-3}}\,\big(\omega\z{n-1}{n-3}\big)^{-2j_{n-2}}
\vspace{2mm}
\\
\dps
\times\big(\z{n-1}{n-3}w_n({\bf{z'}}) + \z{n-3}{n-2}\big)^{-2j_n}\prod_{i = 1}^{n-4}\big(\z{n-1}{n-3}w_i({\bf{z'}}) + \z{n-3}{n-2}\big)^{-2j_i}\cV_{j_1 \cdots j_n\tilde{j}_1 \cdots \tilde{j}_{n-3}}({\bf{w}}({\bf{z}}))\bigg]\,,
\ea
\ee 
we obtain the final expression for the $n$-point AdS vertex function \eqref{n_pt_vert}.

\section{Conformal transformations of the Wilson matrix elements}
\label{app:conf_transf}

As an independent check of the conformal invariance property we derive the transformation rule  \eqref{coftransconf} by considering the  gauge transformation with the gauge element \cite{Besken:2016ooo}
 \be
 \label{gauge_tr}
 g_{j}(x)=e^{-\rho J_0}e^{-2\ln(cz+d)J_0}e^{c(cz+d)J_{-1}}e^{\rho J_0},
 \ee
where $a,b,c,d$ are real parameters subjected to $ad-bc=1$, and $j$ indicates choosing particular module $\cR_j$. The gauge transformation changes the Wilson line according to  \eqref{ft1}:
\be
\label{Wilson_gauge}
W_j[x_1,x_2]\xrightarrow{} g_{j}(x_2)W_j[x_1, x_2]g^{-1}_{j}(x_1)=W_j[x'_1,x'_2]\,,
\ee
where $x'=(\rho,w)$ and  $w=\frac{az+b}{cz+d}$ is the $SL(2, \mathbb{R})$ transformation of $z$-coordinate. Inserting the identity $\mathbb{1} = g^{-1}_{j}(x)g_{j}(x)$ into the AdS vertex function \eqref{vertex_func} yields 
\be
\ba{l}
V_{j\tilde{j}}({\bf{z}})=\bra{a_1}g^{-1}_{j_1}(x_1)g_{j_1}(x_1)W_{{j}_1}[0,x_1]g^{-1}_{j_1}(0)g_{j_1}(0)I_{j_1 j_2 \tilde{j}_1} g^{-1}_{j_2}(0)g_{j_2}(0)W_{{j}_2}[x_2,0]g^{-1}_{j_2}(x_2)g_{j_2}(x_2) 
\vspace{2mm} 
\\ 
\dps \times I_{\tilde{j}_1 j_3 \tilde{j}_2}\ldots I_{\tilde{j}_{{n-3}} j_{n-1} j_n}g^{-1}_{j_n}(0)g_{j_n}(0)W_{{j}_n}[x_n,0]g^{-1}_{j_n}(x_n)g_{j_n}(x_n)\ket{{a}_2}\otimes \ket{{a}_3}\otimes \cdots\otimes \ket{{a}_n}\,.
\ea
\ee
Then, using the invariance property of intertwiners \eqref{invariance}  results in
\be
\ba{l}
V_{j\tilde{j}}({\bf{z}}) = \bra{a_1}g^{-1}_{j_1}(x_1)g_{j_1}(x_1)W_{{j}_1}[0,x_1]g^{-1}_{j_1}(0)I_{j_1 j_2 \tilde{j}_1} g_{j_2}(0)W_{{j}_2}[x_2,0]g^{-1}_{j_2}(x_2)g_{j_2}(x_2) I_{\tilde{j}_1 j_3 \tilde{j}_2}\ldots
\vspace{2mm} 
\\ 
\dps 
\hspace{10mm} \ldots I_{\tilde{j}_{{n-3}} j_{n-1} j_n}g_{j_n}(0)W_{{j}_n}[x_n,0]g^{-1}_{j_n}(x_n)g_{j_n}(x_n)\ket{{a}_2}\otimes \ket{{a}_3}\otimes \cdots\otimes \ket{{a}_n}.
\ea
\ee
By means of the gauge transformation \eqref{Wilson_gauge} the previous relation can be cast into the form
\be
\label{vert_func1}
\ba{l}
V_{j\tilde{j}}({\bf{z}}) =  \bra{a_1}g^{-1}_{j_1}(x_1)W_{{j}_1}[0,x'_1]I_{j_1 j_2 \tilde{j}_1} W_{{j}_2}[x'_2,0]g_{j_2}(x_2) I_{\tilde{j}_1 j_3 \tilde{j}_2}\ldots\vspace{2mm} 
\\ 
\dps 
\hspace{30mm}\ldots I_{\tilde{j}_{{n-3}} j_{n-1} j_n}W_{{j}_n}[x'_n,0]g_{j_n}(x_n)\ket{{a}_2}\otimes \ket{{a}_3}\otimes \cdots\otimes \ket{{a}_n}\,.
\ea\ee
Then, one inserts the identity resolutions $\mathbb{1}=\sum_{m} \ket{j,m}\bra{j,m}$ between the intertwiners and  Wilson line operators and  calculates the Wilson matrix elements for the  Ishibashi states \eqref{cap_states} (in the form derived in Appendix \bref{app:LWME}) by means of the same technique as in Appendix \bref{app:NLO}. In this way, one finds that the left  matrix element is given by 
\be 
\ba{l}\dps
\bra{a}g_{j}^{-1}(x) W_{{j}}[0,x']\ket{j,m} = (-)^{2j+m}\frac{j!}{m!}\left[\frac{(j+m)!}{(j-m)!(2j)!}\right]^\half e^{\rho j} (cz+d)^{2j}\alpha_-^{-m}\alpha_+^{m}
\vspace{2mm}
\\ 
\dps
\hspace{35mm}\times\left(w\alpha_-+\frac{ie^{-\rho}}{cz+d}\right)^{m}\left((w\alpha_-+\frac{ie^{-\rho}}{cz+d})\alpha_+-2ie^{-\rho}\right)^{j}
\vspace{2mm}
\\ 
\dps
\hspace{35mm}\times{}_2F_1\Big(-j, m-j; m+1\Big|\frac{(w\alpha_-+\frac{ie^{-\rho}}{cz+d})\alpha_+}{(w\alpha_-+\frac{ie^{-\rho}}{cz+d})\alpha_+-2ie^{-\rho}}\Big)\;,
\ea
\ee 
where we introduced the notation 
\be 
\alpha_\pm = (cz+d)\pm ice^{-\rho}\;.
\ee 
The right matrix element is given by
\be 
\ba{l}
\dps
\bra{j,m} W_{{j}}[x',0]g_j(x)\ket{a}  = \frac{j!}{m!}\left[\frac{(j+m)!}{(j-m)!(2j)!}\right]^\half e^{\rho j}(cz+d)^{2j}\alpha_-^{j}\alpha_+^{m}
\vspace{2mm}
\\ 
\dps
\hspace{35mm}\times\left(w\alpha_+-\frac{ie^{-\rho}}{cz+d}\right)^{j-m}
{}_2F_1\left(-j, m-j; m+1\Big|\frac{w\alpha_++\frac{ie^{-\rho}}{cz+d}}{w\alpha_+-\frac{ie^{-\rho}}{cz+d}}\right).
\ea
\ee 

Now, one can expand the resulting  expressions near $\rho = \infty$ and see that up to higher-order $\rho$-dependent terms acting with the gauge elements on the cap states  boils down to the standard Jacobians prefactors, 
\be
\label{confmatrix1}
\bra{a}g_{j}^{-1}(x) W_{{j}}[0,x']\ket{j,m}=\Big(\frac{\partial w}{\partial z}\Big)^{-j}\bra{a}W_{{j}}[0,x']\ket{j,m}+\cO(e^{\rho (j-1)})\,,
\ee
\be
\label{confmatrix2}
 \bra{j,m}W_{{j}}[x',0]\,g_{j}(x)\ket{a}=
\Big(\frac{\partial w}{\partial z}\Big)^{-j} \bra{j,m}W_{{j}}[x',0]\ket{a} + \cO(e^{\rho (j-1)})\,.
\ee
Substituting  the asymptotics \eqref{confmatrix1} and \eqref{confmatrix2} into the AdS vertex function \eqref{vert_func1} we finally find the desired conformal transformation law \eqref{coftransconf}:    
\be
\ba{l}
\dps
\cV_{ j\tilde{j}}({\bf z},\rho)=\Big(\frac{\partial w}{\partial z}\Big)^{-j_1}\Big|_{z=z_1}\cdots\Big(\frac{\partial w}{\partial z}\Big)^{-j_n}\Big|_{z=z_n}
\vspace{2mm} 
\\ 
\dps
\times \bra{a_1}W_{{j}_1}[0,x'_1]I_{j_1 j_2 \tilde{j}_1} W_{{j}_2}[x'_2,0] I_{\tilde{j}_1 j_3 \tilde{j}_2}\ldots I_{\tilde{j}_{{n-3}} j_{n-1} j_n}W_{{j}_n}[x'_n,0]\ket{{a}_2}\otimes \ket{{a}_3}\otimes \cdots\otimes \ket{{a}_n}
\vspace{2mm} 
\\ 
\dps
\hspace{25mm} =\Big(\frac{\partial w}{\partial z}\Big)^{-j_1}\Big|_{z=z_1}\cdots\Big(\frac{\partial w}{\partial z}\Big)^{-j_n}\Big|_{z=z_n}\cV_{ j\tilde{j}}({\bf w},\rho)+ \cO\big(e^{\rho \left(\sum_{i=1}^nj_i-1\right)}\big)\,.
\ea
\ee

Note that  the relations \eqref{confmatrix1} and \eqref{confmatrix2} are equally true for the rotated HW \eqref{cap_states2} and LW/HW vectors \eqref{cap_states3}. The latter was proven in \cite{Besken:2016ooo}. In the former case one finds 
\be 
\ba{l}\dps
\bra{a}g_{j}^{-1}(x) W_{{j}}[0,x']\ket{j,m} = \left[\frac{(2j)!}{(j-m)!(j+m)!}\right]^\half(cz+d)^{2j}e^{\rho j}(-)^{j+m}
\vspace{2mm} 
\\ 
\dps
\hspace{42mm} \times\left(1-\frac{ce^{-\rho}}{cz+d}\right)^{j-m}\left(w\left(1+\frac{ce^{-\rho}}{cz+d}\right)-\frac{e^{-\rho}}{(cz+d)^2}\right)^{j+m},
\ea
\ee  

\be 
\ba{l}\dps
\bra{j,m} W_{{j}}[x',0]g_j(x)\ket{a}= \left[\frac{(2j)!}{(j-m)!(j+m)!}\right]^\half(cz+d)^{2j}e^{\rho j}
\vspace{2mm} 
\\ 
\dps
\hspace{39mm}\times\left(1-\frac{ce^{-\rho}}{cz+d}\right)^{2j}\left(w-e^{-\rho}+\frac{ce^{-2\rho}}{cz+d}\right)^{j-m}.
\ea
\ee 
Expanding these matrix elements near $\rho = \infty$ one again obtains  \eqref{confmatrix1} and \eqref{confmatrix2}.

\section{Lower-point AdS vertex  and CFT correlation functions}
\label{sec:3-5}

Even though the near-boundary lower-point functions were explicitly considered in the literature \cite{Bhatta:2016hpz,Besken:2016ooo} in this section, for completeness, we rephrase those calculations in our terms. Everything is calculated in the case of finite-dimensional modules by reasons explained in the very end of Section \bref{sec:matrix}. 

\subsection{$3$-point  functions}

The $3$-point AdS vertex function \eqref{vertex_func}  is given by 
\be 
\label{3point}
\cV_{j_1 j_2 j_3}(\rho,{\bf z})=\sum_{m_i\in \sJ_i}[I_{j_1j_2j_3}]^{m_1}{}_{m_2m_3}\braket{\tilde{a}_1|j_1, m_1}\braket{j_2, m_2|\tilde{a}_2}\braket{j_3, m_3|\tilde{a}_3}\,.
\ee
To simplify the calculation we use the technique from Section \bref{sec:n-p} and make the following $SL(2, \mathbb{R})$ transformation\footnote{Note that this  transformation differs from that used in calculating the $n$-point AdS vertex function \eqref{n-transfrom} where it was crucial to set the two last points to $0$ and $\infty$.} 
\be
\label{transform_3} 
w(z)=\frac{z_{_{23}}z-z_{_{23}}z_{_1}}{z_{_{21}}z-z_{_{21}}z_{_3}}\;,
\ee 
which maps  $z_1,z_2,z_3$ into 
\be
\label{param3}
\ba{l}
\dps w_1:= w(z_1)=0\,,
\quad 
w_2:= w(z_2)=1\,,
\quad 
\dps
w_3:= w\Big(z_3+\frac{z_{_{31}}z_{_{23}}}{z_{_{21}} }\,\omega^{-1}\Big)=\omega+\frac{z_{_{23}}}{z_{_{21}}}=\omega+\cO(1)\,,
\ea
\ee
where we introduced a large coordinate parameter $\omega\to \infty$ to regularize the pole $z = z_3$ in \eqref{transform_3}. Then, the Wilson matrix elements \eqref{LHW_fin} associated to the three boundary  points in $w$-coordinates  are getting simplified (see also \eqref{n-Wilson}):
\be
\label{wilson3}
\ba{l}
\dps
\braket{\tilde{a}_1|j_1,m_1}= e^{\rho j_1}0^{j_1+m_1}\left[\frac{(2j_1)!}{(j_1+m_1)!(j_1-m_1)!}\right]^{\half}=e^{\rho j_1}\delta_{-j_1 ,m_1}\,,
\vspace{3mm}
\\
\dps\braket{j_2,m_2|\tilde{a}_2}=e^{\rho j_2}(-1)^{j_2-m_2}\left[\frac{(2j_2)!}{(j_2+m_2)!(j_2-m_2)!}\right]^{\half},
\vspace{3mm}
\\
\dps
\braket{j_3,m_3|\tilde{a}_3}=e^{\rho j_3}\delta_{-j_3, m_3}\omega^{2j_3}+\cO(\omega^{2j_3-1})\,.
\ea
\ee
The  3-point AdS vertex function \eqref{3point} in $w$-coordinates takes the form 
\be
\label{3pt1}
\ba{l}
\dps
\cV_{j_1 j_2 j_3}(\rho,0, 1, \omega)=e^{\rho (j_1+j_2+j_3)}\sum_{m_i\in \sJ_i}(-1)^{j_1-m_1}\begin{pmatrix}j_1&j_2&j_3\\-m_1&m_2&m_3\end{pmatrix}\delta_{-j_1 ,m_1}\delta_{-j_3, m_3}
\vspace{2mm} 
\\
\dps
\times (-1)^{j_2-m_2}\left[\frac{(2j_2)!}{(j_2+m_2)!(j_2-m_2)!}\right]^{\half}\omega^{2j_3}+\cO(\omega^{2j_3-1})=e^{\rho (j_1+j_2+j_3)}\sum_{m_2\in J_2}(-1)^{2j_1}(-1)^{j_2-m_2}
\vspace{2mm} 
\\
\dps\times\begin{pmatrix}j_1&j_2&j_3\\j_1&m_2&-j_3\end{pmatrix}\left[\frac{(2j_2)!}{(j_2+m_2)!(j_2-m_2)!}\right]^{\half}\omega^{2j_3}+\cO(\omega^{2j_3-1})\,.
\ea
\ee
Note that after resolving the Kronecker symbols we obtained the simplified $3j$ symbol that can be found from  \eqref{s3-j}: 
\be
\label{simpleint}
\begin{pmatrix}j_1&j_2&j_3\\j_1&m_2&-j_3\end{pmatrix}=
\dps
(-1)^{-j_1+j_2-j_3} \delta_{j_3-j_1 m_2}\left[\frac{(2j_1)!(2j_3)!}{(j_1+j_2+j_3+1)!(j_1-j_2+j_3)!}\right]^{\half}.
\ee
Substituting this expression into \eqref{3pt1} and summing over $m_2$ yields 
\be
\label{3pt2}
\cV_{j_1 j_2 j_3}(\rho,0, 1, \omega)=e^{\rho (j_1+j_2+j_3)}(-1)^{j_1+j_2-j_3} \pref_{j_1j_2j_3}\omega^{2j_3}+\cO(\omega^{2j_3-1})\,,\ee
where the coefficient $\pref_{j_1j_2j_3}$ is given by  \eqref{prefactor}.
Then, using the transformation inverse to \eqref{transform_3},  
\be 
\label{reverse}
z(w)=\frac{-z_{_3}z_{_{21}}w+z_{_1} z_{_{23}}}{-z_{_{21}}w+z_{_{23}}}\,, 
\ee
which maps  $0,1,\infty$ to $z_1,z_2,z_3$, the AdS vertex function in $z$-coordinates is obtained by virtue of the conformal transformation  law \eqref{coftransconf}:
\be
\ba{l}
\dps\cV_{j_1 j_2 j_3}(\rho,z_1, z_2, z_3)=\lim_{\omega\to\infty}\left[\Big(\frac{\partial z}{\partial w}\Big)^{j_1}\Big|_{w=0}\Big(\frac{\partial z}{\partial w}\Big)^{j_2}\Big|_{w=1}\Big(\frac{\partial z}{\partial w}\Big)^{j_3}\Big|_{w=\omega}\cV_{j_1 j_2 j_3}(\rho,0, 1, \omega)\right]
\vspace{3mm} 
\\
\dps
\hspace{1.8mm}=\lim_{\omega\to\infty}\left[\Big(\frac{z_{_{13}}z_{_{21}}z_{_{23}}}{z_{_{23}}^2}\Big)^{j_1}\Big(\frac{z_{_{13}}z_{_{21}}z_{_{23}}}{z_{_{13}}^2}\Big)^{j_2}\Big(\frac{z_{_{13}}z_{_{21}}z_{_{23}}}{(z_{_{12}}\omega+z_{_{23}})^2}\Big)^{j_3}e^{\rho (j_1+j_2+j_3)}(-1)^{j_1+j_2-j_3} \pref_{j_1j_2j_3}\omega^{2j_3}\right]
\vspace{3mm} 
\\
\dps
\hspace{29mm}=\pref_{j_1j_2j_3} e^{\rho (j_1+j_2+j_3)}z_{_{12}}^{j_1+j_2-j_3}z_{_{13}}^{j_1+j_3-j_2}z_{_{23}}^{j_2+j_3-j_1}.
\ea
\ee
Using  the extrapolate dictionary   \eqref{vert_conf} one finds the $3$-point correlation function in CFT$_2$. Of course, the same result can be achieved without fixing the points and using the $SL(2,\mathbb{R})$ transformation \eqref{reverse} but at the cost of much more lengthy and less technically transparent calculations.

\subsection{$4$-point  functions}
\label{app:4pt}

The $4$-point AdS vertex function \eqref{vertex_func} reads  
\be
\label{4point}
\ba{l}
\dps
\cV_{j_1j_2j_3j_4\tilde{j}_1}(\rho,{\bf z})= \sum_{m_i\in \sJ_i,\, p\in \stJ_1} [I_{j_1j_2\tilde{j}_1}]^{m_1}{}_{m_2 p} [I_{\tilde{j}_1j_3j_4}]^{p}{}_{m_3 m_4}
\vspace{3mm} 
\\
\dps 
\hspace{40mm}\times \braket{\tilde{a}_1|j_1, m_1}\braket{j_2, m_2|\tilde{a}_2}\braket{j_3, m_3|\tilde{a}_3}\braket{j_4, m_4|\tilde{a}_4}\,.
\ea
\ee
In order to simplify calculations one makes the transformation \eqref{transform_3} again. As we saw earlier, $z_1,z_2,z_3$ are mapped into $0,1,\infty$, while $z_4$ goes to
\be
\label{param4}
w_4:=w(z_4)=\frac{z_{_{23}}z_{_{14}}}{z_{_{21}}z_{_{34}}}\equiv \frac{\chi_4-1}{\chi_4}\,,
\ee
where $\chi_4$ is the cross-ratio \eqref{ratio}.

The Wilson matrix elements in $w$-coordinates $\braket{\tilde{a}_i|j_i,m_i}$ for $i=1,2,3$ are given by \eqref{wilson3}, while  $\braket{j_4,m_4|\tilde{a}_4}$ can be obtained from \eqref{LHW_fin} and \eqref{param4}
\be
\label{wilson4}
\braket{j_4,m_4|\tilde{a}_4}=e^{\rho j_4}\left[\frac{(2j_4)!}{(j_4+m_4)!(j_4-m_4)!}\right]^{\half}\Big(\frac{1-\chi_4}{\chi_4}\Big)^{j_4-m_4}.
\ee
One can write the expression for the $4$-point AdS vertex function in $w$-coordinates, resolve the Kronecker symbols from matrix elements \eqref{wilson3} and obtain 
\be
\label{4pt1}
\ba{c}
\dps
\cV_{j_1j_2j_3j_4\tilde{j}_1}(\rho,0, 1, \omega, w_4)=e^{\rho (j_1+j_2+j_3+j_4)}\sum_{m_2\in \sJ_2,\, m_4\in \sJ_4,\, p\in \stJ_1} (-1)^{2j_1}(-1)^{\tilde{j}_1-p}(-1)^{j_2-m_2}
\vspace{2mm} 
\\
\dps 
\hspace{10mm}\times \,\begin{pmatrix}j_1&j_2&\tilde{j}_1\\j_1&m_2&p\end{pmatrix} \begin{pmatrix}\tilde{j}_1&j_3&j_4\\-p&-j_3&m_4\end{pmatrix}
\left[\frac{(2j_2)!}{(j_2+m_2)!(j_2-m_2)!}\right]^{\half}\vspace{2mm} 
\\
\dps 
\hspace{22mm}\times \,\left[\frac{(2j_4)!}{(j_4+m_4)!(j_4-m_4)!}\right]^{\half}\omega^{2j_3}\,\left(\frac{1-\chi_4}{\chi_4}\right)^{j_4-m_4}+\cO(\omega^{2j_3-1})\,.
\ea
\ee
The rest of the calculation can be done in three simple steps: (1) substitute $3j$ symbols \eqref{s3-j} into \eqref{4pt1}; (2) resolve  the Kronecker symbols from \eqref{s3-j} by summing  over $m_2$ and $m_4$; (3) change $p=k-\tilde{j}_1$. The resulting expression reads
\be
\label{4pt2}
\ba{l}
\dps \cV_{j_1j_2j_3j_4\tilde{j}_1}\big(\rho,0, 1, \omega, w_4\big)=e^{\rho (j_1+j_2+j_3+j_4)}
\Bigg[(j_2-j_1+\tilde{j}_1)!(-j_3+j_4+\tilde{j}_1)!\frac{\pref_{j_1 j_2 \tilde{j}_1}\pref_{\tilde{j}_1 j_3 j_4}}{(2\tilde{j}_1)!}
\vspace{2.5mm}
\\
\dps
\times\sum_{k\in K}\frac{(2\tilde{j}_1-k)!}{(j_2-j_1+\tilde{j}_1-k)!(-j_3+j_4+\tilde{j}_1-k)!k!}\left(\frac{\chi_4}{1-\chi_4}\right)^{k+j_3+j_4-\tilde{j}_1}\Bigg]\omega^{2j_3}+\cO(\omega^{2j_3-1})\,,
\ea
\ee
where the summation domain is
\be
K:=[\![0,\min(\tilde{j}_1+j_4-j_3, \tilde{j}_1+j_2-j_1)]\!]\;.
\ee
This expression can be conveniently represented in terms of the  hypergeometric series \eqref{def_hyper} in the cross-ratio $\chi_4/(1-\chi_4)$ which can be mapped to that of $\chi_4$ by means of the Pfaff transformation  \eqref{4kummer2f1}:
\be
\label{4p3}
\ba{l}
\dps\cV_{j_1j_2j_3j_4\tilde{j}_1}\big(\rho,0, 1, \omega, w_4\big)= e^{\rho (j_1+j_2+j_3+j_4)}\omega^{2j_3}\pref_{j_1j_2j_3j_4\tilde{j}_1}\chi_4^{j_3-j_4-\tilde{j}_1}
\vspace{2mm}
\\
\dps
\hspace{30mm} \times\F\left(-\tilde{j}_1-j_1+j_2,-\tilde{j}_1-j_4+j_3; -2\tilde{j}_1\big|\chi_4\right)+\cO(\omega^{2j_3-1})\;,
\ea
\ee
where $\pref_{j_1j_2j_3j_4\tilde{j}_1} \equiv  \pref_{j_1 j_2 \tilde{j}_1}\pref_{\tilde{j}_1 j_3 j_4}$ stands for the product of the structure constants  \eqref{prefactor}.
Finally, making the inverse transformation \eqref{reverse} back to $z$-coordinates and using the transformation rule \eqref{coftransconf}  one obtains  
\be
\label{final4pt}
\ba{l}
\dps 
\cV_{j_1 j_2 j_3 j_4\tilde{j}_1}(\rho,{\bf z})= e^{\rho \sum_{i=1}^4 j_i}\,\pref_{j_1j_2j_3j_4\tilde{j}_1}\, 
\vspace{2mm}
\\
\dps
\hspace{20mm}\times\mathcal{L}_{j_1j_2j_3j_4}({\bf z})\,\chi_4^{-\tilde{j}_1} \F\left(-\tilde{j}_1-j_1+j_2,-\tilde{j}_1-j_4+j_3; -2\tilde{j}_1\big|\chi_4\right)\,,
\ea
\ee
where the leg factor is given by 
\be 
\mathcal{L}_{j_1j_2j_3j_4}({\bf z}) = z_{_{12}}^{j_1+j_2}z_{_{34}}^{j_3+j_4}\left(\frac{z_{_{23}}}{z_{_{13}}}\right)^{j_2-j_1}\left(\frac{z_{_{24}}}{z_{_{23}}}\right)^{j_4-j_3}\,.
\ee 
Note that the conformal transformation properties are encoded in the leg factor only as the rest depends only on the cross-ratio. Using the extrapolate dictionary relation  \eqref{vert_conf} one finds the 4-point conformal block (the second line in \eqref{final4pt}). This is in complete agreement with the known results of reproducing the 4-point conformal block  from the Wilson network as in  \cite{Bhatta:2016hpz,Besken:2016ooo}. The novel feature of using the extrapolate dictionary here is that we have also obtained the structure constant prefactor.

\subsection{$5$-point functions}
\label{app:5pt}

In this Appendix we calculate the 5-point conformal block from the AdS vertex function by two methods as a demonstration of the difference between the recursive procedure and the straightforward calculation. The former method immediately gives the answer based on the known relation between the Gauss and Appell functions (in the higher-point case this relation is between $n-1$-point and $n$-point comb functions). The later method produces the second Appell only after additional resummations. In this case a straightforward generalization  to $n$ points would be laborious and technically difficult.

Previously, the 5-point block  as the near-boundary AdS vertex function was found in \cite{Bhatta:2016hpz} by straightforward calculation which resulted in the 5-point block in the form of the hypergeometric-type series  found in \cite{Alkalaev:2015fbw}. Here, we show that the present technique yields the 5-point block as the second Appell function found in \cite{Rosenhaus:2018zqn}. The two expressions can be translated to each other by choosing different leg factors and cross-ratios along with numerous resummations.

\paragraph{Direct calculation.} The 5-point AdS vertex function \eqref{vertex_func} reads
\be
\label{5point}
\ba{l}
\dps
\cV_{j_1\ldots j_5\tilde{j}_1\tilde{j}_2}(\rho,{\bf z})= \sum_{m_i\in J_i}\sum_{p_k\in \tilde{J}_k} [I_{j_1j_2\tilde{j}_1}]^{m_1}{}_{m_2 p_1} [I_{\tilde{j}_1j_3\tilde{j}_2}]^{p_1}{}_{m_3 p_2}[I_{\tilde{j}_2 j_4 j_5}]^{p_2}{}_{m_4 m_5}
\vspace{2mm}
\\
\dps
\hspace{40mm}\times\braket{\tilde{a}_1|j_1, m_1}\braket{j_2, m_2|\tilde{a}_2}\braket{j_3, m_3|\tilde{a}_3}\braket{j_4, m_4|\tilde{a}_4}\braket{j_5, m_5|\tilde{a}_5}\,.
\ea
\ee
After making the transformation \eqref{transform_3} points $z_i$, $i=1,...,4$ are mapped into  \eqref{param3}, \eqref{param4}, while $z_5$ goes to
\be
\label{param5}
w_5:=w(z_5)=\frac{z_{_{23}}z_{_{15}}}{z_{_{21}}z_{_{35}}} \equiv \frac{\chi_4+\chi_5-1}{\chi_4}\,.
\ee
The Wilson matrix elements $\braket{\tilde{a}_i|j_i,m_i}$ for $i=1,...,4$ are given by  \eqref{wilson3} and  \eqref{wilson4}, while $\braket{j_5,m_5|\tilde{a}_5}$ can be obtained from \eqref{LHW_fin} and \eqref{param5} 
\be
\label{wilson5}
\braket{j_5,m_5|\tilde{a}_5}=e^{\rho j_5}\left[\frac{(2j_5)!}{(j_5+m_5)!(j_5-m_5)!}\right]^{\half}\left(-\frac{\chi_4+\chi_5-1}{\chi_4}\right)^{j_5-m_5},
\ee
As in the 4-point case, one can write the AdS vertex function in $w$-coordinates, resolve the Kronecker symbols and get
\be
\label{5pt1}
\ba{l}
\dps
\cV_{j_1\ldots j_5\tilde{j}_1\tilde{j}_2}(\rho,0,1,\omega,w_4,w_5)=e^{\rho \sum_{i=1}^5 j_i}\sum_{m_2\in J_2}\sum_{m_4\in J_4}\sum_{m_5\in J_5}\sum_{p_1\in \tilde{J}_1}\sum_{p_2\in \tilde{J}_2} (-1)^{2j_1}(-1)^{\tilde{j}_1-p_1}(-1)^{\tilde{j}_2-p_2}
\vspace{2mm}
\\
\dps
\hspace{15mm}\times \begin{pmatrix}j_1&j_2&\tilde{j}_1\\j_1&m_2&p_1\end{pmatrix} 
\begin{pmatrix}\tilde{j}_1&j_3&\tilde{j}_2\\-p_1&-j_3&p_2\end{pmatrix}
\begin{pmatrix}\tilde{j}_2&j_4&j_5\\-p_2&m_4&m_5\end{pmatrix}
\vspace{2mm}
\\
\dps
\hspace{15mm} \times\left[\frac{(2j_2)!}{(j_2+m_2)!(j_2-m_2)!}\right]^{\half}(-1)^{j_2-m_2}
\omega^{2j_3}\left[\frac{(2j_4)!}{(j_4+m_4)!(j_4-m_4)!}\right]^{\half}\left(-\frac{1-\chi_4}{\chi_4}\right)^{j_4-m_4}
\vspace{2mm}
\\
\hspace{15mm}\dps\times\,\left[\frac{(2j_5)!}{(j_5+m_5)!(j_5-m_5)!}\right]^{\half}\left(-\frac{\chi_4+\chi_5-1}{\chi_4}\right)^{j_5-m_5}+\cO(\omega^{2j_3-1})\;.
\ea
\ee
Here, the  last $3j$ symbol has the general form \eqref{3-j}, the first two have the particular form \eqref{s3-j}. Substituting them into  \eqref{5pt1}, summing over $m_2, m_5, p_2$ (thereby  resolving the Kronecker symbols contained in \eqref{3-j} and \eqref{s3-j}) and renaming $p_1=n_1$, $m=n_2$ (the parameter from \eqref{3-j}), $m_4=n_3$ one finds 
$$
\ba{l}
\dps
\cV_{j_1\ldots j_5\tilde{j}_1\tilde{j}_2}(\rho,0,1,\omega,w_4,w_5)=e^{\rho \sum_{i=1}^5 j_i}\omega^{2j_3}
\frac{\pref_{j_1j_2j_3j_4j_5\tilde{j}_1\tilde{j}_2}}{(2\tilde{j}_1)!(2\tilde{j}_2)!}
\vspace{3mm} 
\\
\dps
\times\,(-j_1+j_2+\tilde{j}_2)!(-j_3+\tilde{j}_1+\tilde{j}_2)!(\tilde{j}_2+j_4-j_5)!
(\tilde{j}_2-j_4+j_5)!(-\tilde{j}_2+j_4+j_5)!
\vspace{3mm} 
\\
\dps
\times\sum_{n_1\in N_1}\,\sum_{n_2\in N_2}\,\sum_{n_3\in N_3}(-1)^{-j_1+2j_2-j_4+\tilde{j}_1+n_2}
\frac{(j_3+\tilde{j}_2+n_1)!}{(-j_1+j_2-n_1)!}\frac{(\tilde{j}_1-n_1)!}{(\tilde{j}_1+n_1)!}
\ea
$$
\be
\ba{l}
\dps
\hspace{-23mm}\times\frac{\left(-\frac{1-\chi_4}{\chi_4}\right)^{j_4}(-\frac{\chi_4+\chi_5-1}{\chi_4})^{j_5-j_3-n_1}}{n_2!(\tilde{j}_2+j_4-j_5-n_2)!(\tilde{j}_2+j_3+n_1-n_2)!(j_5-j_4-j_3-n_1+n_2)!}
\vspace{3mm}
\\
\dps
\hspace{-23mm}\times\,\frac{\left(\chi_4-1\right)^{-n_3}(1-\chi_4-\chi_5)^{n_3}}{(j_4+n_3-n_2)!(j_5-\tilde{j}_2-n_3+n_2)!}+\cO(\omega^{2j_3-1})\;,
\ea\ee
where the summation domains are 
\be
\ba{l}
\dps
N_1 \;:= [\![-\tilde{j}_1,\min(j_2-j_1, \tilde{j}_2 - j_3)]\!]\;,
\vspace{2mm}
\\
N_2\;:= [\![\max(0,n_1+j_3-j_5+j_4),\min(n_1+j_3+\tilde{j}_2, j_4-j_5+\tilde{j}_2)]\!]\;,
\vspace{2mm}
\\
\dps
N_3\;:= [\![n_2-j_4, j_5-\tilde{j}_2+n_2]\!] \;,
\ea
\ee
and $\pref_{j_1j_2j_3j_4j_5\tilde{j}_1\tilde{j}_2} \equiv  \pref_{j_1j_2\tilde{j}_1}\pref_{\tilde{j}_1 j_3 \tilde{j}_2}\pref_{\tilde{j}_2 j_4 j_5}$ is the product of the structure constants \eqref{prefactor}. The last summation can be done by observing that it is exactly the generalized Newton binomial  
\be
\sum_{n_3\in N_3}\frac{\left(\chi_4-1\right)^{-n_3}(1-\chi_4-\chi_5)^{n_3}}{(j_4+n_3-n_2)!(j_5-\tilde{j}_2-n_3+n_2)!}
=
\frac{1}{(j_4+j_5-\tilde{j}_2)!}\Big(\frac{1-\chi_4-\chi_5}{\chi_4-1}\Big)^{-j_4+n_2}\Big(\frac{\chi_5}{1-\chi_4}\Big)^{j_4+j_5-\tilde{j}_2}\,,
\ee
so that one obtains  
\be
\ba{l}
\dps
\cV_{j_1\ldots j_5\tilde{j}_1\tilde{j}_2}(\rho,0,1,\omega,w_4,w_5)=
\vspace{3mm} 
\\
\dps
\hspace{2mm}e^{\rho \sum_{i=1}^5 j_i}\omega^{2j_3}(-j_1+j_2+\tilde{j}_2)!(-j_3+\tilde{j}_1+\tilde{j}_2)!(\tilde{j}_2+j_4-j_5)!(\tilde{j}_2-j_4+j_5)!\frac{\pref_{j_1j_2j_3j_4j_5\tilde{j}_1\tilde{j}_2}}{(2\tilde{j}_1)!(2\tilde{j}_2)!}
\vspace{3mm} 
\\
\dps
\hspace{2mm} \times\,\Big(\frac{\chi_5}{1-\chi_4}\Big)^{j_4+j_5-\tilde{j}_2}\sum_{n_1\in N_1}\sum_{n_2\in N_2}

(-1)^{-j_1+2j_2-j_4+\tilde{j}_1+n_2}
\frac{(j_3+\tilde{j}_2+n_1)!}{(-j_1+j_2-n_1)!}\frac{(\tilde{j}_1-n_1)!}{(\tilde{j}_1+n_1)!}
\vspace{3mm}
\\ 
\dps
\hspace{2mm} \times\frac{\left(-\frac{1-\chi_4}{\chi_4}\right)^{j_4}(-\frac{\chi_4+\chi_5-1}{\chi_4})^{j_5-j_3-n_1}\Big(\frac{1-\chi_4-\chi_5}{\chi_4-1}\Big)^{-j_4+n_2}}{n_2!(\tilde{j}_2+j_4-j_5-n_2)!(\tilde{j}_2+j_3+n_1-n_2)!(j_5-j_4-j_3-n_1+n_2)!} +\cO(\omega^{2j_3-1})\,.
\ea\ee
The remaining calculation is straightforward: (1) change $k_1=-n_1-j_3$; (2) use the Lemma \eqref{lemma} when summing over $n_2$ and $j_a=\tilde{j}_2$, $j_b=j_5$, $j_c=j_4$, $n=k_1$; (3) change $l_1=\tilde{j}_1-j_3-k_1$, $l_2=\tilde{j}_2+j_4-j_5-n_2$; (4) express the summation coefficients in terms of the Pochhammer  symbols. The resulting expression takes the form 
\be
\ba{l}
\dps
\cV_{j_1\ldots j_5\tilde{j}_1\tilde{j}_2}(\rho, 0,1,\omega,w_4,w_5)=\omega^{2j_3}\pref_{j_1j_2j_3j_4j_5\tilde{j}_1\tilde{j}_2}\left(\frac{1-\chi_4-\chi_5}{\chi_4}\right)^{j_5-j_3+\tilde{j}_1}\Big(\frac{1-\chi_4-\chi_5}{\chi_5}\Big)^{\tilde{j}_2-j_4-j_5}
\vspace{3mm}
\\
\dps
\hspace{2mm}\times e^{\rho \sum_{i=1}^5 j_i}\sum_{l_{1}\in L_{1}}\sum_{l_{2}\in L_{2}}(-1)^{-j_1+2j_2-j_5+\tilde{j}_1}
\\
\dps
\hspace{2mm}\times
\frac{(j_1-j_2-\tilde{j}_2)_{l_1}(j_3-\tilde{j}_1-\tilde{j}_2)_{l_1+l_2}(j_5-j_4-\tilde{j}_2)_{l_2}}{(-2\tilde{j}_1)_{l_1}(-2\tilde{j}_2)_{l_2}}\frac{\Big(\frac{\chi_4}{\chi_4+\chi_5-1}\Big)^{l_1}}{l_1!}\frac{\Big(\frac{\chi_5}{\chi_4+\chi_5-1}\Big)^{l_2}}{l_2!}
+\cO(\omega^{2j_3-1})\,,
\ea
\ee
where the summation domains are 
\be
\ba{l}
\dps
L_1:= [\![0,\min(j_1-j_2+\tilde{j}_1,\tilde{j}_2+\tilde{j}_1-j_3)]\!]\;,
\vspace{2mm}
\\
\dps 
L_2:=[\![0,\min(j_4-j_5+\tilde{j}_2, \tilde{j}_2+\tilde{j}_1-j_3-l_1)]\!]\;.
\ea
\ee
The double power series here is exactly the second Appell function \eqref{def_appell} with the summation domain \eqref{comb-domains}:
\be
\label{5pt2}
\ba{l}
\dps
\cV_{j_1\ldots j_5\tilde{j}_1\tilde{j}_2}(\rho,0,1,\omega,w_4,w_5)=e^{\rho \sum_{i=1}^5 j_i}\omega^{2j_3}\,\pref_{j_1j_2j_3j_4j_5\tilde{j}_1\tilde{j}_2}
\vspace{2mm} 
\\
\dps
\hspace{2mm}
\times
(-1)^{-j_1+2j_2-j_5+\tilde{j}_1}\left(\frac{1-\chi_4-\chi_5}{\chi_4}\right)^{j_5-j_3+\tilde{j}_1}\Big(\frac{1-\chi_4-\chi_5}{\chi_5}\Big)^{\tilde{j}_2-j_4-j_5}
\vspace{2mm} 
\\
\dps
\hspace{2mm}\times F_2 \left[\begin{array}{ccc}
     j_1-j_2-\tilde{j}_1,&j_3-\tilde{j}_1-\tilde{j}_2, &j_5-j_4-\tilde{j}_2  \\
     &-2\tilde{j}_1, -2\tilde{j}_2 \end{array};\frac{\chi_4}{\chi_4+\chi_5-1},\frac{\chi_5}{\chi_4+\chi_5-1}\right]+\,\cO(\omega^{2j_3-1})\,.

\ea
\ee
Using the Pfaff-type transformation   \eqref{3kummerf2}, making the inverse transformation \eqref{reverse} and using the conformal transformation  \eqref{coftransconf} we can finally  represent the 5-point vertex function as
\be 
\label{final5pt}
\ba{l}\dps
\cV_{j_1\ldots j_5\tilde{j}_1\tilde{j}_2}(\rho,{\bf z})=e^{\rho \sum_{i=1}^5}\pref_{j_1\ldots j_5\tilde{j}_1\tilde{j}_2}
\vspace{2mm} 
\\
\dps
\hspace{10mm}\times\mathcal{L}_{j_1\cdots j_5}({\bf z})\chi_4^{-\tilde{j}_1}\chi_5^{-\tilde{j}_2}\,  F_2 \left[\begin{array}{ccc}
     j_2-j_1-\tilde{j}_1,&j_3-\tilde{j}_1-\tilde{j}_2, &j_4-j_5-\tilde{j}_2  \\
     &-2\tilde{j}_1, -2\tilde{j}_2 \end{array};\chi_4,\chi_5\right],
\ea
\ee 
where the leg factor is given by 
\be 
\mathcal{L}_{j_1\cdots j_5}({\bf z}) =z_{_{12}}^{j_2+j_1}\Big(\frac{z_{_{13}}}{z_{_{23}}}\Big)^{j_1-j_2}\Big(\frac{z_{_{23}}z_{_{34}}}{z_{_{24}}}\Big)^{j_3}\Big(\frac{z_{_{34}}}{z_{_{35}}}\Big)^{j_4-j_5}z_{_{45}}^{j_4+j_5}\,.
\ee 
The 5-point global block is given in the second line of \eqref{final5pt}. 
 
\paragraph{Recursion.} On the other hand, one may observe that using 4-point and 5-point AdS vertex functions \eqref{final4pt} and  \eqref{final5pt} as well as definitions \eqref{def_hyper}  and \eqref{def_appell}  the $5$-point  function can be rewritten as a sum over $4$-point  functions with running forth external weight, 
\be
\label{4pt_double1}
\ba{c}
\dps
\cV_{j_1j_2j_3j_4 j_5\tilde{j}_1\tilde{j}_2}(\rho,{\bf z}) = 
\cV_{\tilde{j}_2j_4 j_5}(\rho,z_3, z_4, z_5)z_{_{34}}^{-2\tilde{j}_2}
\vspace{2mm} 
\\
\dps
\times\sum_{l\in L}e^{\rho(-2\tilde{j_2}+l)}\frac{\pref_{\tilde{j_1}j_3\tilde{j_2}}}{\pref_{\tilde{j_1}j_3\tilde{j_2}-l}}\cV_{j_1j_2j_3(\tilde{j}_2-l)\tilde{j}_1}(\rho,{\bf z})\left(\frac{z_{_{34}}z_{_{45}}}{z_{_{35}}}\right)^{l}\frac{(j_3-\tilde{j}_1-\tilde{j}_2)_l(-j_5+j_4-\tilde{j}_2)_l}{l!(-2\tilde{j}_2)_l},
\ea
\ee
or, equivalently, 
\be
\label{4pt_double2}
\ba{c}
\dps
\cV_{j_1j_2j_3j_4 j_5\tilde{j}_1\tilde{j}_2}(\rho,{\bf z}) =\frac{\pref_{\tilde{j_2}j_4j_5}}{\pref_{j_3j_4j_5}}
\cV_{j_3j_4 j_5}(\rho,z_3, z_4, z_5)e^{-\rho(j_3+\tilde{j_2})}
\vspace{2mm} 
\\
\dps
\times\sum_{l\in L}e^{\rho l}\frac{\pref_{\tilde{j_1}j_3\tilde{j_2}}}{\pref_{\tilde{j_1}j_3\tilde{j_2}-l}}\cV_{j_1j_2j_3(\tilde{j}_2-l)\tilde{j}_1}(\rho,{\bf z})\left(\frac{z_{_{45}}}{z_{_{35}}}\right)^{l-\tilde{j}_2+j_3}z_{_{34}}^{l-\tilde{j}_2-j_3}\frac{(j_3-\tilde{j}_1-\tilde{j}_2)_l(-j_5+j_4-\tilde{j}_2)_l}{l!(-2\tilde{j}_2)_l},
\ea
\ee
where $L:= [\![0, \min(-j_4+j_5+\tilde{j}_2,\tilde{j}_2-j_3+\tilde{j}_1)]\!]$. Of course, this is a particular case of the recursion relation of  Section \bref{sec:rec}.   
The relation \eqref{4pt_double2} is essentially that one  between $5$-point and $4$-point conformal global blocks  obtained in \cite{Rosenhaus:2018zqn} (see eq. (2.32) therein). Based on this observation, we conclude that the above direct  calculation of the 5-point block is  not really  necessary. Moreover, this opens a way to effectively  find the $n$-point conformal block just by extending the recursive relation to $n$ points. This is the subject of Section \bref{sec:n-p}.

\providecommand{\href}[2]{#2}\begingroup\raggedright\endgroup

\end{document}